\documentclass[12pt]{article}
\usepackage{amsmath,amsfonts,amssymb,graphicx,fullpage}
\usepackage[tight]{subfigure}
\usepackage{rotating}
\usepackage{ifpdf}
\ifpdf
  \usepackage{color}
  \definecolor{darkblue}{rgb}{0.3,0.3,0.6}
  \usepackage[pdftitle={Complete Thresholds for Intersecting Fractional D6-Branes},pdfauthor={Florian Gmeiner and Gabriele Honecker},pdfsubject={String theory},pdfkeywords={string theory, intersecting branes, threshold corrections},colorlinks=true,linkcolor=black,urlcolor=darkblue,filecolor=darkblue,citecolor=darkblue,menucolor=darkblue,breaklinks=true,bookmarks=true,anchorcolor=black,unicode=true]{hyperref}
\else
  \def\texorpdfstring#1#2{#1}
\fi
\usepackage[sort,compress]{cite}

\newcommand{\sgn}{{\rm sgn}}
\newcommand{\unity}{{\footnotesize\mbox{1\!\!I}}}
\def\muc{\multicolumn}

\def\bZ{\mathbb{Z}}
\def\Z{\mathbb{Z}}
\def\unity{1\!\!{\rm I}}
\def\ov{\overline}
\def\N{\mathbf{N}}
\def\Sym{\mathbf{Sym}}
\def\Anti{\mathbf{Anti}}
\def\Adj{\mathbf{Adj}}
\def\cc{c.c.}
\def\ov{\overline}
\def\1{{\bf 1}}
\def\2{{\bf 2}}
\def\3{{\bf 3}}
\def\4{{\bf 4}}
\def\6{{\bf 6}}
\def\OR{\Omega\mathcal{R}}
\def\half{\frac{1}{2}}

\def\targ#1#2{\genfrac{[}{]}{0pt}{}{#1}{#2}}
\def\half{{\textstyle\frac{1}{2}}}

\newcommand{\captionfonts}{\small}
\makeatletter
\long\def\@makecaption#1#2{%
  \vskip\abovecaptionskip
  \sbox\@tempboxa{{\captionfonts #1: #2}}%
  \ifdim \wd\@tempboxa >\hsize
    {\captionfonts #1: #2\par}
  \else
    \hbox to\hsize{\hfil\box\@tempboxa\hfil}%
  \fi
  \vskip\belowcaptionskip}
\makeatother

\def\incfig#1#2{\includegraphics[width=#2\linewidth]{#1}}
\def\fig#1#2{\begin{figure}[th]\centering\incfig{#1}{0.75}\caption{#2}\label{fig:#1}\end{figure}}
\def\wfig#1#2{\begin{figure}[th]\centering\incfig{#1}{0.99}\caption{#2}\label{fig:#1}\end{figure}}
\def\twofig#1#2#3#4{\begin{figure}[th]\centering\subfigure[\label{fig:#1}]{\incfig{#1}{0.49}}\hfill\subfigure[\label{fig:#2}]{\incfig{#2}{0.49}}\caption{#3}\label{#4}\end{figure}}

\def\mathtab#1#2#3{\begin{table}[th]\centering$#1$\caption{#3}\label{tab:#2}\end{table}}
\def\mathtabfix#1#2#3{\begin{table}[th]\centering\resizebox{\linewidth}{!}{$#1$}\caption{#3}\label{tab:#2}\end{table}}

\def\mathsidetabfix#1#2#3{\begin{sidewaystable}[H]\centering\resizebox{\linewidth}{!}{$#1$}\caption{#3}\label{tab:#2}\end{sidewaystable}}

\renewcommand{\arraystretch}{1.3}

\begin{document}
%\DeclareGraphicsExtensions{.pdf}
\begin{center}
\begin{flushright}
{\small KUL-TF-09/12\\NIKHEF/2009-001\\NSF-KITP-09-101\\October 2009}
\end{flushright}

\vspace{25mm}
{\Large\bf Complete Gauge Threshold Corrections}\\[.7ex]{\Large\bf for Intersecting Fractional D6-Branes:}\\[.7ex]{\Large\bf The $\Z_6$ and $\Z_6'$ Standard Models}

\vspace{15mm}
{\large Florian Gmeiner$^1$ and Gabriele Honecker$^2$}

\vspace{10mm}
{~$^1$\it Nikhef, Sciencepark 105, 1098 XG Amsterdam, The Netherlands, {\tt\small fgmeiner@nikhef.nl}}\\[1ex]
{~$^2$\it Institute of Theoretical Physics, K.U.Leuven, Celestijnenlaan 200D, 3001 Leuven, Belgium, {\tt\small honecker@itf.fys.kuleuven.be}}

\vspace{20mm}{\bf Abstract}\\[2ex]\parbox{140mm}{
We perform a complete analysis of one-loop threshold corrections to the gauge couplings of fractional D6-branes.
This includes besides $SU(N)$ also symplectic, orthogonal and massless Abelian gauge factors and the full computation of contributions from discrete and continuous Wilson lines and brane displacements. Two classes of globally consistent  supersymmetric compactifications with Standard Model spectra 
on $T^6/\Z_6$ and $T^6/\Z_6'$ are presented in detail with the latter exhibiting the potential of supersymmetry breaking via a hidden sector gaugino condensate. 
The  $T^6/\Z_6'$ Standard Models are completely classified, and it turns out that out of 768 distinct D6-brane configurations only 16 different sets of massless spectra and ten distinct values of gauge couplings at one-loop arise. The gauge threshold corrections enhance the diversity to 196 nonequivalent models.
 }
\end{center}
%\vfill
%\begin{flushleft}\small\tt
%fgmeiner@nikhef.nl\\
%honecker@itf.fys.kuleuven.be
%\end{flushleft}

\thispagestyle{empty}
\clearpage

\tableofcontents
%\newpage
\vspace*{1cm}\hrulefill\vspace*{1cm}

\setlength{\parskip}{1em plus1ex minus.5ex}
%%%%%%%%%%%%%%%%%%%%%%%%%%%%%%%%%%%%%%%%%%%%%%%%%%%%%%%%%%%%%%%%%%%%%%%%
\section{Introduction}\label{sec:intro}

Over the past years, many supersymmetric vacua with Standard Model gauge group on intersecting D6-branes have been
constructed, for an extensive list see the reports~\cite{Uranga:2003pz,Blumenhagen:2005mu,Blumenhagen:2006ci}\footnote{Early non-supersymmetric models
with intersecting or magnetised branes can e.g. be found
in~\cite{Bachas:1995ik,Berkooz:1996km,Blumenhagen:2000wh,Angelantonj:2000hi,Forste:2001gb,Angelantonj:2000rw,Aldazabal:2000cn,Ibanez:2001nd,Honecker:2002hp} and some early supersymmetric intersecting brane vacua are presented in~\cite{Ohta:1997fr,Blumenhagen:1999md,Pradisi:1999ii,Blumenhagen:1999ev,Forste:2000hx} with non-chiral spectra and with chiral spectra in~\cite{Cvetic:2001tj,Cvetic:2001nr,Blumenhagen:2002gw,Honecker:2003vq,Honecker:2003vw,Cvetic:2004ui,Honecker:2004kb,Cvetic:2004nk}. For a recent supersymmetric model on a smooth Calabi-Yau manifold see~\cite{Palti:2009bt}.}, and it has been argued
that gauge coupling unification can occur if the volumes of the cycles
wrapped by the Standard Model branes fulfill some relations~\cite{Blumenhagen:2003jy}.
In~\cite{Gmeiner:2008xq} however, we found that at tree level at the string
scale, the Standard Model gauge couplings of the three generation models without chiral exotics on the background
$T^6/\Z_6'$ do not fulfill this relation.
Threshold corrections to the gauge couplings can change the setting in a
favorable way, especially if the string scale is close to the Planck scale. 

At string tree level, in~\cite{Gmeiner:2008xq} we found only three different
ratios of gauge couplings at the string scale and associated chiral spectra
supporting the Standard Model. Taking into account stringy one-loop effects
takes us a step further to understanding if these vacua are indeed identical
within the string landscape or do display different phenomenological properties
and four dimensional effective actions, 
and if any pattern of correlations is visible. We will address this question by
performing a complete survey of three generation Standard Model-like vacua 
without chiral exotics in the $T^6/\Z_6'$ background.
Especially in the view of results like~\cite{Gmeiner:2005nh,Gmeiner:2005vz,Gmeiner:2008xq}
that suggest large amounts of Standard Model-like solutions the question
how many of them can actually be distinguished phenomenologically becomes
very relevant. It is to be expected that one-loop corrections will
show differences between models that look identical from the perspective
of chiral matter content and gauge groups. We will see that this is
indeed the case.

From a more technical point of view, threshold corrections provide the first
non-trivial contributions to couplings from orbifold singularities.
They furnish thus checks if K\"ahler metrics and field theoretic quantities are computed correctly
using other methods such as scattering amplitudes. Furthermore, threshold
corrections reappear as prefactors in certain instanton calculations.

Gauge threshold corrections for intersecting D6-branes have been considered
before in to\-roi\-dal
backgrounds~\cite{Lust:2003ky,Akerblom:2007np,SchmidtSommerfeld:2009gs} for
vanishing Wilson line and displacement moduli. In~\cite{Blumenhagen:2006ci},
continuous Wilson lines in the T-dual picture with D9-branes have been
included for the annulus amplitude, and
in~\cite{Blumenhagen:2007ip,Akerblom:2007np,SchmidtSommerfeld:2009gs} gauge
thresholds have been computed for rigid D6-branes at vanishing angles and
three non-trivial angles, for the T-dual formulation see
also~\cite{Angelantonj:2009yj}. In this article, we complete the
classification of gauge threshold contributions by including the annulus and M\"obius
strip amplitudes for arbitrary values of displacement and Wilson line moduli
and computing the missing amplitudes with one vanishing angle in a background
which contains a $\Z_2$ subgroup. We apply these formulae to the explicitly
known $T^6/\Z_6$ and $T^6/\Z_6'$ Standard Model-like
spectra~\cite{Honecker:2004kb,Gmeiner:2007we,Gmeiner:2007zz,Gmeiner:2008xq}
and discuss their dependence on K\"ahler and open string moduli.
We also clarify how RR tadpole cancellation is reformulated in a short and
elegant way in the gauge thresholds for orbifold backgrounds and how the missing assignments of
symplectic and orthogonal gauge groups to orientifold invariant D6-branes is
performed using their beta-function coefficients, which are obtained in the
string loop computation of gauge thresholds.

Gauge threshold corrections for D3-branes at singularities have recently been
discussed in~\cite{Conlon:2009xf,Conlon:2009kt}.
These results, as well as the gauge corrections on heterotic
orbifolds~\cite{Kaplunovsky:1995jw}, differ from our results in that the
discrete angles appearing in the amplitudes are solely due to the orbifold
rotation, whereas the amplitudes in the present context contain relative
angles among different D6-branes as well as $\Z_2$ orbifold rotations.

%\subsection{Outline}

This paper is organized as follows. In section~\ref{sec:notation} we set the
notation for fractional D6-branes in the toroidal orientifold backgrounds
$T^6/(\Z_{2N} \times \OR)$ under consideration and comment on modifications for
$T^6/(\Z_{2N} \times \Z_{2M} \times \OR)$ with discrete torsion.
We proceed by outlining the general procedure of threshold computations for
$SU(N)$ gauge groups in section~\ref{sec:Thresholds}.
In section~\ref{sec:ThreshRes} we summarize the results of the computation for
$SU(N)$ groups and treat the special cases of $Sp(2M)$, $SO(2M)$ and  $U(1)$ factors in
detail.
To illustrate our results we will apply them to some explicit examples with
Standard Model-like spectra on the orbifold backgrounds $T^6/\Z_6$ and
$T^6/\Z_6'$ in section~\ref{S:Examples}.
In section~\ref{S:Statistics} the results of a statistical analysis of the
ensemble of models with Standard Model spectrum on $T^6/\Z_6'$ are presented.
We finish with our conclusions in section~\ref{S:Conclusions} and collect
various technical details of the computations in the
appendices~\ref{App:Spectra-Constraints} to~\ref{AppS:Tables}.

%%%%%%%%%%%%%%%%%%%%%%%%%%%%%%%%%%%%%%%%%%%%%%%%%%%%%%%%%%%%%%%%%%%%%%%%5
\section{Fractional D6-branes}\label{sec:notation}

\subsection{Geometry and bulk three-cycles}

\begin{figure}[ht]
\begin{center}
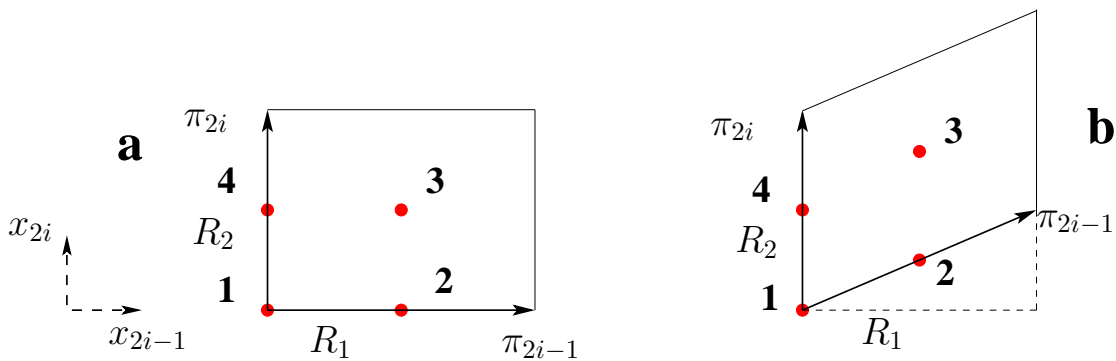
\end{center}
\caption{The two lattices $T^2_i$ which are invariant under a $\Z_2$ symmetry. In both cases, the geometric part of the orientifold action is  $\pi_{2i} \stackrel{{\cal R}}{\rightarrow} -\pi_{2i}$, and by introducing $b=0,1/2$ for the rectangular and tilted torus, respectively, 
$\pi_{2i-1}  \stackrel{{\cal R}}{\rightarrow}\pi_{2i-1} -(2 b) \,\pi_{2i}$. On the {\bf b}-type or tilted lattice, two
$\Z_2$ fixed points (1,4) are invariant under the orientifold projection, whereas the other two (2,3) are exchanged.
The tori are also invariant under a $\Z_4$ symmetry if the radii are related by $R_1=(1-b) \,R_2$.}
\label{Fig:Z2-Z4lattice}
\end{figure}

\begin{figure}[ht]
\begin{center}
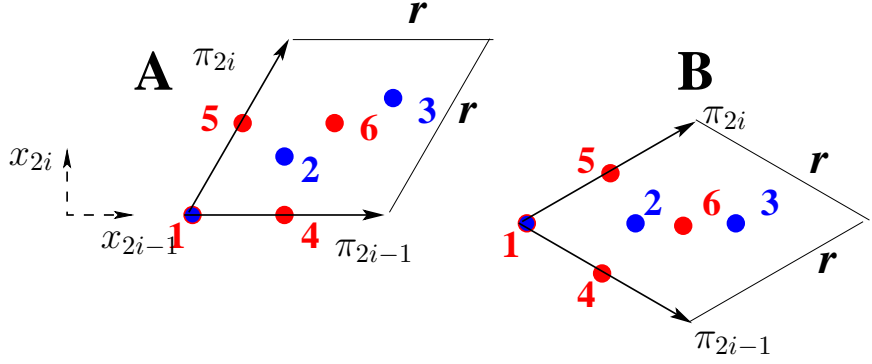
\end{center}
\caption{The two lattices $T^2_i$ which are invariant under a $\Z_3$ or $\Z_6$ rotation. The only $\Z_6$ invariant point 
is the origin. There are  two more $\Z_3$ invariant points 2 and 3 which are exchanged under the $\Z_2$ sub-symmetry,
and a triplet of $\Z_2$ fixed points labeled by 4,5,6. On the {\bf A}-type lattice, the orientifold projection acts
on the one-cycles as $(\pi_{2i-1},\pi_{2i}) \stackrel{\cal R}{\rightarrow} (\pi_{2i-1},\pi_{2i-1} -\pi_{2i})$.
On the {\bf B}-type lattice, the orientifold symmetry exchanges the basic one-cycles, $\pi_{2i-1} \stackrel{{\cal R}}{\leftrightarrow}
\pi_{2i}$. The $\Z_2$ fixed points $(4,5,6)$ are mapped under the orientifold projection to $(4,6,5)$ for the {\bf A}- and $(5,4,6)$ for the {\bf B}-type lattice, respectively.}
\label{Fig:Z3-Z6lattice}
\end{figure}

We consider D6-branes wrapped on fractional three-cycles of a $T^6/\Z_{2N}$ background\footnote{The generalization to  $T^6/\Z_{2N} \times \Z_{2M}$ orbifolds with discrete torsion is straightforward as discussed in the text.}  with a factorized six-torus, $T^6 \simeq T^2_1 \times T^2_2 \times T^2_3$.
Each two-torus $T^2_i$ must respect the orbifold as well as the orientifold symmetry. The situation for a $\Z_2$ or $\Z_4$ symmetry is depicted in figure~\ref{Fig:Z2-Z4lattice}, 
for a $\Z_3$ or $\Z_6$ symmetry see figure~\ref{Fig:Z3-Z6lattice}.

%%%%%%%%%%%%%%%%%%%%%%%%%%%%%%%%%%%%%%%%%%%%%%%%%%%%%
%   figure:
%%%%%%%%%%%%%%%%%%%%%%%%%%%%%%%%%%%%%%%%%%%%%%%%%%%%%
%%%%%%%%%%%%%%%%%%%%%%%%%%%%%%%%%%%%%%%%%%%%%%%%%%%%%

%%%%%%%%%%%%%%%%%%%%%%%%%%%%%%%%%%%%%%%%%%%%%%%%%%%%%
%   figure:
%%%%%%%%%%%%%%%%%%%%%%%%%%%%%%%%%%%%%%%%%%%%%%%%%%%%%
%%%%%%%%%%%%%%%%%%%%%%%%%%%%%%%%%%%%%%%%%%%%%%%%%%%%%

The basic one-cycles along the $i^{th}$ two-torus $T^2_i$ are labeled by $\pi_{2i-1}$ and $\pi_{2i}$. 
A factorizable  \emph{toroidal} three-cycle $a$ is fully specified by its wrapping numbers $(n_i^a,m_i^a)$ along these basic one-cycles.
The $\Z_{2N}$ symmetry acts on the complex torus coordinates, $z_i=x_{2i-1}+ i \, x_{2i}$ for $i\in\{1,2,3\}$, as
\begin{equation}\label{Eq:Z2Naction}
\theta: \, z_i \longrightarrow e^{2 \pi i v_i} z_i,
\end{equation}
with some shift vector $\vec{v}$ such that $2N \vec{v} \in \Z^3$. $\theta$
permutes the basic one-cycles on $T^2_i$ leading to orbifold image cycles $(\theta^k a)_{k=0\ldots N-1}$
for some cycle $a$ with wrapping numbers of the orbifold images $(n_i^{(\theta^k a)},m_i^{(\theta^k a)})$.
One can construct an orbifold invariant {\it bulk} three-cycle as
\begin{equation}
\begin{aligned}
\Pi^{\rm bulk}_a &= \sum_{k=0}^{N-1} \theta^k \left(\otimes_{i=1}^3 \left(n_i^a \pi_{2i-1} + m_i^a \pi_{2i}   \right) \right)
\\
&=  \sum_{k=0}^{N-1}\left(\otimes_{i=1}^3 \left(n_i^{(\theta^k a)} \pi_{2i-1} + m_i^{(\theta^k a)} \pi_{2i}   \right) 
\right). 
\end{aligned}
\end{equation}
For intersecting D6-branes on the $T^6/\Z_4$ orbifold, bulk cycles have been constructed in~\cite{Blumenhagen:2002gw}, for $T^6/\Z_6$ in~\cite{Honecker:2004kb} and for $T^6/\Z_6'$ in~\cite{Bailin:2006zf,Gmeiner:2007zz}. For the
$T^6/\Z_4 \times \Z_2$ orbifold the bulk cycles can be found in~\cite{Honecker:2003vw}.

Furthermore, the orientifold symmetry $\OR$ containing a geometric involution on the six-torus,
\begin{equation}\label{Eq:OmegaAction}
\mathcal{R}:z_i \rightarrow \overline{z}_i,
\end{equation}
acts non-trivially on the lattices and provides an image cycle $a'$.  
As detailed examples, the wrapping numbers for all image cycles in the $T^6/\Z_6$ background on the {\bf AAB} lattice are
explicitly given in section~\ref{AppSs:Z6} and for $T^6/\Z_6'$ on the {\bf ABa} lattice
in section~\ref{AppSs:Z6p}.

\subsection{Intersection numbers and fractional cycles}

The toroidal cycle intersection numbers for cycles $a$ and $b$ are given by
\begin{equation}
I_{ab} =\prod_{i=1}^3 I_{ab}^{(i)}
= \prod_{i=1}^3 \left(n^a_i m^b_i - m^a_i n^b_i   \right),
\end{equation}
and orbifold invariant combinations, i.e. bulk cycle intersection numbers, on $T^6/\Z_{2N}$ are obtained by a superposition,
\begin{equation}\label{Eq:bulk-Inters-Z2N}
\Pi^{\rm bulk}_a \circ \Pi^{\rm bulk}_b  = - 2 \sum_{k=0}^{N-1} I_{a(\theta^k b)}  
.
\end{equation}
For a $T^6/\Z_M$ orbifold with $M$ odd, the orbifold images have to be counted from $k=0$ to $M-1$, the 
factor two is absent in~\eqref{Eq:bulk-Inters-Z2N}, and the toroidal cycle intersection numbers
fully determine the (chiral plus non-chiral) massless spectrum and prefactors of the
open string annulus and M\"obius strip one-loop amplitudes.  

In the presence of two $\Z_2$ sub-symmetries such as on $T^6/\Z_{2N} \times \Z_{2M}$,
 the factor 2 in~\eqref{Eq:bulk-Inters-Z2N} is replaced by a factor of $4=2^2$ and a double sum $\sum_{k=0}^{N-1} \sum_{l=0}^{M-1}$ occurs, see e.g. the discussion in~\cite{Blumenhagen:2005tn} for the case $N=M=1$ and~\cite{Honecker:2003vw} for $(N,M)=(2,1)$.

$T^6/\Z_{2N}$ (and $T^6/\Z_{2N} \times \Z_{2M}$ with torsion) backgrounds
also have contributions from $\Z_2$  twisted sectors: 
in a $T^6/\Z_{2N}$ background, there exist also \emph{exceptional} three-cycles which 
are stuck at the $\Z_2$ fixed points of the orbifold group element $\theta^N$ along a four-torus and wrap a one-cycle
along  the remaining two-torus. For concreteness, throughout this article we choose
the $\Z_2$ subgroup as $\vec{v}_{\Z_2}=\frac{1}{2}(1,0,-1)$ which agrees with the $T^6/\Z_6'$ case discussed
in sections~\ref{Ss:Ex2} and~\ref{S:Statistics}. For any other orbifold, e.g. the
$T^6/\Z_6$ case also considered in this article in section~\ref{Ss:Ex1}, the labels of the two-tori have to be permuted in an appropriate way. 

Labeling $\Z_2$ fixed points along $T^2_1 \times T^2_3$ by $\tilde{\alpha}\tilde{\beta}$  
as depicted in figures~\ref{Fig:Z2-Z4lattice} and~\ref{Fig:Z3-Z6lattice},
the orbifold invariant exceptional three-cycles take the form
\begin{equation}
\begin{aligned}
\Pi^{\rm ex}_a &= (-1)^{\tau^0_a} \sum_{k=0}^{N-1} \theta^k \left(\sum_{\tilde{\alpha}\tilde{\beta}} 
(-1)^{\tau_{\tilde{\alpha}}^1+\tau_{\tilde{\beta}}^3} \, e_{\tilde{\alpha}\tilde{\beta}} \otimes \left(n_2^a \pi_3 + m_2^a \pi_4 \right)  \right)
\\
&= (-1)^{\tau^0_a} 
\,
\sum_{\tilde{\alpha}\tilde{\beta}}  \, 
(-1)^{\tau_{\tilde{\alpha}}^1+\tau_{\tilde{\beta}}^3} 
\sum_{k=0}^{N-1} \, 
e_{\theta^k(\tilde{\alpha})\theta^k(\tilde{\beta})} \otimes \left(n_2^{(\theta^k a)} \pi_3 + m_2^{(\theta^k a)} \pi_4 \right)
,
\end{aligned}
\end{equation}
where $\tau^0_a \in \{0,1\}$ parameterizes an overall sign (``$\Z_2$ eigenvalue''), and $\tau^i_{\tilde{\alpha}} \in \{0,1\}$ encode possible discrete Wilson lines on the two-torus $T_i^2$ where $\Z_2$ acts, and $e_{\tilde{\alpha}\tilde{\beta}}$ denote exceptional  two-cycles at the $\Z_2$ orbifold fixed points $\tilde{\alpha}\tilde{\beta}$. For more details on the assignment 
of $\Z_2$ fixed points and Wilson lines see appendix~\ref{AppSs:Z2-numbers}.
Exceptional three-cycles for intersecting D6-branes on the $T^6/\Z_4$ orbifold were constructed in~\cite{Blumenhagen:2002gw}, for $T^6/\Z_6$ in~\cite{Honecker:2004kb} and for $T^6/\Z_6'$ in~\cite{Bailin:2006zf,Gmeiner:2007zz}. 

The intersection number among exceptional three-cycles on $T^6/\Z_{2N}$ is given by 
\begin{equation}\label{Eq:Z2-Intersections}
\begin{aligned}
\Pi^{\rm ex}_a \circ \Pi^{\rm ex}_b &=  -2 \, 
 \sum_{k=0}^{N-1} 
\left[
\sum_{\tilde{\alpha}\tilde{\beta}}
(-1)^{ \tau^0_a + \tau^0_b+ \tau^1_{\tilde{\alpha}_a}+\tau^1_{\tilde{\alpha}_b} + \tau^3_{\tilde{\beta}_a}+\tau^3_{\tilde{\beta}_b}} \, 
\delta_{\tilde{\alpha}_a, (\theta^k \tilde{\alpha}_b)} \, \delta_{\tilde{\beta}_a, (\theta^k \tilde{\beta}_b)} \,  I_{a(\theta^k b)}^{(2)}
\right]
\\
& \equiv -2 \, \sum_{k=0}^{N-1} I_{a(\theta^k b)}^{\Z_2}
.
  \end{aligned}
\end{equation}
 More details on the computation of the $\Z_2$ invariant intersection numbers $I_{a(\theta^k b)}^{\Z_2}$ are displayed
  in appendix~\ref{AppSs:Z2-numbers}.  When generalizing to a $T^6/\Z_{2N} \times \Z_{2M}$ orbifold with torsion, three different twisted 
    sectors associated to the three $\Z_2$ sub-symmetries have to be taken into account, the  factor of two in~\eqref{Eq:Z2-Intersections}
    is replaced by four and a double sum occurs just like for the bulk part, and the over-all signs per twist sector will depend on the choice of torsion.

The intersection numbers for \emph{fractional} three-cycles on $T^6/\Z_{2N}$
\begin{equation}
\Pi^{\rm frac} \equiv \frac{1}{2} \left(\Pi^{\rm bulk} + \Pi^{\rm ex}  \right)
\end{equation}
are  given by a linear combination of bulk and exceptional intersections,
\begin{equation}
\Pi^{\rm frac}_a \circ\Pi^{\rm frac}_b = -
\sum_{k=0}^{N-1} \frac{  I_{a(\theta^k b)}  + I_{a(\theta^k b)}^{\Z_2}}{2}.
\end{equation}
Fractional three-cycles for intersecting D6-branes on the $T^6/\Z_4$ orbifold were constructed in~\cite{Blumenhagen:2002gw}, for $T^6/\Z_6$ in~\cite{Honecker:2004kb} and for $T^6/\Z_6'$ in~\cite{Bailin:2006zf,Gmeiner:2007zz}. 
Again, for $T^6/\Z_{2N} \times \Z_{2M}$ with discrete torsion the formula is modified by summing over three $\Z_2$ twisted sectors and dividing by four instead of two as discussed for the most simple case with $N=M=1$ in~\cite{Blumenhagen:2005tn}.

In~\cite{Gmeiner:2008xq}, we showed how to derive the complete massless spectrum from the individual intersection numbers including all special cases when the intersection number along some two-torus vanishes. Since the multiplicities of states, which are  reproduced for convenience in appendix~\ref{AppSs:Spectrum}, 
appear in the beta-function coefficients, they serve as an indicator for the correct prefactors in the computation of the threshold corrections, which are then
cross-checked by verifying the cancellation of tadpoles in the threshold computation and for the RR vacuum amplitudes.

\subsection{Consistency conditions}
A consistent string vacuum fulfills the RR tadpole cancellation condition which can be written in terms of three-cycles $\Pi_a, \Pi_{a'}$ and $\Pi_{O6}$ wrapped by D6-branes $a$, their orientifold images $a'$ and by the O6-planes,
\begin{equation}\label{Eq:RRtcc}
\sum_a N_a \left( \Pi_a + \Pi_{a'} \right) - 4 \, \Pi_{O6} =0.
\end{equation}
Since it will be necessary later on in sections~\ref{Ss:ZeroAngle} and~\ref{Ss:MS-ZeroAngle} to show the absence of tadpoles in the expressions
for one-loop threshold corrections to the 
gauge couplings, we  recall here that as a consequence of the RR tadpole condition~\eqref{Eq:RRtcc},
any model is automatically free of non-Abelian
$SU(N_a)^3$ gauge anomalies,
\begin{equation}\label{Eq:no-Anomaly}
\begin{aligned}
0 &= \Pi_a  \circ \left(\sum_b N_b \left( \Pi_b + \Pi_{b'} \right) - 4 \,  \Pi_{O6}\right) 
\\
&=
-\frac{1}{2} \left(\sum_b N_b \sum_{k=0}^{N-1}\left( I_{a(\theta^k b)}+ I_{a(\theta^k b')}\right)- 4 \, \sum_{k=0}^{2N-1} \tilde{I}_a^{\OR\theta^{-k}}  \right)\\
&\phantom{=}
-\frac{1}{2} \sum_b N_b  \sum_{k=0}^{N-1} \left( I_{a(\theta^k b)}^{\Z_2}+ I_{a(\theta^k b')}^{\Z_2}\right) ,
\end{aligned}
\end{equation}
where the bulk and the exceptional sums in the last two lines vanish separately.
Similarly, when replacing one one-cycle intersection number $I^{(i)}$ by a symmetric expression $V^{(i)}$   per term
where 
\begin{equation}\label{Eq:Def-Vab}
V_{ab} =\left\{\begin{array}{cc}
\frac{R_1}{R_2} n^an^b + \frac{R_2}{R_1} (m^a+b n^a) (m^b + b n^b)  & {\bf a} \; {\rm or }  \;  {\bf b}-{\rm torus}
\\
\frac{1}{\sqrt{3}}( 2 n^an^b + n^am^b+m^an^b+2 m^am^b)& {\bf A}\;  {\rm or }\;  {\bf B}-{\rm torus}
\end{array}
\right.
,
\end{equation}
and defining
\begin{equation}
\begin{aligned}
\kappa_{ab} &\equiv \sum_{k=0}^{N-1} \left( I_{a(\theta^k b)}^{(1\cdot 2)} \, V_{a(\theta^k b)}^{(3)} 
+ I_{a(\theta^k b)}^{(1\cdot 3)} \, V_{a(\theta^k b)}^{(2)} 
+ I_{a(\theta^k b)}^{(2\cdot 3)} \, V_{a(\theta^k b)}^{(1)} 
\right),
\\
\tilde{\kappa}_{a,\OR} & \equiv \sum_{k=0}^{2N-1} \left(\tilde{ I}_a^{\OR\theta^{-k},(1\cdot 2)} \, \tilde{V}_{a,\OR\theta^{-k}}^{(3)} 
+ \tilde{I}_a^{\OR\theta^{-k},(1\cdot 3)} \, \tilde{V}_{a,\OR\theta^{-k}}^{(2)} 
+ \tilde{I}_a^{\OR\theta^{-k},(2\cdot 3)} \, \tilde{V}_{a,\OR\theta^{-k}}^{(1)} 
\right),
\\
\lambda_{ab} &\equiv\sum_{k=0}^{N-1} \, I_{a(\theta^k b)}^{\Z_2,(1\cdot 3)} \, V^{(2)}_{a(\theta^k b)}
, 
\end{aligned}
\end{equation}
the following sums vanish upon RR tadpole cancellation~\eqref{Eq:RRtcc},
\begin{equation}\label{Eq:RRtcc-rewritten}
\begin{aligned}
0 &= \sum_{b} N_b \left( \kappa_{ab} +\kappa_{ab'}\right) - 4 \, \tilde{\kappa}_{a,\OR},
\\
0&= \sum_{b} N_b \left( \lambda_{ab} + \lambda_{ab'} \right)
.
\end{aligned}
\end{equation}
More details on the relation between the RR tadpole cancellation, anomaly condition and the latter expressions are given
in appendix~\ref{AppSs:RRtcc-rewrite}.

For $a=b$ along some two-torus $T^2$, $V_{ab}$ as defined in~\eqref{Eq:Def-Vab} measures the length $L_a$ of a one-cycle in units of $\alpha'$,
\begin{equation}\label{Eq:Def-Vaa}
V_{aa} = \frac{(L_a)^2}{{\rm Vol}(T^2)}
\quad
{\rm with}
\quad 
v \equiv \frac{{\rm Vol}(T^2)}{\alpha'}  =
\left\{\begin{array}{cc}
\frac{R_1R_2}{\alpha'} & {\bf a} \; {\rm or} \; {\bf b}-{\rm torus}
\\
\frac{\sqrt{3}}{2} \frac{r^2}{\alpha'} & {\bf A} \;  {\rm or} \; {\bf B}-{\rm torus}
\end{array}\right. 
.
\end{equation}
The expressions involving some $\OR\theta^{-k}$ invariant plane are analogous to those for the D6-branes where a quantity with a tilde contains
the number $N_{O6_i} = 2(1-b_i)$ of identical O6-planes per two-torus $T^2_i$,
\begin{equation}\label{Eq:Def-tildes}
\tilde{I}_a^{\OR\theta^{-k},(i)} \equiv 2 (1-b_i) I_a^{\OR\theta^{-k},(i)} ,
 \qquad
\tilde{V}_{a,\OR\theta^{-k}} \equiv 2 (1-b_i) V_{a,\OR\theta^{-k}}^{(i)} .
\end{equation}
Finally, the computation of untwisted RR tadpole cancellation in terms of tree level amplitudes takes the form
\begin{equation}\label{Eq:RRtcc-2nd-rewrite}
\begin{aligned}
&\sum_{a,b} N_a N_b \sum_{k=0}^{N-1} \left(  \prod_{i=1}^3 V_{a(\theta^k b)}^{(i)} + \prod_{i=1}^3 V_{a(\theta^k b')}^{(i)}\right)\\
&\qquad -4 \sum_a N_a \sum_{k=}^{2N-1}  \prod_{i=1}^3 \tilde{V}_{a,\OR\theta^{-k}}^{(i)}
+ 8 \, \sum_{k=0}^{2N-1}  \prod_{i=1}^3 \tilde{V}_{\OR,\OR\theta^{-k}}^{(i)}
=0
,
\end{aligned}
\end{equation}
where more details can be found in Tables~\ref{tab:tadosc1} and~\ref{tab:tadosc2} in the appendix.
The twisted part is the same as the second line in~\eqref{Eq:RRtcc-rewritten}.

In order to generalize the above equations to the $T^6/\Z_{2N} \times \Z_{2M}$ cases with discrete torsion, three independent types of twisted contributions, 
double sums, different normalizations and signs of exotic O6-planes appearing in models with torsion have to be taken care of.

A generic D6-brane configuration must fulfill a second topological condition besides the RR tadpole cancellation~\eqref{Eq:RRtcc}, the K-theory constraint.
For compactifications on smooth Calabi-Yau manifolds, the constraint is equivalent to the probe brane argument~\cite{Uranga:2000xp}, which states that for any probe D6-brane
carrying an $Sp(2)$-gauge factor, the following condition must hold
\begin{equation}\label{Eq:K-theory}
\sum_a N_a \Pi_a \circ \Pi_{\rm probe}  \in 2 \, \Z.
\end{equation}
For orbifold backgrounds $T^6/\Z_{2N}$, this condition is necessary, but at present it is unclear if it is also sufficient. 
Since it has been shown in~\cite{Gmeiner:2007we} for $T^6/\Z_6$ and in~\cite{Gmeiner:2007zz} for $T^6/\Z_{6'}$ that the probe brane constraint~\eqref{Eq:K-theory} is fulfilled for any solution to the 
RR tadpole cancellation conditions, we will not discuss this second topological constraint further in this article.

A \emph{toroidal} three-cycle is ${\cal N}=1$ supersymmetric if the sum over the three angles $\pi \, \phi^{(i)}$ 
with respect to the $\OR$ invariant axis along the two-tori $T^2_i$ is zero,
\begin{equation}
\sum_{i=1}^3 \phi^{(i)}_{a} = 0 \quad {\rm mod} \quad 2,
\end{equation}
and ${\cal N}=2$ supersymmetric sectors arise if one of the angles vanishes.
A generic \emph{fractional} three-cycle preserve ${\cal N}=1$ supersymmetry if the bulk part is supersymmetric and the 
exceptional cycles stem only from the four $\Z_2$ fixed points traversed by the toroidal cycle with only three independent
signs corresponding to the $\Z_2$ eigenvalue and two discrete Wilson lines. More details on these assignments can be found in
appendix~\ref{AppSs:Z2-numbers}. ${\cal N}=2$ supersymmetry on \emph{fractional} branes arises only if the angle on the $\Z_2$ invariant two-torus vanishes, 
vanishing  angles on the four-torus where $\Z_2$ acts cannot lead to any supersymmetry enhancement.
In the case of a $T^6/\Z_{2N} \times \Z_{2M}$ background, all sectors are at most ${\cal N}=1$ supersymmetric, and $2^3=8$ fixed points with 
three choices of discrete Wilson lines and two independent $\Z_2$ eigenvalues appear for a given choice  of discrete torsion.

When the toroidal three-cycles are given in terms of wrapping numbers, the angles with respect to the $\OR$ invariant $x_{2i-1}$-axis along $T^2_i$ are obtained from
\begin{equation}
\tan \left( \pi \phi^{(i)}_{a}\right) = \left\{\begin{array}{cc}
\frac{m_i + b_i n_i}{n_i} \frac{R_2}{R_1} & {\bf a} \; {\rm or} \; {\bf b}-{\rm torus}
\\
\sqrt{3} \frac{m_i}{2 n_i + m_i} & {\bf A}
\\
\frac{1}{\sqrt{3}} \frac{m_i-n_i}{m_i+n_i} & {\bf B}
\end{array}\right.
,
\end{equation}
leading to the relation
\begin{equation}\label{Eq:Rel-I-V-Angle}
I^{(i)}_{ab} \, \cot \left( \pi \phi^{(i)}_{ab} \right)= V^{(i)}_{ab} ,
\end{equation}
which will be needed later on in section~\ref{Ss:ZeroAngle}.
Here, we used the fact that all relative angles among different three-cycles are obtained from the ones relative to the $\OR$-invariant $x_{2i-1}$-axis,
\begin{equation}
\phi^{(i)}_{ab} \equiv \phi^{(i)}_b - \phi^{(i)}_a .
\end{equation}
The angles appearing in the M\"obius strip amplitudes are defined in a similar way,
\begin{equation}
\phi^{(i)}_{a,\OR\theta^{-k}} \equiv \phi^{(i)}_{\OR\theta^{-k}} - \phi^{(i)}_a .
\end{equation}
Observe that since $\phi^{(i)}_{\OR}=0$ we have $\phi^{(i)}_{a,\OR} = - \phi^{(i)}_a$ with a minus sign.

%%%%%%%%%%%%%%%%%%%%%%%%%%%%%%%%%%%%%%%%%%%%%%%%%%%%%%%%%%%%%%%%%%%%%%%%
\section{Threshold corrections to the gauge couplings}\label{sec:Thresholds}

The gauge couplings of an $SU(N_a)$ gauge factor at energy scale $\mu$ are up to one-loop given by
\begin{equation}\label{Eq:Def-gauge}
\frac{8 \pi^2}{g_a^2(\mu)} = \frac{8 \pi^2}{g_{a,{\rm string}}^2}
+\frac{b_a}{2} \ln\left(\frac{M_{\rm string}^2}{\mu^2}  \right) + \frac{\Delta_a}{2},
\end{equation}
where the tree level value for a non-Abelian gauge group supported on a stack of D$6_a$-branes
 is obtained from the dimensionless length of the wrapped three-cycle,
\begin{equation}\label{eq:tree-gauge-value}
\frac{1}{\alpha_{a,{\rm string}}} \equiv \frac{4\pi}{g^2_{a,{\rm string}}} = \frac{M_{\rm Planck}}{2 \sqrt{2} k_a M_{\rm string} }
\frac{\prod_{i=1}^3 \sqrt{V_{aa}^{(i)}}}{c},
\end{equation}
with $V_{aa}^{(i)}$ defined in~\eqref{Eq:Def-Vab}, $c=1$ for toroidal D6-branes and $c=2$ for the fractional D6-branes on $T^6/\Z_{2N}$ considered in this article ($c=4$ for rigid D6-branes on $T^6/\Z_{2N} \times \Z_{2M}$ such as e.g. for $N=M=1$ in~\cite{Blumenhagen:2005tn}) and $k_a=1$ for $SU(N_a)$ gauge factors
(for $SO(2N_a)$ and $Sp(2N_a)$ gauge groups one has $k_a=2$ since the orientifold images of the corresponding branes are not counted separately).\footnote{In order to obtain the canonical formulation in terms of ${\cal N}=1$ supergravity
theory and extract the moduli dependence of the gauge kinetic function, it is necessary to redefine the dilaton and complex structure moduli appropriately
as e.g. done for the $T^6/\Z_2 \times \Z_2$ backgrounds in~\cite{Akerblom:2007uc,Blumenhagen:2007ip} and for magnetized branes in~\cite{Bertolini:2005qh,Billo:2007sw,Billo:2007py}. For our purpose of obtaining concrete values for the gauge couplings, however, it is sufficient
to work with the original string theoretic quantities.}

The one-loop running due to \emph{massless} open string states is described by the beta function coefficients 
 which were derived in~\cite{Gmeiner:2008xq} by field theory considerations to be of the form
\begin{equation}\label{Eq:beta-SU(N)}
b_{SU(N_a)} =
-  N_a \left( 3 - \varphi^{\Adj_a}\right) +\sum_{b\neq a} \frac{N_b}{2} \left( \varphi^{ab} + \varphi^{ab'}\right)  
+ \frac{N_a-2 }{2} \, \varphi^{\Anti_a} + \frac{N_a+2}{2} \, \varphi^{\Sym_a}
\end{equation}
for  an $SU(N_a)$ gauge group factor. 
The multiplicities $\varphi$ of the various allowed representations  are discussed in detail in appendix~\ref{AppSs:Spectrum}.

For symplectic gauge factors $Sp(2M_x)$ of rank $M_x$, the one-loop running is determined by
\begin{equation}\label{Eq:beta-Sp(2M)}
b_{Sp(2M_x)} = (M_x + 1) \left(-3 + \varphi^{\Sym_x}\right) + (M_x -1)  \, \varphi^{\Anti_x} + \sum_{a \neq x} \frac{N_a}{2}  \varphi^{ax}
\end{equation}
as discussed in more detail in appendix~\ref{App:SO-Sp}. For $Sp(2)_x \simeq SU(2)_a$, formula~\eqref{Eq:beta-Sp(2M)} with $M_x=1$ 
matches~\eqref{Eq:beta-SU(N)} with $N_a=2$ as required.

Although $SO(2M_y)$ gauge groups do not appear in the explicitly treated examples in this article, for completeness we note that their
field theoretically derived beta function coefficients take the form
\begin{equation}\label{Eq:beta-SO(2M)}
b_{SO(2M_y)} = (M_y - 1) \left(-3 + \varphi^{\Anti_y}\right) + (M_y+1)  \, \varphi^{\Sym_x} + \sum_{a \neq y} \frac{N_a}{2}  \varphi^{ay}
.
\end{equation}

For Abelian  $U(1)_a$ factors inside a $U(N_a)$ gauge group, the beta function coefficients take the form
\begin{equation}\label{Eq:beta-U(1)a}
b_{U(1)_a} =
N_a \left(\sum_{b\neq a}  N_b \left( \varphi^{ab} + \varphi^{ab'}\right) 
+ 2  \, (N_a +1) \, \varphi^{\Sym_a} + 2 \, (N_a -1) \, \varphi^{\Anti_a} \right)
,
\end{equation}
and for a massless $U(1)_X$ gauge group defined by
\begin{equation}\label{Eq:Def-U(1)-massless}
  U(1)_X= \sum_i x_i U(1)_i
\end{equation}
with some numerical coefficients $x_i$,
the beta function coefficient is given by
\begin{equation}\label{Eq:beta-U(1)X}
b_{U(1)_X} =
\sum_i  x_i^2 \, b_{U(1)_i} + 2 \, \sum_{i < j} N_i N_j x_i x_j \left(-\varphi^{ij} + \varphi^{ij'}\right)
.
\end{equation}

\emph{Massive} string states contribute to the gauge threshold correction $\Delta_a$. 
The CFT computation below will take into account \emph{all}, massless and massive, string excitations,
\begin{equation}\label{Eq:b-Delta}
b_a + \Delta_a = \sum_b \left(\mathcal{T}^A(D6_a,D6_b) + \mathcal{T}^A(D6_a,D6_{b'})   \right) + \mathcal{T}^M (D6_a,O6),
\end{equation}
where $\mathcal{T}^A$ and $\mathcal{T}^M$ denote the threshold amplitudes with annulus and M\"obius strip topology, respectively, and 
the sum runs over all D6-branes in the model. For orbifold actions other than $\Z_2$, each amplitude decomposes into a sum over orbifold 
images just as the multiplicities $\varphi$ of representations in table~\ref{NonChiralSpectrum} do. For example, on a $T^6/\Z_{2N}$ background
the annulus contribution to the gauge threshold of $SU(N_a)$ from strings stretching between branes $a$ and $b$ is
\begin{equation}
\begin{aligned}
\mathcal{T}^A(D6_a,D6_b) =&  \sum_{k=0}^{N-1}  \left(  T^{A,\unity}_{a(\theta^k b)} + T^{A,\Z_2}_{a(\theta^k b)} \right)
,\qquad
\mathcal{T}^M(D6_a,O6) =&  \sum_{k=0}^{2N-1} T^M_a
,
\end{aligned}
\end{equation}
where the first term descends from the torus and the second term arises at $\Z_2$ fixed points.
The  individual contributions $T^{A,{\rm insertion}}_{a(\theta^k b)}$ are discussed below.

%%%%%%%%%%%%%%%%%%%%%
\subsection{Background field method and known results}\label{Ss:background}

In~\cite{Lust:2003ky,Akerblom:2007np}, threshold corrections for \emph{toroidal} intersecting D6-branes {\it without}  (continuous) Wilson lines and displacements were computed using the method
of introducing a magnetic background field $B$ along the non-compact directions and expanding the oscillator contributions to the closed string tree channel amplitudes to second order in $B$.
The resulting non-compact oscillator contributions to the amplitudes for $SU(N_a)$ gauge factors are of the form
\begin{equation}\label{Eq:B-expansion}
\begin{aligned}
&\pi q_a \frac{\partial^2}{\partial B^2} B \frac{\vartheta\targ{\alpha}{\beta}}
{\vartheta\targ{1/2}{1/2}}( \frac{ {\rm arctan} ( \pi q_a B) }{\pi}
%\epsilon_{ab}
,\tau)\Big|_{B=0}\\
=&
-\pi^2 q_a^2 \left\{\left(\frac{1}{3} + \frac{1}{6}E_2(\tau)\right)
\frac{\vartheta\targ{\alpha}{\beta} (0,\tau)}{\eta^3(\tau)} 
+
\frac{1}{2\pi^2}
\frac{\vartheta^{\prime\prime}\targ{\alpha}{\beta} (0,\tau)}{\eta^3(\tau)} 
\right\}
,
\end{aligned}
\end{equation}
where primes denote derivatives with respect to the first argument of the Jacobi theta functions, $E_2$ is an Eisenstein series and $\tau=2 il$ for the annulus and $2il -\frac{1}{2}$ for the M\"obius strip. $q_a$ denotes the charge of an open string endpoint, and if contributions from $aa'$ strings are computed, one has to replace $q_a^2 \rightarrow q_a^2 + q_{a'}^2$.  $\alpha \in \{0,1/2\}$ labels the NS-NS and RR sector contributions, $\beta \in \{0,1/2\}$ the 
$\unity$ and $(-1)^F$ insertions of the GSO projector.
Our conventions on Jacobi theta functions as well as useful identities are summarized in appendix~\ref{AppSs:Thetas}.
As discussed below, for supersymmetric configurations the first term in the last line of~\eqref{Eq:B-expansion} vanishes when summing over $\alpha$ and $\beta$, and only the second term contributes non-trivially to 
the computation of threshold corrections to the gauge couplings.

A discussion of the most simple orbifold background,  $T^6/\Z_2 \times \Z_2'$, admitting \emph{rigid} branes  for three non-vanishing angles or all angles vanishing, but not one zero-angle, appears in~\cite{Blumenhagen:2007ip}.

For {\it anomalous} Abelian gauge factors, besides~\eqref{Eq:B-expansion}, there is  a second kind of contribution~\cite{Lust:2003ky} 
 \begin{equation}
 -\pi^2 q_a q_{a'}  \left\{\frac{1}{3}\left[1 - E_2(\tau)\right]
\frac{\vartheta\targ{\alpha}{\beta} (0,\tau)}{\eta^3(\tau)} 
-
\frac{1}{\pi^2}
\frac{\vartheta^{\prime\prime}\targ{\alpha}{\beta} (0,\tau)}{\eta^3(\tau)} 
\right\}
,
 \end{equation}
where again only the last term can give a non-vanishing contribution, which is proportional to the universally present term in~\eqref{Eq:B-expansion},
when the sum over spin structures in a supersymmetric set-up is performed.

The vacuum annulus and M\"obius strip diagrams from which the RR tadpole cancellation conditions and NS-NS tadpoles arise 
are in the tree channel of the form 
 \begin{equation}
 \begin{aligned}
{\cal A}(D6_a,D6_b) \sim & \int_{0}^{\infty} dl \sum_{(\alpha,\beta)} (-1)^{2(\alpha+\beta)} 
\frac{\vartheta\targ{\alpha}{\beta}(0,2il)}{\eta^3(2il)}
 A^{\rm insertion}_{\rm compact} (\alpha,\beta;\{\phi^{(i)}\};2il)
 ,
\\
 {\cal M} (D6_a,O6) \sim& \int_{0}^{\infty} dl \sum_{(\alpha,\beta)} (-1)^{2(\alpha+\beta)} 
\frac{\vartheta\targ{\alpha}{\beta}
(0,2il-\frac{1}{2})}{\eta^3(2il-\frac{1}{2})} M^{\rm insertion}_{\rm compact} (\alpha,\beta;\{\phi^{(i)}\};2il-\half)
,
\end{aligned}
\end{equation}
where $\alpha,\beta \in \{ 0,1/2\} $ label the different spin structures, $\{\phi^{(i)}\}=\{\phi^{(1)},\phi^{(2)},\phi^{(3)}\}$ are the angles on the three two-tori, and
$A^{\rm insertion}_{\rm compact} $ and $M^{\rm insertion}_{\rm compact}$ are
the contributions from the three complex compact dimensions. Since $\unity$ and the $\Z_2$ subgroup preserve any D6-brane configuration, these are the two 
non-vanishing possible insertions in the loop channel annulus amplitude. For the M\"obius strip, an open string with endpoints on brane  $a$ and its orientifold image $(\theta^k a')$
is invariant under the loop channel insertions $\OR\theta^{-k}$ and $\OR\theta^{-k+N}$. For a supersymmetric brane configuration, $\sum_{i=1}^3 \phi^{(i)} =0$ on toroidal orbifolds,  the sum over all spin structures of the vacuum amplitudes vanishes.

Since the first term in~\eqref{Eq:B-expansion} is proportional to the non-compact part of the vacuum amplitude,  the sum over spin structures in the threshold computation involving this expression  vanishes upon supersymmetry, and only
the non-compact contributions containing second derivatives of some Jacobi theta functions need to be considered.
Furthermore, $\vartheta\targ{1/2}{1/2} (0,\tau) \equiv-\vartheta_1(0,\tau)=0$
and %$\partial_{\nu}^2
$\vartheta^{\prime\prime}\targ{1/2}{1/2} (0,\tau) \equiv- \vartheta_1^{\prime\prime}(0,\tau)=0$,
and it suffices again to compute the sum over three spin structures \mbox{$(\alpha,\beta)\in\{(0,0),(1/2,0),(0,1/2)\}$}.

In the following, we discuss the contributions per two-torus to the vacuum and threshold amplitudes for the cases with vanishing and non-vanishing angles
as well as the role of the $\Z_2$ loop channel insertion.  Special attention goes to the so far neglected cases of continuous Wilson lines and distances since 
the latter can render ${\cal N}=2$ matter multiplets massive.

%%%%%%%%%%%%%%%%%%%%%
\subsection{Contributions to the annulus amplitudes per two-torus}

There exist four different possible types of contributions per two-torus $T^2_i$ to the annulus amplitudes:
\begin{enumerate}
\item[[i\!]]
the two branes $a$ and $b$ under consideration are parallel and the loop channel amplitude has a $\unity$ insertion,
\item[[ii\!]]
the branes are parallel and the loop channel amplitude has a non-trivially acting 
$\Z_2$ insertion,
\item[[iii\!]]
the branes intersect at some non-trivial angle $\phi^{(i)}$ and the loop channel insertion is $\unity$,
\item[[iv\!]]
the branes intersect at some non-trivial angle $\phi^{(i)}$ and the amplitude has a $\Z_2$ insertion 
in the loop channel which acts non-trivially on $T^2_i$.
\end{enumerate}
Cases [i] and [iii] appear for all kinds of bulk, fractional or rigid branes,  see e.g. ~\cite{Lust:2003ky,Akerblom:2007np,Blumenhagen:2007ip}, whereas [ii] and [iv] only appear for fractional or rigid branes, see e.g.~\cite{Blumenhagen:2007ip}. Our special attention goes to branes with continuous Wilson lines or parallel displacements as well as to the case with one vanishing angle which have  to our knowledge not been studied in detail before in the literature.

Up to some constant prefactors which will be determined later, the various oscillator and lattice contributions per two-torus in the tree channel are 
\begin{equation}\label{Eq:Annulus-per-T2}
\begin{aligned}
\mbox{[i]}\qquad & V_{ab}^{(i)} \, 
\tilde{\mathcal L}_{\mathcal{A},ab}^{(i)}(l) \, 
\frac{\vartheta\targ{\alpha}{\beta} (0,2il)}{\eta^3 (2il)}
,
\\
\mbox{[ii]} \qquad & \delta_{\sigma_{ab}^i,0} \, \delta_{\tau_{ab}^i,0} \, 
\frac{\vartheta\targ{ \alpha + 1/2 }{ \beta }}
{\vartheta\targ{ 0 }{ 1/2}}(0,2il)
,
\\
\mbox{[iii]} \qquad & 
I_{ab}^{(i)} \frac{\vartheta\targ{ \alpha }{ \beta }}
{\vartheta \targ{ 1/2 }{ 1/2 }} (\phi_{ab}^{(i)},2il)
,
\\
\mbox{[iv]}\qquad &
 I_{ab}^{\Z_2,(i)} \frac{\vartheta\targ{ \alpha+1/2  }{ \beta }}
{\vartheta \targ{ 0 }{  1/2 }} (\phi_{ab}^{(i)},2il)
.
\end{aligned}
\end{equation}
The lattice contributions $\tilde{\mathcal L}_{ab}^{(i)}(l)$ in [i]  will be discussed in some detail in 
section~\ref{Ss:Lattice}. The contribution [ii] is only present if there is no relative Wilson line $\tau_{ab}^i=|\tau_a^i-\tau_b^i| \in [0,1]$ since this amounts
to having a relative minus sign between two otherwise identical contributions from the two $\Z_2$ invariant points
traversed by the one-cycle. Similarly, a relative distance $\sigma_{ab}^i=|\sigma_a^i-\sigma_b^i| \in [0,1]$
among two branes implies that they do not pass through the same 
$\Z_2$ fixed points, and the corresponding amplitude vanishes.
For case [iv], one has to distinguish the case $(n^a_i,m^a_i) = (n^b_i,m^b_i)$ mod 2 or not. In the first case,
the reasoning is the same as for vanishing angles, $ I_{ab}^{\Z_2,(i)} =2 \, \delta_{\sigma_{ab}^i,0} \, \delta_{\tau_{ab}^i,0}$, whereas otherwise two one-cycles intersect in exactly one $\Z_2$ invariant point such that $I_{ab}^{\Z_2,(i)}= \pm 1$. The derivation of the correct signs and some more details on the counting of 
$\Z_2$ invariant intersection numbers are given in appendix~\ref{AppSs:Z2-numbers}.

In the following section, we will discuss the lattice contributions $\tilde{\mathcal L}_{ab}^{(i)}(l)$ in some detail. In section~\ref{Ss:ZeroAngle}, the complete  contribution of an annulus amplitude with D6-branes at one vanishing angle, i.e. $(\phi,-\phi,0)$ or some permutation of tori, will be discussed in detail and the missing combinatorial prefactor computed by comparison with the field theory result for  the beta function coefficient. 
The remaining amplitudes are listed in detail in appendix~\ref{AppS:Tables} in tables~\ref{tab:Annulus-Amplitudes-thresholds} and~\ref{tab:Mobius-Amplitudes-thresholds}, and their contributions to the beta function coefficients and gauge thresholds is given in table~\ref{tab:Amplitudes-thresholds}, where also the vanishing of tadpoles in the gauge threshold amplitudes upon summation over all D6-branes is demonstrated.

%%%%%%
\subsubsection{Computation of lattice contributions for continuous Wilson lines and displacements}\label{Ss:Lattice}

The computations of gauge thresholds for \emph{bulk} branes in~\cite{Lust:2003ky,Akerblom:2007np} have been performed for vanishing distances and 
Wilson lines. In~\cite{Blumenhagen:2007ip}, for the case of \emph{rigid} branes, \emph{discrete} distances and Wilson lines were taken into account. We will now discuss the case with \emph{continuous} distances and Wilson lines which can appear both for \emph{bulk} and \emph{fractional} branes, but not \emph{rigid} ones.

Open strings stretching between stacks of branes parallel along a two-torus
with relative distance $\sigma_{ab} = |\sigma_a - \sigma_b | \in [0,1]$ and 
relative Wilson line $\tau_{ab} = |\tau_a - \tau_b| \in [0,1]$ have masses proportional to
\begin{equation}\label{Eq:Open-Massformula}
M^2_{mn}(a,b) = \frac{1}{V_{ab}}\left[\frac{1}{v} \left(m+\frac{\tau_{ab}}{2} \right)^2   + v \left(n+\frac{\sigma_{ab}}{2}\right)^2\right]
,
\end{equation}
where $m$ and $n$ are momentum and winding numbers, respectively,
$V_{ab}$  the dimensionless (length)${}^2$ of the one-cycle where the
branes are parallel, as defined in equation~\eqref{Eq:Def-Vab},
and the volume $v$ of the corresponding two-torus in units of $\alpha'$ is
given in~\eqref{Eq:Def-Vaa} for the various shapes of the lattice.
The index $(i)$ for the $i^{th}$ two-torus will be suppressed throughout the whole section.

In the loop channel, the annulus lattice sum per two-torus is given by
\begin{equation}\label{Eq:lattice-Annulus-open}
\mathcal{L}^{\mathcal{A}}_{ab} (t) = \sum_{m,n \in \Z} \exp \left(-2 \pi t \, M^2_{mn} (a,b)  \right),
\end{equation}
for which a modular transformation to the tree channel is performed by means of the Poisson resummation formula,
\begin{equation}
\sum_{k \in \Z} e^{-\pi x (k+\frac{y}{2})^2} = \frac{1}{\sqrt{x}} \sum_{n\in \Z} e^{-\frac{\pi}{x} n^2 + \pi i n y},
\end{equation}
with  $(x,y)=(\frac{2t}{vV_{ab}},\tau_{ab})$ and $(\frac{2t v}{V_{ab}},\sigma_{ab})$ and 
the modular transformation parameter for the annulus $t=\frac{1}{2l}$ .

The tree channel annulus lattice sum per two-torus takes thus the form
\begin{equation}\label{Eq:Tree-Lattice-Sum}
\tilde{\mathcal L}^{\mathcal{A}}_{ab}(l) =
\sum_{m,n \in \Z} \exp \left\{ -\pi l  \, V_{ab} \left( v \,  m^2 
+ \frac{1}{v} \, n^2 \right)
+ \pi i \left(m \, \tau_{ab} + n \, \sigma_{ab} \right) 
\right\}
.
\end{equation}

For vanishing relative distances $\sigma_{ab} = 0 $ and no relative Wilson line 
$\tau_{ab} = 0$, this expression agrees with previous computations, see e.g. table 4 in~\cite{Forste:2001gb}. 

In order to integrate the lattice sum, it is convenient to decompose~\eqref{Eq:Tree-Lattice-Sum} into a constant term 
and three sums,
\begin{equation}\label{Eq:Tree-Lattice-Sum-2}
\begin{aligned}
\tilde{\mathcal L}^{\mathcal{A}}_{ab}(l) & \equiv 1 + L_{\mathcal{A}}(V_{ab},\tau_{ab},\sigma_{ab},v;l)
\\
& = 1 + 2 \, \sum_{m=1}^{\infty} \cos (\pi m \tau_{ab} ) e^{-\pi l \, V_{ab} v \,  m^2}
+ 2 \,  \sum_{n=1}^{\infty} \cos (\pi n \sigma_{ab} ) e^{-\pi l \, V_{ab}  n^2/v}
\\
& \quad\quad
+ 4 \, \sum_{m,n=1}^{\infty} \cos (\pi m \tau_{ab} ) \cos (\pi n \sigma_{ab} ) e^{-\pi l \,  V_{ab} (v \, m^2 +n^2/v)}
.
\end{aligned}
\end{equation}
The constant term contributes to the tadpole, and the three sums to the gauge threshold corrections. For 
$\tau_{ab}=\sigma_{ab}=0$, they also contribute to the beta function coefficients. 

As discussed in the following section~\ref{Ss:ZeroAngle}, the computation of the threshold amplitudes necessitates
to integrate the lattice sums~\eqref{Eq:Tree-Lattice-Sum-2}.
Using dimensional regularization as in~\cite{Akerblom:2007np}, these  integrals are of the form
\begin{equation}
\int_0^{\infty} dl \, l^{\varepsilon} e^{-l \cdot A} = \frac{\Gamma(1+\varepsilon)}{A^{1+\varepsilon}}
\qquad
{\rm with}
\qquad 
A = \pi V_{ab} \times \left\{\begin{array}{c}
v \, m^2 \\ n^2 / v \\ v \, m^2 + n^2 / v
\end{array}
\right.
,
\end{equation}
resulting in 
\begin{equation}\label{Eq:integrated-sums}
\begin{aligned}
 V_{ab}
\int_0^{\infty}dl l^{\varepsilon} L_{\mathcal{A}}(V_{ab},\tau_{ab},\sigma_{ab},v;l)
& = \frac{\Gamma(1+\varepsilon)}{(\pi V_{ab})^{\varepsilon}}
\left(\frac{1}{v^{1+\varepsilon}} \frac{2}{\pi} \sum_{m=1}^{\infty} \frac{\cos(\pi m \tau_{ab})}{m^{2+2\varepsilon}}\right.\\
& \phantom{= \frac{\Gamma(1+\varepsilon)}{(\pi V_{ab}^{(i)})^{\varepsilon}}}\;
+ v^{1+\varepsilon} \,  \frac{2}{\pi} \sum_{n=1}^{\infty} \frac{\cos(\pi n \sigma_{ab})}{n^{2+2\varepsilon}}\\
& \phantom{= \frac{\Gamma(1+\varepsilon)}{(\pi V_{ab}^{(i)})^{\varepsilon}}}\;
+ \left.\frac{4}{\pi} \sum_{m,n=1}^{\infty} \frac{\cos(\pi m \tau_{ab})\cos(\pi n \sigma_{ab})}{(v \, m^2 + n^2 /v)^{1+\varepsilon}}  
\right) 
.
\end{aligned}
\end{equation}
The first two sums are finite for $\varepsilon \rightarrow 0$ and are in this limit Fourier cosine series,
\begin{equation}\label{Eq:Cos-Fouriers}
\frac{2}{\pi} \sum_{k=1}^{\infty} \frac{\cos(\pi k x)}{k^2}
=\pi \left(\frac{1}{3} - x\left(1-\frac{x}{2}\right)
\right)
\quad
{\rm for}
\quad
0 \leq x \leq 2 
.
\end{equation}
The third sum is divergent for $\tau_{ab}=\sigma_{ab}=0$ and $\varepsilon \rightarrow 0$, but finite otherwise.
In order to find a closed expression, we make use of the following resummation formulas for $C >0$  (see e.g. the appendix in~\cite{Foerger:1998kw} for more general cases),
\begin{equation}\label{Eq:sum-simplifications}
\begin{aligned}
4 \, \sum_{j=1}^{\infty} \frac{\cos(\theta j)}{j^2 + C^2} &=-\frac{2}{C^2} + \frac{2\pi}{C} 
\frac{\cosh((\pi-\theta)C)}{\sinh(\pi C)}
,
\\
\sum_{j=1}^{\infty} \frac{1}{(j^2+C^2)^{1+\varepsilon}} &= -\frac{1}{2  C^{2+2\varepsilon} }
+ \frac{1}{2 \Gamma(1+\varepsilon)} \left\{
\frac{4 \, \pi^{1+\varepsilon}}{C^{\frac{1}{2}+\varepsilon}} \sum_{r = 1 }^{\infty} |r|^{\frac{1}{2}+\varepsilon} K_{\frac{1}{2}+\varepsilon} \left[2\pi C r   \right]\right.\\
&\phantom{= -\frac{1}{2  C^{2+2\varepsilon} }+ \frac{1}{2 \Gamma(1+\varepsilon)}}\quad\left.
+\sqrt{\pi} \frac{\Gamma(\frac{1}{2}+\varepsilon)}{C^{1+2\varepsilon}}
\right\}
,
\end{aligned}
\end{equation}
with $K_s[z]$ a modified Bessel function of the third kind and $\Gamma(z)$ the Gamma function defined in appendix~\ref{App:Technical}, equation~\eqref{Eq:App-Gamma-Zeta}.
Furthermore, formula~\eqref{Eq:integrated-sums} fulfills a modular invariance property under the transformation
$(\tau_{ab},\sigma_{ab},v) \leftrightarrow (\sigma_{ab},\tau_{ab},\frac{1}{v})$. 

For $(\tau_{ab},\sigma_{ab}) \neq (0,0)$ and $\varepsilon \rightarrow 0$, the integrated lattice sum~\eqref{Eq:integrated-sums} can be simplified to 
include only one summation using the first formula in~\eqref{Eq:sum-simplifications},
\begin{equation}\label{Eq:Lattice-tau-sigma}
\begin{aligned}
V_{ab}
\int_0^{\infty} dl L_{\mathcal{A}}(V_{ab},\tau_{ab},\sigma_{ab},v;l) 
& = \frac{\pi}{v} \left( \frac{\tau^2_{ab}}{2} -\tau_{ab} + \frac{1}{3} \right) + 
2 \, \sum_{k=1}^{\infty} \frac{\cos(\pi k \sigma_{ab}) \cosh(\pi k \frac{\tau_{ab}-1}{v})}{k \, \sinh(\frac{\pi k}{v})} 
\\
& = - \ln \left| e^{-\pi \tau^2_{ab}/(4v)} \frac{\vartheta_1(\frac{\sigma_{ab}-i\tau_{ab}/v}{2},\frac{i}{v})}{\eta(\frac{i}{v})}
\right|^2
\\ 
&= 
- \ln \left| e^{-\pi \sigma^2_{ab} v/4} \frac{\vartheta_1(\frac{\tau_{ab}}{2}-\frac{i \, v \, \sigma_{ab}}{2},iv)}{\eta(iv)}
\right|^2
,
\end{aligned}
\end{equation}
where in the second equality the product expansion~\eqref{Eq:App-ThetaProduct} of the Jacobi theta functions
was used and in the third line the modular transformation properties~\eqref{Eq:App-Theta-modular} for the first Jacobi theta and the Dedekind eta function.

Using the second formula in~\eqref{Eq:sum-simplifications} with $C= v \, m$ for $\tau_{ab}=\sigma_{ab}=0$
and summing also over $m$ leads to
\begin{equation}\label{Eq:Sum3-Evaluate}
\begin{aligned}
\frac{4\, \Gamma(1+\varepsilon)v^{1+\varepsilon} }{\pi(\pi V_{ab})^{\varepsilon}}  \, \sum_{j,m=1}^{\infty} \frac{1}{(j^2+(v\, m)^2)^{1+\varepsilon}}
= &
- \frac{2 \, \Gamma(1+\varepsilon)}{\pi \,(\pi V_{ab})^{\varepsilon}} \frac{1}{v^{1+\varepsilon}}
\zeta (2+2\varepsilon)
\\
&+ \frac{8 \sqrt{v} }{V_{ab}^{\varepsilon}} \sum_{m=1}^{\infty}\frac{1}{m^{1/2+\varepsilon}} \sum_{r = 1 }^{\infty} |r|^{\frac{1}{2}+\varepsilon} K_{\frac{1}{2}+\varepsilon} \left[2\pi m v |r|   \right]
\\
&+ \frac{2}{\sqrt{\pi}} \frac{\Gamma(\frac{1}{2}+\varepsilon)}{(\pi v V)^{\varepsilon}}  
\zeta(1+2\varepsilon)
\\
= & \left( \frac{1}{\varepsilon} + \gamma -\ln 2 \right)-\left( \frac{1}{v} + v \right) \frac{\pi}{3}\\
& - \ln\left( 2 \pi v V_{ab}  \,  \eta^4(iv) \right) %- \ln (2 \pi v V) 
+\mathcal{O}(\varepsilon) ,
\end{aligned}
\end{equation}
with the Riemann zeta function $\zeta(z)$ given in appendix~\ref{App:Technical} equation~\eqref{Eq:App-Gamma-Zeta} and 
the special values $\Gamma(1)=1$, $\zeta(2)= \frac{\pi^2}{6}$, $ K_{\frac{1}{2}} \left[ 2 \pi x   \right] = \frac{1}{2 \, \sqrt{x}} e^{-2 \pi x}$ and well as $\frac{\Gamma (\frac{1}{2} +\varepsilon)}{\sqrt{\pi}} =1 - (\gamma + 2 \ln 2 ) \, \varepsilon   + \mathcal{O}(\varepsilon^2)$.

In order to extract the one-loop running due to massless strings, the dimensional regularization is given the interpretation of a ratio of scales,
\begin{equation}\label{Eq:Def-Scales}
\frac{1}{\varepsilon} +\gamma - \ln 2  \equiv  \ln \left( \frac{M_{\rm string}}{\mu}  \right)^2
.
\end{equation}
Our choice of regularization differs from the one  in~\cite{Akerblom:2007np} slightly.
With respect to their definition, our ratio of scales is replaced by $\frac{M_{\rm string}}{\mu} \rightarrow
\sqrt{\frac{e^{\gamma}}{2}}\frac{M_{\rm string}}{\mu} \thickapprox 0.94 \cdot  \frac{M_{\rm string}}{\mu}$,  
and a constant proportional to $ \gamma -\ln 2 \thickapprox -0.116$ is removed from the threshold contributions.
Combining the explicitly evaluated sums~\eqref{Eq:Cos-Fouriers},~\eqref{Eq:Lattice-tau-sigma} and~\eqref{Eq:Sum3-Evaluate} in~\eqref{Eq:integrated-sums} leads to the lattice contributions to 
the gauge threshold corrections,
\begin{equation}\label{Eq:FullLattice-tau-sigma}
\begin{aligned}
V_{ab}
\int_0^{\infty} dl L_{\mathcal{A}}(V_{ab},\tau_{ab},\sigma_{ab},v;l) 
=& \;\delta_{\tau_{ab},0} \,  \delta_{\sigma_{ab},0} \, \ln \left( \frac{M_{\rm string}}{\mu}  \right)^2
\\
&
-  \delta_{\tau_{ab},0} \,  \delta_{\sigma_{ab},0} \, \ln \left( 2 \pi v V_{ab} \,  \eta^4(iv) \right) 
\\
&
- \left( 1 -  \delta_{\tau_{ab},0} \,  \delta_{\sigma_{ab},0} \right)
\ln \left| e^{-\pi \sigma_{ab}^2 v/4} \frac{\vartheta_1(\frac{\tau_{ab}}{2}-\frac{i \, v \, \sigma_{ab}}{2},iv)}{\eta(iv)}
\right|^2
.
\end{aligned}
\end{equation}
For the {\it discrete} Wilson line and displacement values $\tau_{ab},\sigma_{ab} \in \{0,1\}$, the last line can be rewritten 
to match the results in~\cite{Blumenhagen:2007ip}, and upon T-duality it matches the formula in~\cite{Blumenhagen:2006ci} for D9-branes with continuous Wilson lines.

In the following section, the analogous discussion for the M\"obius strip is presented, and afterwards in section~\ref{Ss:ZeroAngle}
the threshold computation for D6-branes parallel along a two-torus are presented in 
some detail, as well as how the integrals discussed in this section arise and  the correct prefactor for each annulus and M\"obius strip amplitude 
will be determined by comparison with the field theoretically derived beta function coefficients and by imposing tadpole cancellation among all
threshold amplitudes. In appendix~\ref{AppS:Thresholds} tables~\ref{tab:tadosc1} and~\ref{tab:tadosc2}, we verify that these prefactors are indeed those obtained by RR tadpole cancellation
of the vacuum amplitudes.

%%%%%%%%%%%%%%%%%%%%%
\subsection{The M\"obius strip contributions per two-torus}\label{Ss:MS-per-T2}

The M\"obius strip contributions to the gauge threshold corrections are computed in a similar manner to the annulus
except for some subtleties concerning the lattice contributions. Strings stretching between brane $a$ and the orientifold
image $(\theta^k a')$ are invariant under the two insertions of $\OR\theta^{-k}$ and $\OR\theta^{-k+N}$. In contrast to
the annulus amplitude, both insertions lead to untwisted tree channel amplitudes. This is already known from the 
RR tadpole computation since the O6-planes do not wrap any exceptional cycles. One therefore has to distinguish only 
the contributions per two-torus for the cases with vanishing angle and  at angle $\phi_a^{\OR\theta^{-k},(i)}$:
\begin{enumerate}
\item[[1\!]]
brane $a$ is parallel to the $\OR\theta^{-k}$ invariant O6-plane on the $i^{th}$ two-torus
\begin{equation}
\tilde{V}^{(i)}_{a,\OR\theta^{-k}} \tilde{\mathcal{L}}_a^{\mathcal{M}}(\tau_{aa'}^{(i)},\sigma_{aa'}^{(i)},v_i;l) 
\frac{\vartheta\targ{\alpha}{\beta} (0,2il-\frac{1}{2})}{\eta^3(2il -\frac{1}{2})}
,
\end{equation}
\item[[2\!]]
brane $a$ is at angle $\phi_{a,\OR\theta^{-k}}^{(i)}$ to the $\OR\theta^{-k}$ invariant O6-plane on $T^2_i$
\begin{equation}
I_a^{\OR\theta^{-k},(i)} \frac{\vartheta\targ{\alpha}{\beta} (0,2il-\frac{1}{2})}{\vartheta\targ{1/2}{1/2}}
(\phi_{a,\OR\theta^{-k}}^{(i)},2il-\frac{1}{2})
.
\end{equation}
\end{enumerate}
For vanishing Wilson lines and displacements, $\tau_a = \sigma_a=0$, the lattice contributions have been studied before, see e.g. 
table~ in the appendix of~\cite{Forste:2001gb}. In this case, the annulus loop channel expression~\eqref{Eq:lattice-Annulus-open} is the same as the M\"obius strip one
for a square torus, but for a tilted torus, only half of the winding states are invariant under the orientifold projection
while all momentum states remain invariant,
\begin{equation}\label{Eq:lattice-Moebius-open}
\mathcal{L}^{\mathcal{M}}_{a,0} (t) = \sum_{m,n \in \Z} \exp \left(-2 \pi t \, \tilde{M}^2_{mn,0} (a,a')  \right)
\quad
{\rm with}
\quad 
\tilde{M}^2_{mn,0}(a,a') = \frac{2}{\tilde{V}_{aa'}} \, \left[\frac{1}{\tilde{v}} m^2 + \tilde{v} n^2 \right]
,
\end{equation}
with $\tilde{V}_{aa'} = 2 (1-b) V_{aa'}$ and $\tilde{v} = \frac{v}{1-b}$ and $b=0,1/2$ for the rectangular and tilted torus as before.
The modular transformation with M\"{o}bius strip parameter $t=\frac{1}{8l}$ leads to the tree channel expression
\begin{equation}
\tilde{\mathcal L}^{\mathcal{M}}_{a,0}(l) =
\sum_{m,n \in \Z} \exp \left\{ -2 \pi l  \, \tilde{V}_{aa'} \left( \tilde{v} \,  m^2 
+ \frac{1}{\tilde{v}} \, n^2 \right)
\right\}
,
\end{equation}
which has the same shape as the annulus sum when replacing $(v,V) \rightarrow (\tilde{v}, 2 \tilde{V})$.
Pictorially, one can read off that the relative distance between a brane $a$ and its orientifold image $a'$ is 
$\sigma_{aa'} = 2 \sigma_a$, and T-duality arguments similarly give the relative Wilson line of orientifold image branes $\tau_{aa'}=2\tau_a$. The invariance properties
are unchanged when these continuous parameters are switched on,
\begin{equation}\label{Eq:lattice-M-general}
\tilde{\mathcal L}^{\mathcal{M}}_{a}(l) =
\sum_{m,n \in \Z} \exp \left\{ -2 \pi l  \, \tilde{V}_{aa'} \left( \tilde{v} \,  m^2 
+ \frac{1}{\tilde{v}} \, n^2 \right)
+ \pi i \left(m \sigma_{aa'} + n \tau_{aa'}   \right)
\right\}
.
\end{equation}
The M\"obius strip integral relevant for the threshold corrections thus reads
\begin{equation}
\begin{aligned}
2 \, \tilde{V}_{aa'} \int_{0}^{\infty} dl\tilde{\mathcal L}^{\mathcal{M}}_{a}(l)
=& 2 \, \tilde{V}_{aa'} \int_{0}^{\infty}dl + \delta_{\sigma_{aa'},0} \,\delta_{\tau_{aa'},0} \ln \left(
\frac{M_{\rm string}}{\mu}\right)^2
\\
&-\delta_{\sigma_{aa'},0} \,\delta_{\tau_{aa'},0} \, \ln \left( 4 \pi \tilde{v} \tilde{V}_{aa'} \,  \eta^4 (i \tilde{v}) \right)
\\&
- \left(1-\delta_{\sigma_{aa'},0} \delta_{\tau_{aa'},0} \right) \,\ln\left|e^{-\pi \sigma_{aa'}^2 \tilde{v}_3/4}\frac{\vartheta_1 (\frac{\tau_{aa'}}{2}-i\frac{\sigma_{aa'}}{2} \tilde{v},i \tilde{v})}{\eta (i \tilde{v})}\right|^2
.
\end{aligned}
\end{equation}
Since Wilson lines $\tau_{a}$ and distances from the origin $\sigma_{a}$ are defined modulo 2, the cases with discrete values in $\{0,1\}$  
both lead to the known results for $\tau_{aa'}=\sigma_{aa'}=0$, whereas to our knowledge the case for continuous parameters on D6-branes has not been discussed in the literature.
The full M\"obius strip correction from branes parallel to their orientifold images along one two-torus will be computed
in section~\ref{Ss:MS-ZeroAngle}.

%%%%%%%%%%%%%%%%%%%%%
\subsection{The full annulus threshold computation for one vanishing angle}\label{Ss:ZeroAngle}

In this section, the computation of the gauge threshold contribution from D6-branes parallel along one two-torus and at supersymmetric angles on the remaining four-torus is presented in detail. The relevant constant prefactor is derived using the known form of the contribution of massless strings to the beta function coefficients. 
The discussion for D6-branes parallel everywhere or at angles on the whole six-torus is analogous, and the results are collected in appendix~\ref{App:Technical} and~\ref{AppS:Tables} in tables~\ref{tab:Annulus-Amplitudes-thresholds},~\ref{tab:Mobius-Amplitudes-thresholds} and~\ref{tab:Amplitudes-thresholds}. 

Two kinds $a$ and $b$ of D6-branes can be either parallel along the $\Z_2$ invariant direction, or parallel along one of the two complex directions where $\Z_2$ acts non-trivially.
The corresponding contributions to the beta function coefficient  of a $SU(N_a)$ gauge group are according to the massless states in appendix~\ref{AppSs:Spectrum}
\begin{equation}\label{Eq:beta-contribution-Ann}
b_{SU(N_a)} \supset
\frac{N_b}{2} \varphi^{ab} = \left\{\begin{array}{cc}
\frac{N_b}{2} \left| I_{ab}^{(1 \cdot 3)} +  I_{ab}^{\Z_2,(1 \cdot 3)} \right| & \pi (\phi,0,-\phi)
\\
\frac{N_b}{2} \left| I_{ab}^{(1 \cdot 2)} \right| &  \pi (\phi,-\phi,0)
\\
\frac{N_b}{2} \left| I_{ab}^{(2 \cdot 3)} \right| &  \pi (0,\phi,-\phi)
\end{array}
\right.
,
\end{equation}
where $\vec{v}_{\Z_2} = \frac{1}{2}(1,0,-1)$ leaves the second two-torus invariant. Since the number of $\Z_2$ invariant intersections cannot exceed the total number of intersection points, the first line can be written as
\begin{equation}
\frac{N_b}{2} \left| I_{ab}^{(1 \cdot 3)} \right| -\frac{N_b}{2} \ I_{ab}^{\Z_2,(1 \cdot 3)} ,
\end{equation}
because for one vanishing angle, the toroidal intersection numbers on the remaining two two-tori have opposite signs, i.e. $I_{ab}^{(1 \cdot 3)} < 0$.
These formulae will be used to determine the correct prefactors of the amplitudes which are then checked to be consistent with RR tadpole cancellation. 

The first kind of amplitude present for all three cases is the toroidal annulus contribution which in the open string channel has a $\unity$ insertion from the orbifold projector.
 The tree channel amplitude is of the form
\begin{equation}
 T^{A,\unity}_{ab} =
c_{\mathcal{A}}^{\unity} \int_0^{\infty} dl l^{\varepsilon} \tilde{\mathcal{L}}^{\mathcal{A},(1)}_{ab} (l) \cdot \Theta_{\rm oscillator}^{\unity,(0,\phi,-\phi)}(l) ,
\end{equation}
and the oscillator contributions are read off from the general prescription~\eqref{Eq:Annulus-per-T2},
\begin{equation}\label{Eq:Annulus-theta-1}
\begin{aligned}
 \Theta_{\rm oscillator}^{\unity,(0,\phi,-\phi)} &= \sum_{\alpha,\beta \in \{0,\frac{1}{2}\}} (-1)^{2(\alpha+\beta)}
  \frac{\vartheta^{\prime\prime} \targ{\alpha}{\beta}(0,2il) \, \vartheta \targ{\alpha}{\beta}(0,2il) \, \vartheta \targ{\alpha}{\beta}(\phi,2il) \, \vartheta \targ{\alpha}{\beta}(-\phi,2il)}{\eta^6(2il)
\, \vartheta \targ{1/2}{1/2}(\phi,2il) \, \vartheta \targ{1/2}{1/2}(-\phi,2il)  
  }
 \\
 &=\frac{\vartheta_3''(0) \, \vartheta_3(0) \, 
\vartheta_3(\phi) \, \vartheta_3(\!-\phi)
-\vartheta_2''(0) \, \vartheta_2(0) \, 
\vartheta_2(\phi) \, \vartheta_2(\!-\phi)
-\vartheta_4''(0) \, \vartheta_4(0) \, 
\vartheta_4(\phi) \, \vartheta_4(\!-\phi) }
{\eta^6 \, (-\vartheta_1 (\phi)) \, (- \vartheta_1 (-\phi))}
 \\
 &= 4 \pi^2,
\end{aligned}
\end{equation}
where in the second line for the sake of brevity, the second argument of the Jacobi theta and Dedekind eta functions is omitted, e.g. $\vartheta_1(\phi) \equiv \vartheta_1(\phi,2il)$ and $\eta^6 \equiv \eta^6(2il)$. The amplitude contributes therefore 
\begin{equation}
\begin{aligned}
 T^{A,\unity}_{ab} &=
4 \pi^2 \, c_{\mathcal{A}}^{\unity} \int_0^{\infty} dl l^{\varepsilon} \tilde{\mathcal{L}}^{\mathcal{A},(1)}_{ab}(l)\\
&=4 \pi^2 \, c_{\mathcal{A}}^{\unity}  \left(
\int_0^{\infty} dl 
+ \frac{\delta_{\sigma_{ab}^1,0} \delta_{\tau_{ab}^1,0} }{V_{ab}^{(1)} } \ln \left( \frac{M_{\rm string}}{\mu}  \right)^2
- \frac{ \Lambda(\sigma_{ab}^1,\tau_{ab}^1, v_1,V_{ab}^{(1)})}{V_{ab}^{(1)} }
\right)
,
\end{aligned}
\end{equation}
with the compact notation for the lattice contributions to the thresholds,
\begin{equation}\label{Eq:Def-Lambda}
\Lambda(\sigma,\tau,v;V) \equiv \delta_{\sigma,0} \delta_{\tau,0} \, \ln \left( 2 \pi v V \,  \eta^4 (i v) \right)
+ \left(1-\delta_{\sigma,0} \delta_{\tau,0} \right) \,\ln\left|e^{-\pi \sigma^2 v/4}\frac{\vartheta_1 (\frac{\tau}{2}-i\frac{\sigma}{2} \, v,i v)}{\eta (i v)}\right|^2
.
\end{equation}

%%%%%%%%%%%%%%%%%%%%%%%%%%%%%%%%%%%%%%%%%%%%%%%%%%%%%
%   figure:
%%%%%%%%%%%%%%%%%%%%%%%%%%%%%%%%%%%%%%%%%%%%%%%%%%%%%
\twofig{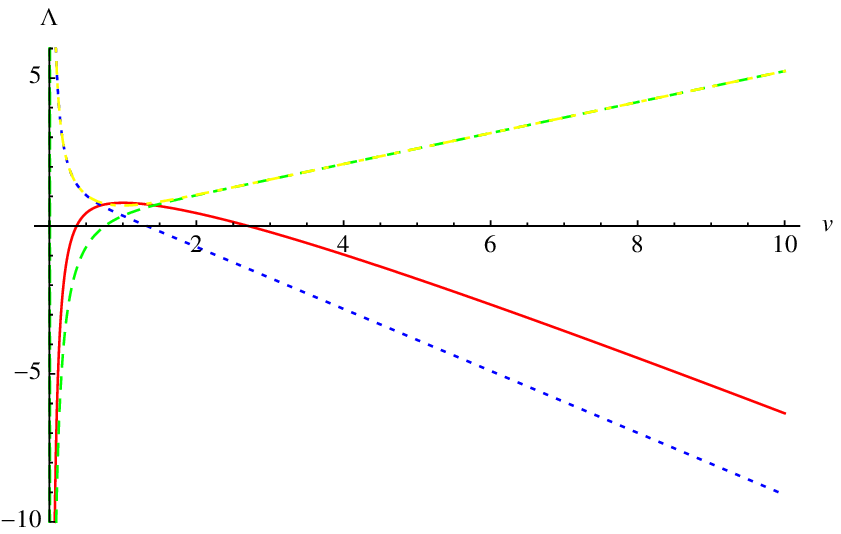}{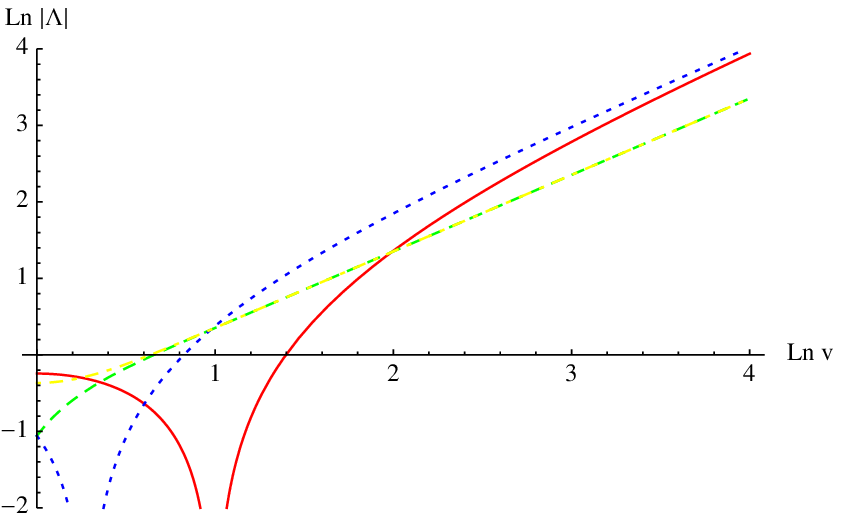}{The lattice contributions $\Lambda$ depending on the two-torus volume $v$ for fixed $V\equiv 1$ and different values of $(\sigma,\tau)$: red, solid: $(0,0)$; green, dashed: $(1,0)$; blue, dotted: $(0,1)$; yellow, dotdashed: $(1,1)$. Note that in figure~(b) the absolute value of $\Lambda$ is displayed double logarithmically.}{fig:lattice_contrib}
%%%%%%%%%%%%%%%%%%%%%%%%%%%%%%%%%%%%%%%%%%%%%%%%%%%%%

In figure~\ref{fig:lattice_contrib}, the lattice contributions
in dependence of the two-torus volume $v$ are plotted for $(\sigma,\tau)=(0,0)$
and fixed value of the brane volume $V\equiv 1$, and for the cases with
discrete distance or Wilson line, $(\sigma,\tau)=(1,0)$, $(0,1)$ and $(1,1)$.

The comparison of the coefficient of $\ln \left( \frac{M_{\rm string}}{\mu}\right)^2$ with the contribution~\eqref{Eq:beta-contribution-Ann} to the one-loop beta function fixes
\begin{equation}
c_{\mathcal{A}}^{\unity}= -\frac{N_b V_{ab}^{(1)} I_{ab}^{(2 \cdot 3)}}{8 \pi^2}
.
\end{equation}
The minus sign is due to the fact that $\left| I_{ab}^{(2 \cdot 3)}\right| = -I_{ab}^{(2 \cdot 3)}$ if the branes are parallel on $T^2_1$. The tadpole contribution from this kind of annulus amplitude is thus
$4 \pi^2 \, c_{\mathcal{A}}^{\unity} \int dl= -\frac{N_b}{2} V_{ab}^{(1)} I_{ab}^{(2 \cdot 3)} \int dl$, which can be compared to the first line in equation~\eqref{Eq:RRtcc-rewritten} bearing in mind that $I_{ab}^{(1 \cdot 2)} = I_{ab}^{(1 \cdot 3)}=0$ for branes parallel long the first two-torus $T^2_1$.

The contributions from $\Z_2$ invariant intersections, or in other words the annulus amplitude with a $\Z_2$ insertion in the loop channel, differ depending on the complex direction
where the branes are parallel:
\begin{equation}
\begin{aligned}
 T^{A,\Z_2,(2)}_{ab} 
&=  c_{\mathcal{A}}^{\Z_2,(2)} \int_0^{\infty} dl l^{\varepsilon} \tilde{\mathcal{L}}^{\mathcal{A},(2)}_{ab} \cdot \Theta_{\rm oscillator}^{\Z_2,(\phi,0,-\phi)} ,
\\
 T^{A,\Z_2,(1)}_{ab} 
&= c_{\mathcal{A}}^{\Z_2,(1)} \int_0^{\infty} dl l^{\varepsilon} \Theta_{\rm oscillator}^{\Z_2,(0,\phi,-\phi)}
.
\end{aligned}
\end{equation}
According to the general rules~\eqref{Eq:Annulus-per-T2}, the oscillator contributions are
\begin{equation}\label{Eq:Oscillators-Z2-two-angles}
\begin{aligned}
\Theta_{\rm oscillator}^{\Z_2,(\phi,0,-\phi)} &=  \sum_{\alpha,\beta \in \{0,\frac{1}{2}\}}\!\!\!\! (-1)^{2(\alpha+\beta)}
 \frac{\vartheta^{\prime\prime} \targ{\alpha}{\beta}(0,2il)
\, \vartheta \targ{\alpha}{\beta}(0,2il) \, \vartheta \targ{\alpha+1/2}{\beta}(\phi,2il) \, \vartheta \targ{\alpha+1/2}{\beta}(-\phi,2il)}{\eta^6 (2il)\, \vartheta \targ{0}{1/2}(\phi,2il) \, \vartheta \targ{0}{1/2}(-\phi,2il)  }
\\
&=\frac{\vartheta_3''(0) \, \vartheta_3(0) \, 
\vartheta_2(\phi) \, \vartheta_2(-\phi)
-\vartheta_2''(0) \, \vartheta_2(0) \, 
\vartheta_3(\phi) \, \vartheta_3(-\phi)}
{\eta^6 \, \vartheta_4 (\phi) \, \vartheta_4 (-\phi)}
\\
&\quad\frac{
-\vartheta_4''(0) \, \vartheta_4(0) \, 
(-\vartheta_1(\phi)) \,(- \vartheta_4(-\phi)) }
{\eta^6 \, \vartheta_4 (\phi) \, \vartheta_4 (-\phi)}
 \\
 &= 4 \pi^2,
\\
\\
\Theta_{\rm oscillator}^{\Z_2,(0,\phi,-\phi)} &= \sum_{\alpha,\beta \in \{0,\frac{1}{2}\}}\!\!\!\! (-1)^{2(\alpha+\beta)}
 \frac{\vartheta^{\prime\prime} \targ{\alpha}{\beta}(0,2il)
\, \vartheta \targ{\alpha+1/2}{\beta}(0,2il) \, \vartheta \targ{\alpha}{\beta}(\phi,2il) \, \vartheta \targ{\alpha+1/2}{\beta}(-\phi,2il)}{\eta^3(2il)\, \vartheta \targ{0}{1/2}(0,2il)\, \vartheta \targ{1/2}{1/2}(\phi,2il) \, \vartheta \targ{0}{1/2}(-\phi,2il) }
\\
&=\frac{\vartheta_3''(0) \, \vartheta_2(0) \, 
\vartheta_3(\phi) \, \vartheta_2(-\phi)
-\vartheta_2''(0) \, \vartheta_3(0) \, 
\vartheta_2(\phi) \, \vartheta_3(-\phi) }
{\eta^3  \, \vartheta_4 (0) \, (-\vartheta_1(\phi))\, \vartheta_4 (-\phi)}
\\
&= - 2 \pi \, 
 \left(  
\frac{\vartheta_1'}{\vartheta_1}(\phi,\tau)
+\frac{\vartheta_4'}{\vartheta_4}(-\phi,\tau)
 \right)
,
\end{aligned}
\end{equation}
where in the second line of each formula the second argument $2il$ has again been omitted.
The integral for the amplitude with angles $(\phi,0,-\phi)$ and $\Z_2$ loop channel insertion is 
therefore completely analogous to the one with $\unity$ loop channel insertion, the only difference arising from 
the prefactors $c_{\mathcal{A}}^{\Z_2,(2)}$ and $c^{\unity}_{\mathcal{A}}$. 

The second identity in~\eqref{Eq:Oscillators-Z2-two-angles} requires the expansions of the Jacobi theta functions given 
in appendix~\ref{AppSs:Thetas} equation~\eqref{Eq:App-ThetaPrime} as well as their regularized integrals~\eqref{Eq:App-Theta-Final-Integrals}. Since these have been discussed before
e.g. in~\cite{Akerblom:2007np}, we delegate their evaluation to appendix~\ref{AppSs:Thetas}. 

The final expression for the twisted annulus contributions with one vanishing angle are 
\begin{equation}
\begin{aligned}
 T^{A,\Z_2,(2)}_{ab} 
&=  4 \pi^2 c_{\mathcal{A}}^{\Z_2,(2)}\left(\int_0^{\infty} dl    + \frac{\delta_{\tau_{ab}^2,0}\delta_{\sigma_{ab}^2,0}}{V_{ab}^{(2)}} \ln \left( \frac{M_{\rm string}}{\mu} \right)^2 
-\frac{\Lambda(\sigma_{ab}^2,\tau_{ab}^2, v_2; V_{ab}^{(2)} )}{V_{ab}^{(2)}} \right)
,
\\ 
T^{A,\Z_2,(1)}_{ab} 
&= -2 \pi^2  c_{\mathcal{A}}^{\Z_2,(1)} \left(  \cot (\pi \phi) \int_0^{\infty} dl - \ln 2 \left( \sgn(\phi) -2 \phi \right)
\right)
,
\end{aligned}
\end{equation}
and by comparison with the beta function coefficients for the case with $(\phi,0,-\phi)$, one obtains
\begin{equation}
 c_{\mathcal{A}}^{\Z_2,(1)} = - \frac{N_b}{8 \pi^2} I_{ab}^{\Z_2,(1 \cdot 3)} V_{ab}^{(2)}
,\qquad
c_{\mathcal{A}}^{\Z_2,(2)} = \frac{N_b}{4 \pi^2}  I_{ab}^{\Z_2} 
, 
\end{equation}
whereas for the case $(0,\phi,-\phi)$ one has to require an identical form of the tadpole as for $(\phi,0,-\phi)$
since the beta function coefficient is vanishing.
In some more detail, the tadpole contribution from D6-branes parallel along the second torus $T^2_2$ is $ 4 \pi^2 c_{\mathcal{A}}^{\Z_2,(2)} \int dl = - \frac{N_b}{2} I_{ab}^{\Z_2,(1 \cdot 3)} V_{ab}^{(2)} \int dl$, and using the twisted part of the rewritten RR tadpole cancellation condition~\eqref{Eq:RRtcc-rewritten}, we have to impose $-2 \pi^2  c_{\mathcal{A}}^{\Z_2,(1)}
 \cot (\pi \phi^{(2)}) \int dl \stackrel{!}{=} - \frac{N_b}{2} I_{ab}^{\Z_2,(1 \cdot 3)} V_{ab}^{(2)} \int dl$, which by using the relation~\eqref{Eq:Rel-I-V-Angle} between angles, intersection numbers and the generalization of lengths gives the above stated prefactor for the amplitude where D6-branes are parallel along the first torus $T^2_1$. The correctness of the relative prefactors is also confirmed by the computation of the RR tadpole cancellation in appendix~\ref{AppS:Thresholds} table~\ref{tab:tadosc1}.

%%%%%%%%%%%%%%%%%%%%%
\subsection{The M\"obius amplitudes for one vanishing angle}\label{Ss:MS-ZeroAngle}

With the information of section~\ref{Ss:MS-per-T2} at hand, the M\"obius strip contribution for a brane with one vanishing angle to the orientifold plane and its orientifold image can be computed. Since the M\"obius strip contains only untwisted tree channel contributions, corresponding to the fact that the $\OR\theta^{-k}$ and $\OR\theta^{-k+N}$ invariant O6-planes only coincide on one torus, but are at  angles $\pm \frac{\pi}{2}$ on the other two two-tori, the only amplitude to be evaluated is
\begin{equation}
\begin{aligned}
T^M_a &=
 c_{\mathcal{M}}
\int_0^{\infty} dl l^{\varepsilon}
\tilde{\mathcal L}_a^{\mathcal{M}}(l) \theta_{\rm oscillator}^{\OR\theta^{-k},(\phi,-\phi,0)} (2il-\frac{1}{2}) 
\\
&= 4 \pi^2 c_{\mathcal{M}}\int_0^{\infty} dl l^{\varepsilon}
\tilde{\mathcal L}_a^{\mathcal{M}} (l)
\\
&= 4 \pi^2 c_{\mathcal{M}}\left( \int_0^{\infty} dl + \frac{\delta_{\sigma_{aa'}^3,0}\delta_{\tau_{aa'}^3,0}}{2 \tilde{V}_{aa'}^{(3)}} \ln \left(\frac{M_{\rm string}}{\mu}\right)^2
-\frac{\Lambda(\sigma_{aa'}^3,\tau_{aa'}^3,\tilde{v}_3;2\tilde{V}^{(3)}_{aa'})}{2 \tilde{V}^{(3)}_{aa'}}
\right)
,
\end{aligned}
\end{equation}
where the oscillator contributions $\theta_{\rm oscillator}^{\OR\theta^{-k},(\phi,-\phi,0)} (2il-\frac{1}{2})$  are the same as for the annulus~\eqref{Eq:Annulus-theta-1}
 up to the change of argument $2il \rightarrow 2il -\frac{1}{2}$.

The global prefactor, 
\begin{equation}
c_{\mathcal{M}}= \frac{ \tilde{I}^{\OR\theta^{-k}(1 \cdot 2)}_a \tilde{V}_{aa'}^{(3)}}{2 \pi^2} 
,
\end{equation}
 is, as for the annulus contribution, determined by demanding that the contribution to the 
beta function coefficient matches the known field theoretic result,
\begin{equation}
\begin{aligned}
b_{SU(N_a)} \supset \;
&\frac{N_a-2}{2} \varphi^{\Anti_a} + \frac{N_a+2}{2} \varphi^{\Sym_a}\\
 &= \frac{N_a-2}{4}\left| I_{a(\theta^k a')}^{(1 \cdot 2)} + \tilde{I}_a^{\OR\theta^{-k}(1 \cdot 2)}  \right|
+ \frac{N_a+2}{4} \left| I_{a(\theta^k a')}^{(1 \cdot 2)} -  \tilde{I}_a^{\OR\theta^{-k}(1 \cdot 2)}  \right|
\\
&= \frac{N_a}{2} \left| I_{a(\theta^k a')}^{(1 \cdot 2)}\right|
+   \tilde{I}_a^{\OR\theta^{-k}(1 \cdot 2)}  
 ,
\end{aligned}
\end{equation}
if, for concreteness, brane $a$ is parallel to the $\OR\theta^{-k}$ invariant plane along $T^2_3$.

The tadpole contribution of this amplitude is 
$4 \pi^2 c_{\mathcal{M}} \int dl= 2 N_b \tilde{I}^{\OR\theta^{-k}(1 \cdot 2)}_a \tilde{V}_{aa'}^{(3)} \int dl$, containing the correct 
relative factor of (-4) with respect to the annulus contributions in section~\ref{Ss:ZeroAngle} so that the tadpoles cancel when the untwisted
part of the rewritten vacuum RR tadpole condition in~\eqref{Eq:RRtcc-rewritten} is satisfied.

The prefactors of the M\"obius strip amplitudes can also be seen to agree with those expected from RR tadpole cancellation of the vacuum amplitudes
in table~\ref{tab:tadosc2}.

\section{Results of the threshold computation}\label{sec:ThreshRes}

%%%%%%%%%%%%%%%%%%%%%
\subsection{The complete thresholds for \texorpdfstring{$SU(N)$}{SU(N)} gauge groups}\label{Ss:SUN}

The contributions from sectors with one vanishing angle have been derived in detail in the previous section. The remaining supersymmetric 
amplitudes with D6-branes either parallel everywhere or at three non-vanishing angles are listed in appendix~\ref{AppS:Tables} in 
table~\ref{tab:Annulus-Amplitudes-thresholds} for the annulus and~\ref{tab:Mobius-Amplitudes-thresholds} for the M\"obius 
strip diagram, respectively. The resulting tadpole, beta function coefficient and threshold correction per sector is listed in table~\ref{tab:Amplitudes-thresholds}.

The entire threshold correction for an $SU(N_a)$ gauge factor arising from a stack of D6-branes in a $T^6/\Z_{2N}$ type IIA orientifold background is 
given by
\begin{equation}\label{Eq:Threh-SU(N)}
\begin{aligned}
\Delta_{SU(N_a)} =&  \sum_b N_b 
\left( \tilde{\Delta}_{ab}^{\rm total} + \tilde{\Delta}_{ab'}^{\rm total}
\right) 
 +  \Delta_{a,\OR}^{\rm total}
 \\
  {\rm with} & \quad
 \tilde{\Delta}_{ab}^{\rm total} =  \tilde{\Delta}_{ba}^{\rm total} 
= \sum _{k=0}^{N-1}\tilde{\Delta}_{a(\theta^k b)}
= \sum _{k=0}^{N-1} \left( \tilde{\Delta}_{a(\theta^k b)}^{\unity} + \tilde{\Delta}_{a(\theta^k b)}^{\Z_2} \right)
\\
{\rm and} & \quad
\Delta_{a,\OR}^{\rm total} = \sum _{k=0}^{2N-1} \Delta_{a,\OR\theta^{-k}}
.
\end{aligned}
\end{equation}
For the convenience of explicit computations in section~\ref{S:Examples}, we defined $\tilde{\Delta}_{ab}^{\rm total} = \Delta_{ab}^{\rm total} / N_b$. 
The newly introduced quantity has several symmetries in its two subscripts,
\begin{equation}
  \tilde{\Delta}_{ab}^{\rm total} = \tilde{\Delta}_{ba}^{\rm total} = 
  \tilde{\Delta}_{a'b'}^{\rm total} = \tilde{\Delta}_{b'a'}^{\rm total}.
\end{equation}

The angle dependent components of $\tilde{\Delta}_{ab}^{(\phi^{(1)}_{ab},\phi^{(2)}_{ab},\phi^{(3)}_{ab})}
\equiv \tilde{\Delta}_{ab}$ where D6-branes $a$ and $b$ are at angles $(\phi^{(1)}_{ab},\phi^{(2)}_{ab},\phi^{(3)}_{ab})$ 
and $\vec{v}_{\Z_2} =\frac{1}{2}(1,0,-1)$ are given by:
\begin{equation}
\begin{aligned}
\tilde{\Delta}_{ab}^{(0,0,0)} & =\frac{ I^{\Z_2,(1 \cdot 3)}_{ab}}{2} \, \Lambda (\sigma^2_{ab},\tau^2_{ab}, v_2; V_{ab}^{(2)})
,
\\
\tilde{\Delta}_{ab}^{(\phi,0,-\phi)} & =\frac{ I^{(1 \cdot 3)}_{ab} +  I^{\Z_2,(1 \cdot 3)}_{ab}}{2} \, \Lambda (\sigma^2_{ab},\tau^2_{ab}, v_2; V_{ab}^{(2)})
,
\\
\tilde{\Delta}_{ab}^{(0,\phi,-\phi)} & =\frac{ I^{(2 \cdot 3)}_{ab} }{2} \, \Lambda (\sigma^1_{ab},\tau^1_{ab}, v_1; V_{ab}^{(1)})
 + \frac{ I^{\Z_2,(2 \cdot 3)}_{ab} }{2} \,
\ln 2 \left( \sgn(\phi) - 2 \, \phi \right)
,
\\
\tilde{\Delta}_{ab}^{(\phi^{(1)},\phi^{(2)},\phi^{(3)})} & =  \, \frac{I_{ab} + I_{ab}^{\Z_2}}{4} \sum_{i=1}^3 \ln \left(\frac{\Gamma(|\phi^{(i)}|)}{\Gamma(1-|\phi^{(i)}|)}\right)^{\sgn(\phi^{(i)})}
\\
&\quad
-  \,I_{ab}^{\Z_2} \, \frac{ \ln (2) }{2} \sum_{j=1,3} \left(\sgn(\phi^{(j)})
 -2 \phi^{(j)}  \right)
,
\end{aligned}
\end{equation}
where the abbreviation $\Lambda(\sigma,\tau,v;V)$ for the lattice contributions in dependence of displacements $\sigma$, Wilson lines $\tau$, two-tori volumes $v$ and one-cycle volumes $\sqrt{V}$ was introduced in~\eqref{Eq:Def-Lambda}.

The beta function coefficients from the various contributions listed in appendix~\ref{AppS:Tables}
in the fourth column of table~\ref{tab:Amplitudes-thresholds} match those computed field theoretically from the massless spectrum in table~\ref{NonChiralSpectrum} by using the 
identities
\begin{equation}
  \sgn(\phi^{(i)}_{ab}) = \sgn(I_{ab}^{(i)})\quad\mbox{and}\quad
  \sum_{i=1}^3 \sgn (\phi^{(i)}_{ab}) = -\prod_{i=1}^3 \sgn(\phi^{(i)}_{ab})
\end{equation}
for supersymmetric D6-brane configurations. Each untwisted amplitude contributes
\begin{equation}
  -\frac{N_b}{2} \sum_{i=1}^3 V_{ab}^{(i)} I_{ab}^{(j \cdot k)} \int dl
\end{equation}
to the tadpoles, and each twisted amplitude contributes
\begin{equation}
  \frac{N_b}{2} V_{ab}^{(2)} I_{ab}^{\Z_2,(1 \cdot 3)} \int dl,
\end{equation}
where the factors $V_{ab}^{(i)}$ arise either from the lattice contributions of a direction where the D6-branes are parallel or as $V_{ab}^{(i)}=I_{ab}^{(i)} \cot (\phi_{ab}^{(i)})$ along a direction where the D6-branes intersect. The twisted part of the tadpoles cancels among all annulus contributions according to the second line in the rewritten RR tadpole cancellation conditions~\eqref{Eq:RRtcc-rewritten}.
In the last column of table~\ref{tab:Amplitudes-thresholds}, the quantities $\Delta_{ab}^{{\rm insertion},(\phi^{(1)},\phi^{(2)},\phi^{(3)})} = N_b \, \tilde{\Delta}_{ab}^{{\rm insertion},(\phi^{(1)},\phi^{(2)},\phi^{(3)})}$ are displayed for all possible angles
and untwisted or twisted contributions separately.

The M\"obius strip contributions in dependence of the angles between D6-brane $a$ and the O6-plane under consideration are
\begin{equation}
\begin{aligned}
 \Delta_{a,\OR\theta^{-m}}^{(0,0,0)} & =0
,
\\
 \Delta_{a,\OR\theta^{-m}}^{(\phi,-\phi,0)} & =- \tilde{I}_a^{\OR\theta^{-m},(1\cdot 2)}\, \Lambda (\sigma^3_{aa'} ,\tau^3_{aa'}, \tilde{v}_3; 2 \tilde{V}^{(3)}_{aa'})
\\
&\qquad \text{and permutations of the two-tori}
,
\\
 \Delta_{a,\OR\theta^{-m}}^{(\phi^{(1)},\phi^{(2)},\phi^{(3)})} & =-\frac{\tilde{I}_a^{\OR\theta^{-m}}}{2} \;
  \sum_{n=1}^3 \ln \left[ \left(\frac{\Gamma(|\phi^{(n)}|)}{\Gamma(1-|\phi^{(n)}|)}\right)^{\sgn(\phi^{(n)})}
\right.\\
&\qquad\qquad\qquad\qquad\qquad \cdot\left.
 \left(\frac{\Gamma(\phi^{(n)}+\frac{1}{2} - \sgn(\phi^{(n)}) \cdot H( |\phi^{(n)}|-\frac{1}{2}))}{\Gamma(\frac{1}{2}-\phi^{(n)} + \sgn(\phi^{(n)}) \cdot H ( |\phi^{(n)}|-\frac{1}{2}) ) }\right)\right]
  \\
  &\quad\; + \tilde{I}_a^{\OR\theta^{-m}}   \, \ln(2) \, \sgn(\phi^{(k)}) \cdot \left[H( |\phi^{(k)}|-\frac{1}{2}) +\frac{1}{2}  \right]
  .
\end{aligned}
\end{equation}
The index $k$ in the third equation belongs to the largest absolute value of the three angles,
$0 \leq |\phi^{(i)}|,|\phi^{(j)}| \leq |\phi^{(k)}| \leq 1$, and $H(z)$ is the Heavyside step
function defined in~\eqref{EqApp:Heavyside}. The M\"obius strip amplitudes
contribute only to the untwisted tadpole with
\begin{equation}
2 \sum_{i=1}^3 \tilde{V}^{(i)}_{a,\OR\theta^{-m}}\tilde{I}_a^{\OR\theta^{-m},(j \cdot k)} \int dl
\end{equation}
and $(i,j,k)$ cyclic permutations of (1,2,3),
which serve to cancel the annulus tadpoles according to the first line of the
rewritten RR tadpole cancellation condition~\eqref{Eq:RRtcc-rewritten}.

%%%%%%%%%%%%%%%%%%%%%
\subsection{The complete thresholds for \texorpdfstring{$Sp(2M)$}{Sp(2M)} and  \texorpdfstring{$SO(2M)$}{SO(2M)} gauge groups}\label{Ss:Sp2M}

The beta function coefficients for $Sp(2M_x)$ gauge factors in~\eqref{Eq:beta-Sp(2M)} together with the simplified rewritten 
RR tadpole cancellation conditions for an orientifold invariant D6-brane $x$, 
\begin{equation}\label{Eq:RRtcc-rewritten-Sp}
\begin{aligned}
0 &= \sum_{b} N_b \,   \kappa_{xb} - 2 \, \tilde{\kappa}_{x,\OR},
\\
0&= \sum_{b} N_b \,  \lambda_{xb} ,
\end{aligned}
\end{equation}
serve as guiding principle to read off the correct prefactors of the threshold contributions to the $Sp(2M_x)$ gauge couplings 
in table~\ref{tab:Amplitudes-thresholds}. 
In terms of the notation introduced in section~\ref{Ss:SUN}, the gauge threshold for a symplectic gauge factor is given by
\begin{equation}
\Delta_{Sp(2M_x)} = \sum_a N_a \tilde{\Delta}_{ax}^{\rm total} + \frac{1}{2} \, \Delta_{x,\OR}^{\rm total},
\end{equation}
where for all other symplectic gauge factors $Sp(2M_y)$, one has to identify $N_a=M_y$.

Even though $SO(2M_y)$ gauge factors do not occur in the explicit examples in section~\ref{S:Examples}, they can also only appear on
orientifold invariant D6-branes and provide the same rewritten RR tadpole cancellation conditions~\eqref{Eq:RRtcc-rewritten-Sp} as the symplectic 
gauge factors. Together with the analogous form of the beta function coefficients~\eqref{Eq:beta-Sp(2M)} and~\eqref{Eq:beta-SO(2M)} for symplectic and orthogonal gauge factors,
we arrive at the gauge threshold correction for an $SO(2M_y)$ gauge group
\begin{equation}
\Delta_{SO(2M_y)} = \sum_a N_a \tilde{\Delta}_{ay}^{\rm total} + \frac{1}{2} \, \Delta_{y,\OR}^{\rm total}
.
\end{equation}

%%%%%%%%%%%%%%%%%%%%%%%%%%%%%%%%%%%%%%%%%%%%%%%%%%%%%%%%%%%%%%%%%%%%%%%%5
\subsection{Threshold corrections for massless \texorpdfstring{$U(1)$}{U(1)} gauge factors}\label{S:U(1)}

The gauge threshold corrections for Abelian groups are computed from the same type of annulus and M\"obius strip amplitudes as for $SU(N)$ groups,
but with different prefactors since $aa$ strings do not carry any $U(1)$ charge, $ab$ strings for $a\neq b$ have charge 1 
under $U(1)_a \subset U(N_a)$, and $aa'$ strings have charge 2. The resulting prefactors can again be read off from the 
beta function coefficients. 
Comparing the beta function coefficient~\eqref{Eq:beta-U(1)a} for an $U(1)_a$ gauge factor with that of an $SU(N_a)$ gauge factor in~\eqref{Eq:beta-SU(N)} leads to the gauge threshold correction for an Abelian subgroup 
\begin{equation}\label{Eq:Threh-U(1)a}
\begin{aligned}
\Delta_{U(1)_a} &=N_a \left(4 \, N_a  \tilde{\Delta}_{aa'}^{\rm total} + 
2\,  \sum_{b\neq a} N_b 
\left( \tilde{\Delta}_{ab}^{\rm total} + \tilde{\Delta}_{ab'}^{\rm total}
\right)  + 2 \,  \Delta_{a,\OR}^{\rm total} \right)
\\
&= 2 \, N_a \, \Delta_{SU(N_a)} + 2 \, N_a^2 \left(\tilde{\Delta}_{aa'}^{\rm total} -  \tilde{\Delta}_{aa}^{\rm total}  \right)
,
\end{aligned}
\end{equation}
where the individual threshold contributions are those defined in equation~\eqref{Eq:Threh-SU(N)}.
The global factor of two arises from the different canonical normalizations of the kinetic terms of $SU(N)$ and $U(1)$ 
gauge factors discussed e.g. in~\cite{Ghilencea:2002da,Gmeiner:2008xq}, and the different factor for the $aa'$ contributions is due to the different charge.
If the individual $U(1)$ factor is anomaly-free and massless, its tadpole contributions to the gauge threshold amplitudes cancel 
upon RR tadpole cancellation.

If on the other hand, the massless Abelian gauge factor $U(1)_X$ is a linear combination as in equation~\eqref{Eq:Def-U(1)-massless}, also the gauge threshold is a superposition which is most easily obtained by comparing the corresponding beta function coefficient~\eqref{Eq:beta-U(1)X} with that of the $U(1)_a$ and $SU(N_a)$ factors~\eqref{Eq:beta-SU(N)}, \eqref{Eq:beta-U(1)a}: 
\begin{equation}\label{Eq:Threh-U(1)X}
\Delta_{U(1)_X} = \sum_i x_i^2 \Delta_{U(1)_i} + 4 \sum_{i<j} N_iN_jx_ix_j\left(-\tilde{\Delta}_{ij}^{\rm total} + \tilde{\Delta}_{ij'}^{\rm total}  \right) 
.
\end{equation}
The linear combination of tadpole contributions to the threshold amplitudes vanishes also for this kind of massless Abelian gauge factor upon RR tadpole cancellation. This can be seen as follows. A massless $U(1)_X$ factor wraps and orientifold invariant cycle
\begin{equation}
\Pi_X = \sum_i N_i x_i \Pi_i \stackrel{!}{=} \Pi_X^{\prime}.
\end{equation}
After using the rewritten RR tadpole cancellation condition~\eqref{Eq:RRtcc-rewritten}, i.e. $\kappa_{SU(N_a)}=0$, a single $U(1)_a$ factor contributes
\begin{equation}
\kappa_{U(1)_a} = 2 \, N_a^2\left( \kappa_{aa'} - \kappa_{aa}\right)
\end{equation}
to the bulk tadpole of the threshold amplitudes. The massless linear combination must therefore fulfill
\begin{equation}\label{Eq:U(1)-no-tad}
\begin{aligned}
0 &\stackrel{!}{=}  \sum_i x_i^2 \kappa_{U(1)_i} + 4 \sum_{i<j} N_iN_jx_ix_j \left( -\kappa_{ij} + \kappa_{ij'} \right)
\\
&=2 \, \sum_{i,j} \left(N_ix_i\right)  \left(N_jx_j\right) \left( -\kappa_{ij} + \kappa_{ij'} \right)
,
\end{aligned}
\end{equation}
which vanishes indeed according to the arguments in appendix~\ref{AppSs:RRtcc-rewrite} since the second line in~\eqref{Eq:U(1)-no-tad} corresponds to the symmetrized version of the intersection number
\begin{equation}\label{Eq:U(1)X-no-tcc}
\begin{aligned}
0 = \Pi_X \circ & \underbrace{\left( \Pi_X^{\prime} - \Pi_X \right)}. 
\\
& \qquad =0 \text{ for a massless } U(1)_X
\end{aligned}
\end{equation}
The vanishing of the twisted part of the tadpoles is shown in the same way.

We have checked explicitly for the examples in the following section~\ref{S:Examples} that all tadpole contributions 
to the massless $U(1)$ factors do indeed vanish.

%%%%%%%%%%%%%%%%%%%%%%%%%%%%%%%%%%%%%%%%%%%%%%%%%%%%%%%%%%%%%%%%%%%%%%%%5
\section{Examples with Standard Model like spectra}\label{S:Examples}

In the following we give four examples of supersymmetric globally consistent intersecting D6-models, one on $T^6/\Z_6$ and three on $T^6/\Z_6'$, 
and use them to demonstrate explicitly how to compute the gauge threshold corrections.
The motivation for choosing these particular discrete orbifold groups is phenomenological -- as it turns out it is possible in these
set-ups to obtain large ensembles of models with the gauge group and chiral matter content of the MSSM~\cite{Honecker:2004kb,Gmeiner:2007we}, in the case of $T^6/\Z_6'$ also without any chiral exotic matter~\cite{Gmeiner:2007zz,Gmeiner:2008xq}.\footnote{For more Standard Model building considerations on $T^6/\Z_6'$ see also 
~\cite{Bailin:2006zf,Bailin:2007va,Bailin:2008xx}.}
Models without \emph{chiral exotic matter}  do not exclude
non-chiral states that are charged under both the Standard Model and some hidden gauge group. 
These (potentially massive) states might even be desirable for
the mediation of supersymmetry breaking.

The $T^6/\Z_6$ model was described first in~\cite{Honecker:2004kb} and was also found in
a systematic computer search for models on this background~\cite{Gmeiner:2007we}.
The two examples on $T^6/\Z_6'$ have been described in~\cite{Gmeiner:2008xq,Gmeiner:2007zz} and are in fact
two variations of the same setup with many common features. The difference between the two models that we
consider lies in the fact that one of them includes a hidden sector (with one or two stacks of D6-branes),
while the other one has just the Standard Model gauge group plus a $B-L$ symmetry. The models also differ in the choice of the 
toroidal complex structure parameter.

For more details on the construction of the models and how they are related to generic supersymmetric solutions to the RR tadpole cancellation conditions on
the respective orbifold backgrounds we refer the reader to the original publications.

%%%%%%%%%%%%%%%%%%%%%
\subsection{Example 1: The \texorpdfstring{$T^6/\Z_6$}{T6/Z6} model}\label{Ss:Ex1}

In this section, we first describe the orbifold action and then give the D6-brane configuration in terms of wrapping numbers, angles, discrete displacements 
and Wilson lines together with toroidal and $\Z_2$ invariant intersection numbers. We proceed by computing the full massless spectrum and the resulting 
beta function coefficients for the unbroken gauge group after the Green-Schwarz mechanism has rendered some $U(1)$ factors massive, and finally we explicitly determine the gauge thresholds for these gauge groups.

\subsubsection{Orbifold and orientifold action for \texorpdfstring{$T^6/\Z_6$}{T^6/Z_6} on {\bf AAB}}\label{AppSs:Z6}

The orbifold action~\eqref{Eq:Z2Naction} is given by
\begin{equation}\label{Eq:Z6action}
\theta: \, z_i \longrightarrow e^{2 \pi i v_i} z_i 
\quad 
{\rm with } 
\quad
\vec{v}=\frac{1}{6}(1,1,-2),
\end{equation}
%%%%%%%%%%%%%%%%%%%%%%%%%%%%%%%%%%%%%%%%%%%%%%%%%%%%%%%%%%%%%%%%%%%%%%%%%%%%%%
and the wrapping numbers $(n_i,m_i)$ along the toroidal one-cycles $(\pi_{2i-1},\pi_{2i})$ 
transform under the $\Z_6$ action $\theta$ given in~\eqref{Eq:Z6action} and the orientifold projection $\OR$ as defined in~\eqref{Eq:OmegaAction}, 
on the  {\bf AAB} lattice as
\begin{equation}
\begin{aligned}
& \left(\begin{array}{cc} 
n_1 & m_1 \\ n_2 & m_2 \\ n_3 & m_3
\end{array}\right)
\stackrel{\theta}{\longrightarrow}
\left(\begin{array}{cc} 
-m_1 & n_1+m_1 \\ -m_2 & n_2 +m_2 \\ m_3 & -(n_3 +m_3)
\end{array}\right)
\stackrel{\theta}{\longrightarrow}
\left(\begin{array}{cc} 
-( n_1+m_1) & n_1 \\ -(n_2 +m_2) & n_2 \\  -(n_3 +m_3) & n_3
\end{array}\right)
\\
& \downarrow \OR
\\
& \left(\begin{array}{cc} 
n_1+m_1 & -m_1 \\ n_2+m_2 & -m_2 \\ m_3 & n_3
\end{array}\right)
\stackrel{\theta}{\longrightarrow}
\left(\begin{array}{cc} 
m_1 & n_1 \\ m_2 & n_2 \\ n_3 & -(n_3 +m_3)
\end{array}\right)
\stackrel{\theta}{\longrightarrow}
\left(\begin{array}{cc} 
-n_1 & n_1+m_1 \\ - n_2 & n_2 +m_2  \\ -(n_3 +m_3) & m_3
\end{array}\right)
.
\end{aligned}
\end{equation}
The three complex structure moduli which descend from the torus are frozen by the requirement of having a $\Z_6$ invariant factorizable lattice,
the $SU(3)^3$ group lattice, see figure~\ref{Fig:Z3-Z6lattice}.

%%%%%%%%%%%%%%%%%%%%%%%%%%%%%%%%%%%%%%%%%%%%%%%%%%%%

\subsubsection{Brane configuration and intersection numbers}

In~\cite{Honecker:2004kb}, a supersymmetric model with three chiral matter generations of $SU(3) \times SU(2) \times U(1)_Y \times U(1)_{B-L}$ was given.
It consists of five stacks of D6-branes, labelled by $a\ldots e$.
We list the angles of the different D6-brane stacks with respect to the $\OR$ plane, their toroidal wrapping numbers $(n_i,m_i)$, discrete
displacements $\vec{\sigma}$ and Wilson lines $(\tau^1,\tau^2)$ on $T^2_1 \times T^2_2$ and $\Z_2$ eigenvalue $\tau^0$ in table~\ref{tab:smmodelz6}.

\mathtab{
\begin{array}{|c|c||c|c|c| }\hline
{\rm brane} & \frac{\rm Angle}{\pi} \text{ w.r.t. } \OR & (n_1,m_1;n_2,m_2;n_3,m_3) & \vec{\sigma} & (\tau^0,\tau^1,\tau^2)
\\\hline\hline
a & (0,-1/3,1/3) & (1,0;1,-1;-1,2) & \vec{0} & (1,0,0)
\\
b &  & & & (1,1,0)
\\
d &  & & & (0,0,0)
\\\hline
c & (-1/3,-1/3,2/3) & (1,-1;1,-1;-2,1) & (0,0;1,0) & (1,0,1)
\\
e &  & & & (1,1,1)
\\\hline
\end{array}
}{smmodelz6}{Geometrical setup of the supersymmetric Standard Model example on the {\bf AAB} lattice in the 
$T^6/\mathbb{Z}_6$ background.}

The toroidal and $\Z_2$ invariant intersection numbers with signs, from which multiplicities of
bifundamental, symmetric and antisymmetric massless matter states of this model are computed along the rules in appendix~\ref{AppSs:Spectrum},
are listed in tables~\ref{tab:smz6inter1} to~\ref{tab:smz6inter5}.
It should be noted that we are keeping the notation of~\cite{Honecker:2004kb}
with $\Z_2$ subgroup $3 \vec{v} = \frac{1}{2}(1,-1,0)$.
Since the third two-torus $T_3^2$ is the $\Z_2$ invariant one,
the labels of the tori in the general formulae for gauge threshold corrections in section~\ref{sec:Thresholds} and~\ref{sec:ThreshRes}  have to be permuted.

The multiplicities of adjoint representations are computed from
\begin{equation}
\begin{aligned}
& I_{x(\theta x)} = I_{x(\theta x)}^{\Z_2} = -3,
\qquad
I_{x(\theta^2 x)} = I_{x(\theta^2 x)}^{\Z_2} =3,
\qquad
{\rm with}
\quad 
x=a,b,d,
\\
&  I_{y(\theta y)} = - I_{y(\theta y)}^{\Z_2} = -3,
\qquad
I_{y(\theta^2 x)} =- I_{y(\theta^2 y)}^{\Z_2} =3,
\qquad
{\rm with}
\quad 
y=c,e.
\end{aligned}
\end{equation}

Branes $c$ and $e$ are orientifold invariant if no continuous displacement or Wilson line on $T^2_3$ is switched on.

\mathtabfix{
\begin{array}{|c||c|c|c||c|c|c||c|c|c||c|c|c|}\hline
& I_{\cdot b} & I_{\cdot (\theta b)} & I_{\cdot (\theta^2 b)}
&  I_{\cdot c} & I_{\cdot (\theta c)} & I_{\cdot (\theta^2 c)}
& I_{\cdot d} & I_{\cdot (\theta d)} & I_{\cdot (\theta^2 d)}
& I_{\cdot e} & I_{\cdot (\theta e)} & I_{\cdot (\theta^2 e)}
\\\hline \hline
a & 0_{123} & -3 & 3
& (-3) \cdot 0_2 & 0_1 \cdot (-3) & (-1) \cdot 0_{3}
& 0_{123} & -3 & 3 
& (-3) \cdot 0_2 & 0_1 \cdot (-3) & (-1) \cdot  0_{3}
\\\hline
b & \multicolumn{3}{|c|}{} & (-3) \cdot 0_2 & 0_1 \cdot (-3) & (-1) \cdot 0_{3}
& 0_{123} & -3 & 3 
& (-3) \cdot 0_{2} & 0_{1}  \cdot (-3) & (- 1) \cdot 0_{3}
\\\hline
c & \multicolumn{6}{|c|}{} & (-3) \cdot 0_{2} &  (-1) \cdot 0_{3} & (-3) \cdot 0_{1} 
& 0_{123} & -3 &  3
\\\hline
d & \multicolumn{9}{|c|}{} & (-3) \cdot 0_{2}  & 0_{1} \cdot (-3) & (-1) \cdot 0_{3}
\\\hline
\end{array}
}{smz6inter1}{Toroidal intersection numbers of the supersymmetric Standard Model example on $T^6/\mathbb{Z}_6$, Part I.
For vanishing intersection numbers, the lower index, e.g. $0_2$, indicates that bulk cycles are parallel on the corresponding two-torus, and 
the intersection number on the remaining four-torus is explicitly given. $0_{123}$ indicates toroidal cycles which are parallel along the whole six-torus.}

\mathtabfix{
\begin{array}{|c||c|c|c||c|c|c||c|c|c||c|c|c|}\hline
& I_{\cdot b}^{\Z_2} & I_{\cdot (\theta b)}^{\Z_2} & I_{\cdot (\theta^2 b)}^{\Z_2}
&  I_{\cdot c}^{\Z_2} & I_{\cdot (\theta c)}^{\Z_2} & I_{\cdot (\theta^2 c)}^{\Z_2}
& I_{\cdot d}^{\Z_2} & I_{\cdot (\theta d)}^{\Z_2} & I_{\cdot (\theta^2 d)}^{\Z_2}
& I_{\cdot e}^{\Z_2} & I_{\cdot (\theta e)}^{\Z_2} & I_{\cdot (\theta^2 e)}^{\Z_2}
\\\hline \hline
a & \emptyset  & -3 & 3
& \emptyset  & 6 & (-1) \cdot 0_{3}
& (-4) \cdot 0_{3} & 3 & - 3
& \emptyset  & \emptyset  & (-1) \cdot 0_{3}
\\\hline
b & \multicolumn{3}{|c|}{} & \emptyset  & \emptyset & (-1) \cdot 0_{3}
& \emptyset  & 3 & - 3
& \emptyset & 6 & (-1) \cdot 0_{3}
\\\hline
c & \multicolumn{6}{|c|}{} & \emptyset \cdot (-3) &  1 \cdot 0_{3} & 6
& \emptyset  & 3 & - 3
\\\hline
d & \multicolumn{9}{|c|}{} & \emptyset  & \emptyset & 1 \cdot 0_{3}
\\\hline
\end{array}
}{smz6inter2}{$\Z_2$ invariant intersection numbers with appropriate signs from relative $\Z_2$ eigenvalues and discrete Wilson lines
of the supersymmetric Standard Model example on $T^6/\mathbb{Z}_6$, Part I.
$\emptyset$ denotes a $\Z_2$ invariant intersection number which vanishes due to a discrete relative Wilson line or displacement.}

\mathtabfix{
\begin{array}{|c||c|c|c||c|c|c||c|c|c||c|c|c|}\hline
& I_{\cdot b'} & I_{\cdot (\theta b')} & I_{\cdot (\theta^2 b')} 
& I_{\cdot d'} & I_{\cdot (\theta d')} & I_{\cdot (\theta^2 d')}
& I_{\cdot b'}^{\Z_2} & I_{\cdot (\theta b')}^{\Z_2} & I_{\cdot (\theta^2 b')}^{\Z_2}
& I_{\cdot d'}^{\Z_2} & I_{\cdot (\theta d')}^{\Z_2} & I_{\cdot (\theta^2 d')}^{\Z_2}
\\\hline \hline
a & 0_{1}  \cdot (-3) & (-3) \cdot 0_{2}  & (-1) \cdot 0_{3}
 & 0_{1}  \cdot (-3) & (-3) \cdot 0_{2}  &  (-1) \cdot 0_{3}
& \emptyset & -6 & (-1) \cdot 0_3
& -6 & 6 & 1 \cdot 0_3
\\\hline
b & \multicolumn{3}{|c|}{} 
 & 0_1 \cdot (-3) & (-3) \cdot  0_{2}  &(-1) \cdot 0_{3}
 & \multicolumn{3}{|c|}{} 
 & \emptyset & 6 & 1 \cdot 0_3
\\\hline
\end{array}
}{smz6inter3}{Intersection numbers of the supersymmetric Standard Model example on $T^6/\mathbb{Z}_6$, Part II. Since branes $y=c,e$
have $y'=(\theta y)$ for no continuous displacement along $T^2_3$, the intersection numbers involving the orientifold images $(\theta^k y')$
can be read off from tables~\ref{tab:smz6inter1} and~\ref{tab:smz6inter2}.}

\mathtab{
\begin{array}{|c||c|c||c|c||c|c|}\hline
x & I_{xx'} & I_{xx'}^{\Z_2} & I_{x(\theta x')} & I_{x(\theta x')}^{\Z_2} & I_{x(\theta^2 x')} & I_{x(\theta^2 x')}^{\Z_2} 
\\\hline\hline
a,b,d & 0_{1}  \cdot (-3 ) &   6
& (-3) \cdot 0_{2}  &  -6 
& (-1) \cdot 0_{3}   & (-1) \cdot 0_{3}
\\\hline
\hline\hline
x & \tilde{I}_x^{\OR}  & \tilde{I}_x^{\OR\theta^{-3}} & \tilde{I}_x^{\OR\theta^{-1}}  & \tilde{I}_x^{\OR\theta^{-4}} & \tilde{I}_
x^{\OR\theta^{-2}}  & \tilde{I}_x^{\OR\theta^{-5}} 
\\\hline\hline
a,b,d & 0_1  \cdot (-3) & 6
& -6 & (-3) \cdot 0_{2} 
& (-1) \cdot 0_{3} & (-1) \cdot 0_{3}
\\\hline
c,e & -3 & 3 & (-4) \cdot 0_3 & 0_{123} & 3 & -3
\\\hline
\end{array}
}{smz6inter5}{Toroidal and $\Z_2$ invariant intersection numbers of the supersymmetric Standard Model example on $T^6/\mathbb{Z}_6$ contributing to antisymmetric and symmetric representations. Branes $y=c,e$ fulfill $y'=(\theta y)$.}

%%%%%%%%%%%%%%%%%%%%%%%%%%%%%%%%%%%%%%%%

\subsubsection{Spectrum}

Using the intersection numbers among the various D6-brane stacks one can compute the complete massless matter
spectrum as described in~\cite{Gmeiner:2008xq} and summarized in appendix~\ref{AppSs:Spectrum}.
The notation is such that we give the representations of $(S){\bf U(3)_a}$ and $(S){\bf U(2)_b}$ in bold and the charges
under the hypercharge and the massless $B-L$ symmetry as subscripts.  
An index $m$ indicates that this multiplet can become massive under a continuous relative displacement of the relevant D6-brane stacks along $T^2_3$,
and the upper indices in parenthesis are the (unphysical) charges under $U(1)_c$ and $U(1)_d$. Brane stack $c$ carries the gauge group $SO(2)_c$ or 
$Sp(2)_c$ if it sits on the orientifold planes, but in the following, the gauge group will be explicitly broken along a flat direction by a continuous parallel displacement from the origin on $T^2_3$. Alternatively, a continuous Wilson line could be switched on along the same two-torus.

The hyper charge and $B-L$ assignments for the Standard Model realized on four stacks of D6-branes $a \ldots d$ are 
\begin{equation}\label{Eq:Ex1-U(1)-charges}
Q_Y= \frac{1}{6} Q_a +  \frac{1}{2} Q_c + \frac{1}{2} Q_d 
,
\qquad
Q_{B-L} = \frac{1}{3} Q_a + Q_d.
\end{equation}
In this case, the spectrum contains three generations of quarks and leptons plus
some exotic fields $\tilde{H}$ which, after $U(1)_b \subset U(2)_b$ has acquired a mass through the generalized Green Schwarz mechanism, are non-chiral.
When $SO/Sp(2)_e$ is broken to $U(1)_e$ through a continuous open string modulus, i.e. a displacement from the origin or a Wilson line, this Abelian group is massless and can be included in the 
hyper charge assignment, $Q_{Y'}=  Q_Y + \frac{1}{2} Q_e$, and the fields $\tilde{H}$ can then be interpreted as non-standard Higgs-up and -down pairs $H_u + H_d$.

The massless open string spectrum consists of the gauge group $SU(3)_a \times SU(2)_b \times U(1)_Y \times U(1)_{B-L} \times SO/Sp(2)_e$ with the Abelian charges defined 
by~\eqref{Eq:Ex1-U(1)-charges} and two kinds of matter spectra $[C]+[V]$: 
\begin{itemize}
\item the `chiral' spectrum stemming from non-vanishing intersection numbers
%%%
\begin{equation}\label{Eq:ex1-explicitspectrum-chiral}
\begin{aligned}
{}
[C] &=3\times\bigg[
    \left(\3,\2\right)_{\bf 1/6, 1/3}^{(0,0)}
  + \left(\ov{\3},\1\right)_{\bf 1/3,-1/3}^{(1,0)}
  + \left(\ov{\3},\1\right)_{\bf -2/3, -1/3}^{(-1,0)}
  + \left(\1,\1\right)_{\bf 1,1}^{(1,1)}
\\
&\qquad\qquad
  +  \left(\1,\1\right)_{\bf 0,1}^{(-1,1)}
  + \left(\1,\ov{\2}\right)_{\bf -1/2,-1}^{(0,-1)}
+ \left(\1,\ov{\2};\2 \right)_{\bf 0,0}^{(0,0)}
\bigg]
\\
& \equiv 3 \times \bigg[Q_L +d_R + u_R + e_R +\nu_R + L +\tilde{H}  \bigg]
.
\end{aligned}
\end{equation}
It contains three quark-lepton generations with right-handed neutrinos and three generations of (exotic) Higgs candidates.
The latter become non-chiral upon the Green-Schwarz mechanism which renders $U(1)_b \subset U(2)_b$ massive.
%%%%%%%
\item
the `truly non-chiral' spectrum 
\begin{equation}\label{Eq:ex1-explicitspectrum-non-chiral}
\begin{aligned}
{}
[V] &= 4 \times \left(\bf{8},\1\right)_{\bf 0,0}^{(0,0)}
  + 4 \times \left(\1,\3\right)_{\bf 0,0}^{(0,0)}
  + 16  \times \left(\1,\1\right)_{\bf 0,0}^{(0,0)}
\\
&\quad
  + z_1 \times (\1,\1;\3_S)_{\bf 0,0}^{(0,0)}
+ z_2  \times (\1,\1;\1_A)_{\bf 0,0}^{(0,0)}
\\
&\quad + \bigg[
  1_m \times   \left(\3,\2\right)_{\bf 1/6,1/3}^{(0,0)}
+3 \times     \left(\3,\ov{\2}\right)_{\bf 1/6,1/3}^{(0,0)}
+  1_m \times  \left(\ov{\3},\1\right)_{\bf 1/3,-1/3}^{(1,0)}
\\
&\qquad\quad
+  1_m \times  \left(\ov{\3},\1\right)_{\bf -2/3,-1/3}^{(-1,0)}
  + 2_m \times \left(\ov{\3},\1\right)_{\bf 1/3,2/3}^{(0,1)}
  + 3 \times \left(\ov{\3},\1\right)_{\bf -2/3,-4/3}^{(0,-1)}
\\
&\qquad\quad
+ 1_m \times \left(\ov{\3},\1;\2 \right)_{\bf -1/6,-1/3}^{(0,0)}
+  1_m \times \left(\1,\2;\2 \right)_{\bf 0,0}^{(0,0)}
+  1_m \times  \left(\1,\2\right)_{\bf 1/2,0}^{(1,0)}
\\
&\qquad\quad
+  1_m \times  \left(\1,\2\right)_{\bf  -1/2,0}^{(-1,0)}
  + (1_m + y ) \times \left(\1,\1 \right)_{\bf 1,0}^{(2,0)}\\
&\qquad\quad
 + \left(3+1_m  \right) \times (\ov{\3}_A,\1)_{\bf 1/3,2/3}^{(0,0)}
  + \left(3+ 1_m\right) \times (\1,\1_A)_{\bf 0,0}^{(0,0)}
   + \;c.c.\;\bigg].
\end{aligned}
\end{equation}
It consists of adjoints, antisymmetric and symmetric representations of the (pseudo)real group $SO/Sp(2)_e$ 
and bifundamental and antisymmetric matter in ${\cal N}=2$ hyper multiplets of the unitary and Abelian groups.
An index $m$ denotes that these multiplets acquire a mass if a continuous relative Wilson line on $T^2_3$ is switched on.
This applies to most of the bifundamental representations in the square bracket.

By a simple counting of massless GSO projected states and Chan-Paton matrices as in appendix~C in~\cite{Gmeiner:2008xq}, we obtain $y \leq 3$ and $z_1+z_2 \leq 4$. 
The analysis of the beta function contributions in appendix~\ref{App:SO-Sp} gives $(z_1,z_2)=(1,0)$ and $y=0$ and that brane $e$ carries an $Sp(2)_e$ gauge group.
%%%%%%%%%%%%%%%%%
\end{itemize}

The beta function coefficients of the Standard Model group are computed from equations~\eqref{Eq:beta-SU(N)} and~\eqref{Eq:beta-U(1)X} to be 
\begin{equation}
\begin{aligned}
b_{SU(3)_a} &= 21 + 9_m
, \qquad\qquad &
b_{SU(2)_b} &= 20 + 7_m
,\\ 
b_{U(1)_{B-L}} &= 60 + 12_m
, \qquad\qquad &
b_{U(1)_Y} &= 17 + 2y + 8_m
,
\end{aligned}
\end{equation}
where an index $m$ indicates that these contributions are absent if the corresponding non-chiral multiplets in~\eqref{Eq:ex1-explicitspectrum-non-chiral} have acquired a mass.

If the hyper charge is replaced by $Q_Y \rightarrow Q_{Y'} = Q_Y + \frac{1}{2} Q_e$, its beta function is 
shifted by $\frac{b_{U(1)_e}}{4} = 3 + 5_m +2z_1$. All other possible contributions are zero due to $-\varphi^{xe} + \varphi^{xe'}=0$ for any brane $x$.

If on the other hand $Sp(2)_e$ remains unbroken, its beta function coefficient is $b_{Sp(2)_e} = -1 +5_m$, which means that in case all brane stacks
have relative displacements on the third torus, the gauge coupling becomes stronger at lower energies.

%%%%%%%%%%%%%%%%%%%%%%%%%%%%%%%%%%%%%%%%%%%%%%%%%%%%%%%%

\subsubsection{Threshold corrections for example 1}\label{Ss:Thresholds-Ex1}

Since there are only two types of toroidal cycles, their contributions to the thresholds are computed in a very economical way.
For $x_i \in \{ a,b,d\}$ and $y_j \in \{c,e\}$, one obtains the following toroidal annulus contributions
\begin{equation}
\begin{aligned}
\tilde{\Delta}^{\unity}_{x_ix_j} &= \tilde{\Delta}^{\unity}_{y_iy_j} = 0 ,
\\
\tilde{\Delta}^{\unity}_{x_i(\theta x_j)} &=\tilde{\Delta}^{\unity}_{x_i(\theta^2 x_j)} =\tilde{\Delta}^{\unity}_{y_i(\theta y_j)} =
\tilde{\Delta}^{\unity}_{y_i(\theta^2 y_j)}
= -\frac{9}{4} \ln 2  ,
\\%%%%
\tilde{\Delta}^{\unity}_{x_ix_j'} &=  -\frac{3}{2}\Lambda ( \sigma^1_{x_ix_j'},\tau^1_{x_ix_j'},v_1;\frac{2}{\sqrt{3}}) ,
\\
\tilde{\Delta}^{\unity}_{x_i(\theta x_j')} &=-\frac{3}{2}\Lambda ( \sigma^2_{x_ix_j'},\tau^2_{x_ix_j'},v_2;\frac{2}{\sqrt{3}}) ,
\\
\tilde{\Delta}^{\unity}_{x_i(\theta^2 x_j')} &= - \frac{1}{2}\Lambda(\sigma^3_{x_ix_j'},\tau^3_{x_ix_j'},v_3;2 \sqrt{3}) ,
\\%%%
\tilde{\Delta}^{\unity}_{x_iy_j} &=-\frac{3}{2}\Lambda ( \sigma^2_{x_iy_j},\tau^2_{x_iy_j},v_2;\frac{2}{\sqrt{3}}) ,
\\
\tilde{\Delta}^{\unity}_{x_i(\theta y_j)} &=-\frac{3}{2}\Lambda ( \sigma^1_{x_iy_j},\tau^1_{x_iy_j},v_1;\frac{2}{\sqrt{3}}) ,
\\
\tilde{\Delta}^{\unity}_{x_i(\theta^2 y_j)} &=- \frac{1}{2}\Lambda(\sigma^3_{x_iy_j},\tau^3_{x_iy_j},v_3; 2 \sqrt{3}) ,
\end{aligned}
\end{equation}
with $\Lambda(\sigma,\tau,v;V)$ defined in~\eqref{Eq:Def-Lambda},
and from the orientifold planes with $x\in \{a,b,d\}$ and $y \in \{c,e\}$ the M\"obius strip contributions
\begin{equation}
\begin{aligned}
\left.\begin{array}{c}
\Delta^{\OR}_x = 3 \, \Lambda(0,0,2v_1;\frac{4}{\sqrt{3}}),
\\
\Delta^{\OR\theta^{-4}}_x = 3 \, \Lambda(0,0,2v_2;\frac{4}{\sqrt{3}}),
\\
\Delta^{\OR\theta^{-1}}_x =\Delta^{\OR\theta^{-3}}_x = 6 \, \ln 2 ,
\\
\Delta^{\OR\theta^{-2}}_x =\Delta^{\OR\theta^{-5}}_x  = \Lambda(\sigma^3_{xx'},\tau^3_{xx'},2v_3;4 \sqrt{3})
\end{array}\right\} & \Rightarrow \, \Delta_{x,\OR}^{\rm total} = 
\left\{ \begin{array}{c}
12 \, \ln 2 \\+  3 \, \Lambda(0,0,2v_1;\frac{4}{\sqrt{3}}) \\+ 3 \, \Lambda(0,0,2v_2;\frac{4}{\sqrt{3}}) \\+ 2 \,  \Lambda(\sigma^3_{xx'},\tau^3_{xx'},2v_3;4 \sqrt{3})
\end{array}\right.
\!,
\\
\left.\begin{array}{c}
\Delta^{\OR}_y = \Delta^{\OR\theta^{-2}}_y= 9 \, \ln 2 - \frac{9}{2} \ln 5 
\\
\Delta^{\OR\theta^{-1}}_y = 4 \, \Lambda(\sigma^3_{yy'}, \tau^3_{yy'},2v_3;4 \sqrt{3}),
\\
\Delta^{\OR\theta^{-4}}_y = 0,
\\
\Delta^{\OR\theta^{-3}}_y = \Delta^{\OR\theta^{-5}}_y = -3 \, \ln 2 +\frac{9}{2} \ln 5
\end{array}\right\} & \Rightarrow \, \Delta_{y,\OR}^{\rm total} = 
\left\{ \begin{array}{c}
12 \, \ln 2 \\ +4 \, \Lambda(\sigma^3_{yy'}, \tau^3_{yy'},2v_3;4 \sqrt{3})
\end{array}\right.
\!.
\end{aligned}
\end{equation}
The twisted contributions to the annulus amplitudes depend on the relative discrete displacements and Wilson lines through the 
$\Z_2$ invariant intersection numbers as well as the relative $\Z_2$ eigenvalues given in tables~\ref{tab:smz6inter2} to~\ref{tab:smz6inter5}. 
Several of them vanish, leaving the non-trivial contributions
\begin{equation}
\begin{aligned}
\tilde{\Delta}^{\Z_2}_{x_ix_j} &=  2 \, \alpha \,  \Lambda(\sigma^3_{x_ix_j},\tau^3_{x_ix_j},v_3;2 \sqrt{3})
\\ &
\qquad {\rm with}
\quad 
\alpha = \left\{ \begin{array}{cc}
1 & x_i=x_j
\\
0 & (x_i,x_j)=(a,b) \; {\rm or} \; (b,d)
\\
-1 & (x_i,x_j)=(a,d)
\end{array}\right.
,
\\
\tilde{\Delta}^{\Z_2}_{x_i(\theta^k x_j)_{k=1,2}} &= -\, \beta \, \frac{5}{4} \, \ln 2 
\\ &
\qquad {\rm with}
\quad 
\beta= \left\{ \begin{array}{cc}
1 & x_i=x_j \; {\rm or}\; (a,b)
\\
-1 & (x_i,x_j) = (a,d) \; {\rm or}\; (b,d)
\end{array}\right.
,
\end{aligned}
\end{equation}

\begin{equation}
\begin{aligned}
\tilde{\Delta}^{\Z_2}_{x_ix_j'} &= \alpha \ln 2 ,
\\
\tilde{\Delta}^{\Z_2}_{x_i(\theta x_j')} &= \beta \ln 2  ,
\\
\tilde{\Delta}^{\Z_2}_{x_i(\theta^2 x_j')} &= -\frac{\beta}{2} \, \Lambda  (\sigma^3_{x_ix_j},\tau^3_{x_ix_j},v_3;2 \sqrt{3})  ,
\end{aligned}
\end{equation}

\begin{equation}
\begin{aligned}
\tilde{\Delta}^{\Z_2}_{y_i y_j} &= 2 \, \delta_{y_i y_j} \, \Lambda(\sigma^3_{y_iy_j},\tau^3_{y_iy_j},v_3;2\sqrt{3})  
,
\\
\tilde{\Delta}^{\Z_2}_{y_i (\theta^k y_j)_{k=1,2}} &= \frac{5}{4} \, \ln 2 ,
\end{aligned}
\end{equation}

\begin{equation}
\begin{aligned}
\tilde{\Delta}^{\Z_2}_{xy} &= 0 ,
\\
\tilde{\Delta}^{\Z_2}_{x(\theta y)} &=  - \zeta \, \ln 2 
\qquad
{\rm with} \quad
\zeta=\left\{\begin{array}{cc} 1 & (x,y)=(a,c) \; {\rm or}\; (b,e) \\ 0 & {\rm else}  \end{array}\right. ,
\\
\tilde{\Delta}^{\Z_2}_{x(\theta^2 y)} &= - \frac{\epsilon}{2} \Lambda(\sigma^3_{xy},\tau^3_{xy},v_3;2\sqrt{3}) 
\qquad
{\rm with} \quad
\epsilon = \left\{\begin{array}{cc}
-1 & (x,y)=(d,e) \\ 1 & {\rm else}
\end{array}\right.
.
\end{aligned}
\end{equation}

In summary, one can write the contributions per fixed stacks of D6-branes as
\begin{equation}
\begin{aligned}
\tilde{\Delta}_{x_ix_j}^{\rm total} &=-\frac{9+5 \beta}{2} \, \ln 2 +2 \, \alpha \, \Lambda(\sigma^3_{x_ix_j},\tau^3_{x_ix_j},v_3;2 \sqrt{3}) ,
\\
\tilde{\Delta}_{x_ix_j'}^{\rm total} &=\left( \alpha + \beta \right) \, \ln 2 -\frac{3}{2} \Lambda(\sigma^1_{x_ix_j'},\tau^1_{x_ix_j'},v_1;\frac{2}{\sqrt{3}})
-\frac{3}{2} \Lambda(\sigma^2_{x_ix_j'},\tau^2_{x_ix_j'},v_2;\frac{2}{\sqrt{3}}) 
\\
& \quad
-\frac{1+\beta}{2} \, \Lambda(\sigma^3_{x_ix_j'},\tau^3_{x_ix_j'},v_3;2 \sqrt{3}) ,
\\
\tilde{\Delta}_{x_iy_j}^{\rm total} &= - \zeta \, \ln 2 - \frac{3}{2}\Lambda(\sigma^1_{x_iy_j},\tau^1_{x_iy_j},v_1;\frac{2}{\sqrt{3}}) 
\\&\quad\;
 - \frac{3}{2}\Lambda(\sigma^2_{x_iy_j},\tau^2_{x_iy_j},v_2;\frac{2}{\sqrt{3}}) 
 -\frac{1+\epsilon}{2} \Lambda(\sigma^3_{x_iy_j},\tau^3_{x_iy_j},v_3;2\sqrt{3})  , 
\\
\tilde{\Delta}_{y_iy_j}^{\rm total} &=-2 \, \ln 2 + 2 \delta_{y_iy_j} \Lambda(\sigma^3_{y_iy_j},\tau^3_{y_iy_j},v_3;2 \sqrt{3}) 
.
\end{aligned}
\end{equation}

The threshold corrections for the non-Abelian gauge factors of the Standard Model are (from here on, for shortness the superscript `total' would belong to every $\tilde{\Delta}$ and is dropped)
\begin{equation}
\begin{aligned}
\Delta_{SU(3)_a} =&\; 3 \, \left( \tilde{\Delta}_{aa} +\tilde{\Delta}_{aa'} \right) + 2 \, \left(  \tilde{\Delta}_{ab} + \tilde{\Delta}_{ab'}\right) 
 + \left( \tilde{\Delta}_{ac} + \tilde{\Delta}_{ac'} \right)
\\
&+ \left( \tilde{\Delta}_{ad} + 
\tilde{\Delta}_{ad'}\right) 
+ \left( \tilde{\Delta}_{ae} + 
\tilde{\Delta}_{ae'}\right)
+\Delta_{a,\OR}
\\
=& -21 \, \ln 2 - 9 \,  \Lambda(0,0,v_1;\frac{2}{\sqrt{3}}) - 6 \,  \Lambda(0,1,v_1;\frac{2}{\sqrt{3}})+ 3 \, \Lambda(0,0,2v_1;\frac{4}{\sqrt{3}})
\\
& - 9 \, \Lambda(0,0,v_2;\frac{2}{\sqrt{3}}) -6 \, \Lambda(1,1,v_2;\frac{2}{\sqrt{3}}) + 3 \, \Lambda(0,0,2v_2;\frac{4}{\sqrt{3}})
\\
&  + 6 \, \Lambda(0,0,v_3;2\sqrt{3})-3 \, \Lambda(\sigma^3_{aa'},\tau^3_{aa'},v_3;2\sqrt{3} )+2 \, \Lambda(\sigma^3_{aa'},\tau^3_{aa'},2v_3;4\sqrt{3})
\\
&
 -2 \, \Lambda(\sigma^3_{ab'},\tau^3_{ab'},v_3; 2\sqrt{3}) 
-\Lambda(\sigma^3_{ac},\tau^3_{ac},v_3;2\sqrt{3})-\Lambda(\sigma^3_{ac'},\tau^3_{ac'},v_3;2\sqrt{3})
\\
&
-2 \, \Lambda(\sigma^3_{ad},\tau^3_{ad},v_3;2\sqrt{3})
- \Lambda(\sigma^3_{ae},\tau^3_{ae},v_3;2\sqrt{3})-\Lambda(\sigma^3_{ae'},\tau^3_{ae'},v_3;2\sqrt{3})
,
\end{aligned}
\end{equation}

\begin{equation}
\begin{aligned}
% \quad
\Delta_{SU(2)_b} =&\; 3 \, \left( \tilde{\Delta}_{ba} +\tilde{\Delta}_{ba'} \right) + 2 \, \left(  \tilde{\Delta}_{bb} + \tilde{\Delta}_{bb'}\right) 
 + \left( \tilde{\Delta}_{bc} + \tilde{\Delta}_{bc'} \right) 
\\
&
+ \left( \tilde{\Delta}_{bd} + \tilde{\Delta}_{bd'}\right) 
+ \left( \tilde{\Delta}_{be} + \tilde{\Delta}_{be'}\right)
 +\Delta_{b,\OR}
\\
=&
- 21 \, \ln 2 - 6 \, \Lambda(0,0,v_1;\frac{2}{\sqrt{3}}) - 9 \, \Lambda(0,1,v_1;\frac{2}{\sqrt{3}}) + 3 \, \Lambda(0,0,2v_1;\frac{4}{\sqrt{3}})
\\
& -9 \, \Lambda(0,0,v_2;\frac{2}{\sqrt{3}}) -6 \, \Lambda(1,1,v_2;\frac{2}{\sqrt{3}}) + 3 \, \Lambda(0,0,2v_2;\frac{4}{\sqrt{3}})
\\
& -3 \, \Lambda(\sigma^3_{ab'},\tau^3_{ab'},v_3; 2\sqrt{3}) -2 \, \Lambda(\sigma^3_{bb'},\tau^3_{bb'},v_3;2\sqrt{3} ) + 4 \, \Lambda(0,0,v_3;2\sqrt{3}) 
\\
&
+2 \, \Lambda(\sigma^3_{bb'},\tau^3_{bb'},2v_3;4\sqrt{3})
-\Lambda(\sigma^3_{bc},\tau^3_{bc},v_3;2\sqrt{3})-\Lambda(\sigma^3_{bc'},\tau^3_{bc'},v_3;2\sqrt{3})
\\
&
-\Lambda(\sigma^3_{be},\tau^3_{be},v_3;2\sqrt{3})-\Lambda(\sigma^3_{be'},\tau^3_{be'},v_3;2\sqrt{3}).
\end{aligned}
\end{equation}

For the massless Abelian groups, one obtains 
\begin{equation}
\begin{aligned}
\Delta_{U(1)_{B-L}}
=&\;\frac{1}{9} \Delta_{U(1)_a} + \Delta_{U(1)_d} +4 \, \left(-\tilde{\Delta}_{ad} + \tilde{\Delta}_{ad'}   \right)
\\
=& - 36 \, \Lambda(0,0,v_1;\frac{2}{\sqrt{3}}) - 16 \, \Lambda(0,1,v_1;\frac{2}{\sqrt{3}}) + 8 \Lambda(0,0,2v_1;\frac{4}{\sqrt{3}})
\\
& - 36 \, \Lambda(0,0,v_2;\frac{2}{\sqrt{3}}) - 16 \, \Lambda(1,1,v_2;\frac{2}{\sqrt{3}}) + 8 \Lambda(0,0,2v_2;\frac{4}{\sqrt{3}})
\\
& -4\, \Lambda(\sigma^3_{aa'},\tau^3_{aa'},v_3;2\sqrt{3} )+ \frac{4}{3} \, \Lambda(\sigma^3_{aa'},\tau^3_{aa'},2v_3;4\sqrt{3})
\\
&
 -\frac{4}{3} \, \Lambda(\sigma^3_{ab'},\tau^3_{ab'},v_3; 2\sqrt{3}) 
-\frac{2}{3} \Lambda(\sigma^3_{ac},\tau^3_{ac},v_3;2\sqrt{3})
\\
&
-\frac{2}{3} \Lambda(\sigma^3_{ac'},\tau^3_{ac'},v_3;2\sqrt{3})
-\frac{16}{3} \, \Lambda(\sigma^3_{ad},\tau^3_{ad},v_3;2\sqrt{3})
\\
&
- \frac{2}{3} \Lambda(\sigma^3_{ae},\tau^3_{ae},v_3;2\sqrt{3})
-\frac{2}{3} \Lambda(\sigma^3_{ae'},\tau^3_{ae'},v_3;2\sqrt{3})
\\
&
-2\, \Lambda(\sigma^3_{dc},\tau^3_{dc},v_3;2\sqrt{3})
-2 \, \Lambda(\sigma^3_{dc'},\tau^3_{dc'},v_3;2\sqrt{3})
\\
&
-4 \, \Lambda(\sigma^3_{dd'},\tau^3_{dd'},v_3;2\sqrt{3})+4 \,  \Lambda(\sigma^3_{dd'},\tau^3_{dd'},2v_3;4\sqrt{3})
,
\end{aligned}
\end{equation}
and
\begin{equation}
\begin{aligned}
\Delta_{U(1)_Y} 
=&\;\frac{1}{4} \Delta_{U(1)_{B-L}} + \frac{1}{4} \Delta_{U(1)_c} + \left(-\tilde{\Delta}_{ac} + \tilde{\Delta}_{ac'} 
-\tilde{\Delta}_{cd} + \tilde{\Delta}_{cd'}   \right)
\\
=& - 3 \, \ln 2 - 15 \, \Lambda(0,0,v_1;\frac{2}{\sqrt{3}}) -7 \, \Lambda(0,1,v_1;\frac{2}{\sqrt{3}}) + 2\, \Lambda(0,0,2v_1;\frac{4}{\sqrt{3}})
\\
&-9 \, \Lambda(0,0,v_2;\frac{2}{\sqrt{3}}) - 13 \, \Lambda(1,1,v_2;\frac{2}{\sqrt{3}}) + 2\, \Lambda(0,0,2v_2;\frac{4}{\sqrt{3}})
\\
& -\Lambda(\sigma^3_{aa'},\tau^3_{aa'},v_3;2\sqrt{3} )+ \frac{1}{3} \, \Lambda(\sigma^3_{aa'},\tau^3_{aa'},2v_3;4\sqrt{3})
 -\frac{1}{3} \, \Lambda(\sigma^3_{ab'},\tau^3_{ab'},v_3; 2\sqrt{3}) 
\\
&
-\frac{2}{3} \Lambda(\sigma^3_{ac},\tau^3_{ac},v_3;2\sqrt{3})-\frac{8}{3} \Lambda(\sigma^3_{ac'},\tau^3_{ac'},v_3;2\sqrt{3})
-\frac{4}{3} \, \Lambda(\sigma^3_{ad},\tau^3_{ad},v_3;2\sqrt{3})
\\
&
- \frac{1}{6} \Lambda(\sigma^3_{ae},\tau^3_{ae},v_3;2\sqrt{3})-\frac{1}{6} \Lambda(\sigma^3_{ae'},\tau^3_{ae'},v_3;2\sqrt{3})
-2 \, \Lambda(\sigma^3_{dc'},\tau^3_{dc'},v_3;2\sqrt{3})
\\
&
- \Lambda(\sigma^3_{dd'},\tau^3_{dd'},v_3;2\sqrt{3})
+  \Lambda(\sigma^3_{dd'},\tau^3_{dd'},2v_3;4\sqrt{3}) 
+ 2 \, \Lambda(0,0,v_3;2\sqrt{3})
\\
&
-\Lambda(\sigma^3_{bc},\tau^3_{bc},v_3;2\sqrt{3}) -\Lambda(\sigma^3_{bc'},\tau^3_{bc'},v_3;2\sqrt{3}) 
\\
&
+2 \, \Lambda(\sigma^3_{cc'},\tau^3_{cc'},v_3;2\sqrt{3}) +2 \, \Lambda(\sigma^3_{cc'},\tau^3_{cc'},2v_3;4\sqrt{3}) 
.
\end{aligned}
\end{equation}

For the shifted definition of the hyper charge, $Q_{Y'} = Q_Y + \frac{1}{2} Q_e$, the corresponding gauge 
threshold is shifted by 
\begin{equation}
\begin{aligned}
\Delta_{U(1)_Y'} =&\; \Delta_{U(1)_Y} + \frac{ \Delta_{U(1)_e}}{4}   -\tilde{\Delta}_{ae} + \tilde{\Delta}_{ae'} 
-\tilde{\Delta}_{ce} + \tilde{\Delta}_{ce'}
-\tilde{\Delta}_{de} + \tilde{\Delta}_{de'}
\\
=&\; \Delta_{U(1)_Y} -2 \, \ln 2 -  \Lambda(0,0,v_1;\frac{2}{\sqrt{3}}) -6 \, \Lambda(0,1,v_1;\frac{2}{\sqrt{3}})
-6 \, \Lambda(1,1,v_2;\frac{2}{\sqrt{3}}) 
\\
&
-\frac{3}{2} \, \Lambda(\sigma^3_{ae},\tau^3_{ae},v_3;2\sqrt{3})-\frac{5}{2} \, \Lambda(\sigma^3_{ae'},\tau^3_{ae'},v_3;2\sqrt{3})
\\
&
- \Lambda(\sigma^3_{be},\tau^3_{be},v_3;2\sqrt{3})-  \Lambda(\sigma^3_{be'},\tau^3_{be'},v_3;2\sqrt{3})
\\
&
+ 2 \,  \Lambda(\sigma^3_{ee'},\tau^3_{ee'},v_3;2\sqrt{3}) + 2 \,\Lambda(\sigma^3_{ee'},\tau^3_{ee'},2v_3;4\sqrt{3})
.
\end{aligned}
\end{equation}
If on the other hand the $Sp(2)_e$ gauge factor remains unbroken, its gauge threshold is given by
\begin{equation}
\begin{aligned}
\Delta_{Sp(2)_e} =\;& 3 \tilde{\Delta}_{ae} + 2 \tilde{\Delta}_{be} + \tilde{\Delta}_{ce} + \tilde{\Delta}_{de} + \tilde{\Delta}_{ee} + \frac{1}{2}\Delta_{e,\OR}
\\
=\;& - 3 \, \Lambda(0,0,v_1;\frac{2}{\sqrt{3}}) -6 \,  \Lambda(0,1,v_1;\frac{2}{\sqrt{3}})-9 \, \Lambda(1,1,v_2;\frac{2}{\sqrt{3}})
\\
&+ 2 \, \Lambda(0,0,v_3;2\sqrt{3}) + 2 \,  \Lambda(0,0,2v_3;4\sqrt{3})-3 \, \Lambda(\sigma^3_{ae},\tau^3_{ae},v_3;2\sqrt{3})
\\
&-2 \, \Lambda(\sigma^3_{be},\tau^3_{be},v_3;2\sqrt{3})
.
\end{aligned}
\end{equation}

As an example, one can take $(\sigma^3_c,\tau^3_c) \neq (0,0)$ in order to break $SO/Sp(2)_c$ down to $U(1)_c$, but $(\sigma^3_x,\tau^3_x)=(0,0)$ for all other branes $x\in \{a,b,d,e\}$,
i.e. $(\sigma^3_{xc},\tau^3_{xc})=(\sigma^3_c,\tau^3_c)$ for $x \neq c$ and $(\sigma^3_{cc'},\tau^3_{cc'}) =(2 \sigma^3_c, 2\tau^3_c)$, 
 and isotropic two-tori volumes $v_1=v_2=v_3 =v$. The gauge threshold corrections simplify significantly in this case,
\begin{equation}
\begin{aligned}
\Delta_{SU(3)_a} 
=&\;26 \ln 2 +\frac{11}{3} \ln 3 
-21 \,  \ln \left(2\pi v \eta^4(iv)\right) + 8 \,  \ln \left(2\pi (2v) \eta^4(i2v)\right) 
\\
&
-6 \, \ln \left|\frac{\vartheta_1(\frac{1}{2},iv) }{\eta(iv)} \right|^2
-6 \, \ln \left|e^{-\pi v/4} \frac{\vartheta_1(\frac{1-iv}{2},iv) }{\eta(iv)} \right|^2
\\
&
-2 \, \ln \left|e^{-\pi (\sigma^3_{c})^2 v/4} \frac{\vartheta_1(\frac{\tau^3_{c}-i\sigma^3_{c} v }{2},iv) }{\eta(iv)} \right|^2
,
\end{aligned}
\end{equation}

\begin{equation}
\begin{aligned}
\Delta_{SU(2)_b}
=&\;-23 \ln 2 +4 \ln 3 
-18 \,\ln \left(2\pi v \eta^4(iv)\right) + 8 \,  \ln \left(2\pi (2v) \eta^4(i2v)\right) 
\\
&
-9 \, \ln \left|\frac{\vartheta_1(\frac{1}{2},iv) }{\eta(iv)} \right|^2
-6 \, \ln \left|e^{-\pi v/4} \frac{\vartheta_1(\frac{1-iv}{2},iv) }{\eta(iv)} \right|^2
\\
&
-2 \, \ln \left|e^{-\pi (\sigma^3_{c})^2 v/4} \frac{\vartheta_1(\frac{\tau^3_{c}-i\sigma^3_{c} v }{2},iv) }{\eta(iv)} \right|^2
,
\end{aligned}
\end{equation}

\begin{equation}
\begin{aligned}
\Delta_{U(1)_{B-L}} 
=&\; - 44 \ln 2 +\frac{70}{3} \ln 3 
-\frac{260}{3} \,  \ln \left(2\pi v \eta^4(iv)\right) + \frac{64}{3} \,  \ln \left(2\pi (2v) \eta^4(i2v)\right) 
\\
&
-16 \, \ln \left|\frac{\vartheta_1(\frac{1}{2},iv) }{\eta(iv)} \right|^2
-16 \, \ln \left|e^{-\pi v/4} \frac{\vartheta_1(\frac{1-iv}{2},iv) }{\eta(iv)} \right|^2
\\
&
-\frac{16}{3}\ln \left|e^{-\pi (\sigma^3_{c})^2 v/4} \frac{\vartheta_1(\frac{\tau^3_{c}-i\sigma^3_{c} v }{2},iv) }{\eta(iv)} \right|^2
,
\end{aligned}
\end{equation}

\begin{equation}
\begin{aligned}
\Delta_{U(1)_Y} 
=&\; \frac{55}{3} \, \ln 2 + \frac{58}{3} \, \ln 3 
-26 \,   \ln \left(2\pi v \eta^4(iv)\right) + \frac{16}{3} \,  \ln \left(2\pi (2v) \eta^4(i2v)\right) 
\\
&
-7 \, \ln \left|\frac{\vartheta_1(\frac{1}{2},iv) }{\eta(iv)} \right|^2
-13 \, \ln \left|e^{-\pi v/4} \frac{\vartheta_1(\frac{1-iv}{2},iv) }{\eta(iv)} \right|^2
\\
&
-\frac{22}{3}\ln \left|e^{-\pi (\sigma^3_{c})^2 v/4} \frac{\vartheta_1(\frac{\tau^3_{c}-i\sigma^3_{c} v }{2},iv) }{\eta(iv)} \right|^2
+ 2 \, \ln \left|e^{-\pi (\sigma^3_{c})^2 v} \frac{\vartheta_1(\tau^3_{c}-i\sigma^3_{c} v ,iv) }{\eta(iv)} \right|^2
\\
&
+ 2 \, \ln \left|e^{-2 \pi (\sigma^3_{c})^2 v} \frac{\vartheta_1(\tau^3_{c}-i\sigma^3_{c}2 v ,i2v) }{\eta(i2v)} \right|^2
,
\end{aligned}
\end{equation}

\begin{equation}
\begin{aligned}
\Delta_{Sp(2)_e} 
= &\; -2 \, \ln 2 + \ln 3 -6 \,  \ln \left(2\pi v \eta^4(iv)\right)   + 2 \,   \ln \left(2\pi (2v) \eta^4(i2v)\right) 
\\
& - 6 \,  \ln \left|\frac{\vartheta_1(\frac{1}{2},iv) }{\eta(iv)} \right|^2
- 9 \, \ln \left|e^{-\pi v/4} \frac{\vartheta_1(\frac{1-iv}{2},iv) }{\eta(iv)} \right|^2
,
\end{aligned}
\end{equation}
where we have inserted the explicit form~\eqref{Eq:Def-Lambda} of the lattice contributions.

A numeric evaluation of these simplified threshold corrections for the choice $(\sigma_c^3,\tau_c^3)=(\frac{1}{2},0)$ in dependence of the two-torus volume parameter $v$
is shown in Figure~\ref{fig:thresholds_ex_z6}. Depending on the value of $v$ the threshold corrections of
the $SU(2)_b$, $U(1)_{B-L}$ and $Sp(2)_e$ gauge groups can be either enhanced (negative $\Delta$) or reduced
(positive $\Delta$).

\fig{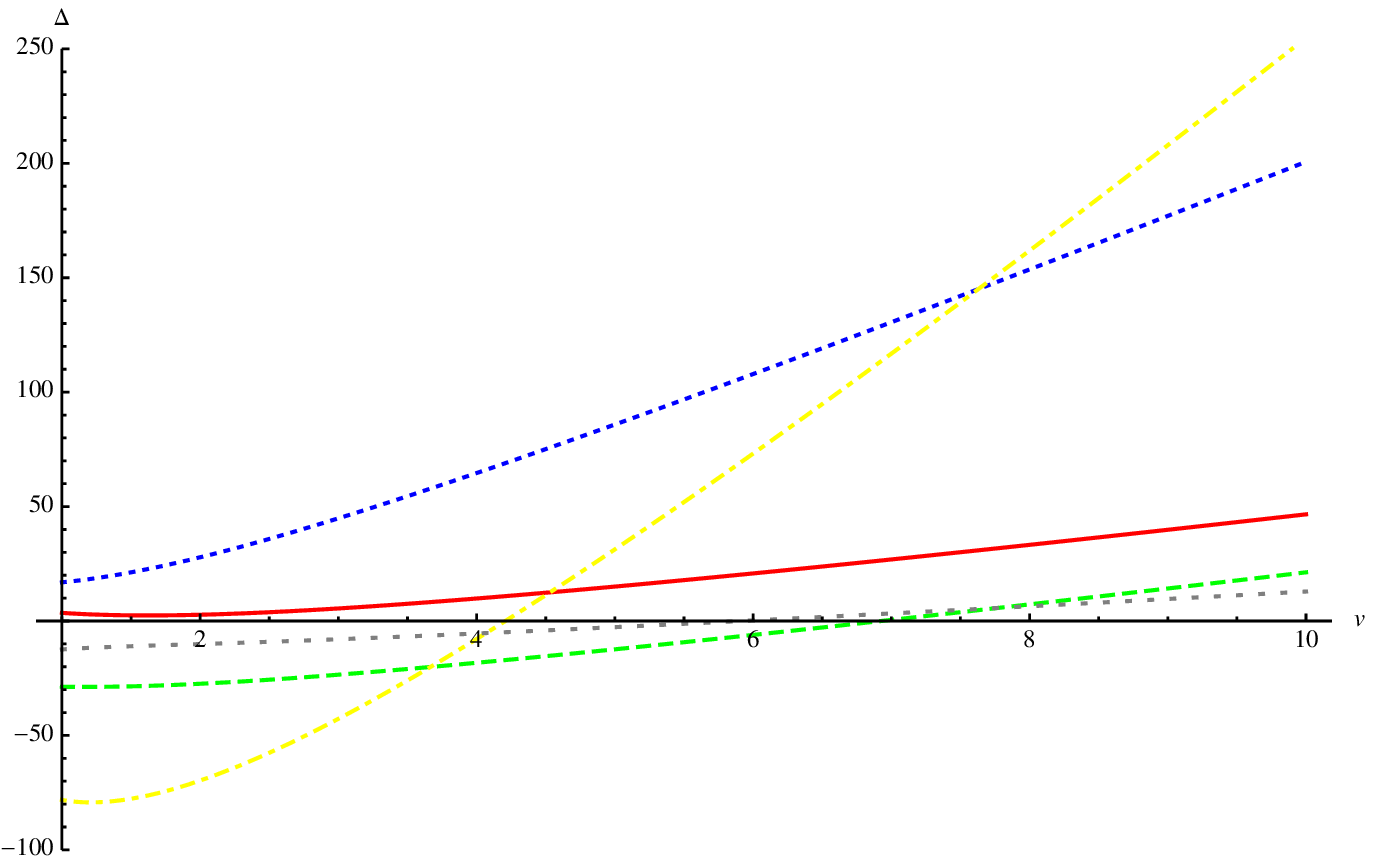}{Threshold contributions in dependence of the two-torus volume $v$ for the $T^6/\Z_6$ example and $(\sigma^3_c,\tau^3_c)=(\frac{1}{2},0)$ and $(\sigma^3_x,\tau^3_x)=(0,0)$ for $x \neq c$.  The contribution to $SU(3)_a$ is shown in solid red, $SU(2)_b$ in dashed green, $U(1)_Y$ in dotted blue, $U(1)_{B-L}$ in dot-dashed yellow and the `hidden' sector $Sp(2)_e$ as sparse dashed grey.}

%%%%%%%%%%%%%%%%%%%%
%%%%%%%%%%%%%%%%%%%%%
\subsection{Example 2: The \texorpdfstring{$T^6/\Z_6'$}{T6/Z6'} models}\label{Ss:Ex2}
%%%%%%%%%%%%%%%%%%%%

In the following, we describe the orbifold action on $T^6/\Z_6'$, give the D6-brane configurations and intersections numbers from which the massless spectrum and 
the beta function coefficients are computed and finally determine the gauge threshold contributions for three examples with Standard Model spectrum and various hidden sectors.

\subsubsection{Orbifold action for $T^6/\Z_6'$ on {\bf ABa}}\label{AppSs:Z6p}

Under the $\Z_6'$ action $\theta'$ with shift vector $\vec{v}^{\prime} = \frac{1}{6}(1,2,-3)$ and the orientifold projection
on the {\bf ABa} lattice, the torus cycle wrapping numbers transform as 
\begin{equation}
\begin{aligned}
& \left(\begin{array}{cc} 
n_1 & m_1 \\ n_2 & m_2 \\ n_3 & m_3
\end{array}\right)
\stackrel{\theta'}{\longrightarrow}
\left(\begin{array}{cc} 
-m_1 & n_1+m_1 \\ -(n_2+m_2) & n_2  \\ -n_3 & -m_3
\end{array}\right)
\stackrel{\theta'}{\longrightarrow}
\left(\begin{array}{cc} 
-( n_1+m_1) & n_1 \\ m_2 & -(n_2+m_2) \\  n_3 & m_3
\end{array}\right)
\\
& \downarrow \OR
\\
& \left(\begin{array}{cc} 
n_1+m_1 & -m_1 \\ m_2 & n_2 \\ n_3 & -m_3
\end{array}\right)
\stackrel{\theta'}{\longrightarrow}
\left(\begin{array}{cc} 
m_1 & n_1 \\ -(n_2+m_2) & m_2 \\ -n_3 & m_3
\end{array}\right)
\stackrel{\theta'}{\longrightarrow}
\left(\begin{array}{cc} 
-n_1 & n_1+m_1 \\ n_2 & -(n_2 +m_2)  \\ n_3 & -m_3
\end{array}\right)
.
\end{aligned}
\end{equation}
In this case, the $\Z_2$ subgroup $3 \,\vec{v}^{\prime} = \frac{1}{2}(1,0,-1)$ leaves the second two-torus invariant, and the generic formulae are taylor-made
for this set-up. The complex structure moduli on $T^2_1 \times T^2_2$ are fixed by the underlying $\Z_3$ symmetry, see figure~\ref{Fig:Z3-Z6lattice}, whereas there is a 
continuous real complex structure parameter on $T^2_3$,
\begin{equation}
\varrho = \frac{\sqrt{3}}{2} \frac{R_2}{R_1}, 
\end{equation}
with the radii $R_i$ defined in figure~\ref{Fig:Z2-Z4lattice}.  $\varrho$ takes a set of discrete values for  supersymmetric solutions 
to the RR tadpole cancellation conditions as discussed in~\cite{Gmeiner:2008xq}.

\subsubsection{D6-brane configuration and intersection numbers}

The configuration of D6-brane stacks used in~\cite{Gmeiner:2008xq} is given in table~\ref{tab:smmodez6p}.
The Standard Model gauge group and matter content is again supported on four
stacks of D6-branes, labelled by $a,b,c,d$ with gauge group
\begin{equation}
SU(3)_a \times SU(2)_b \times U(1)_Y \times U(1)_{B-L} 
\qquad 
{\rm with}
\quad
\left\{\begin{array}{c}
Q_Y = \frac{1}{6} Q_a + \frac{1}{2} Q_c + \frac{1}{2} Q_d
\\
Q_{B-L} =  \frac{1}{3} Q_a +Q_d
\end{array}\right.
.
\end{equation}
Depending on the model under consideration there are also no, one or two stacks of
branes in the hidden sector, labelled by the rank of their gauge groups $h_3, h_1 + h_2$ or $\hat{h}_1$.
These four different models can be treated simultaneously using the parameterization $\omega = 1$ for 
all examples with hidden sectors and $\omega = 2$ for the model without hidden sector. The complex
structure on $T^2_3$ is $\varrho = 1/(2\omega)$, and intersection numbers involving brane $b$ have 
some $\omega$ dependence.
The intersection numbers between the various D6-brane stacks are listed in
tables~\ref{tab:smz6pinter1} to~\ref{tab:smz6pinter6}.

\mathtabfix{
\begin{array}{|c|c||c||c|c||c|c| }\hline
{\rm brane} & \frac{\rm Angle}{\pi} & (n_1,m_1;n_2,m_2;n_3,m_3) 
& \vec{\sigma}_{\rm w.h.} & (\tau^0,\tau^1,\tau^3)_{\rm w.h.}
& \vec{\sigma}_{\rm no-h} & (\tau^0,\tau^1,\tau^3)_{\rm no-h}
\\\hline\hline
a & (-\frac{1}{3},-\frac{1}{6},\frac{1}{2})
& (1,-1;1,0;0,1) & (1,0;1,0) & (0,1,1) 
& (1,0;1,0) & (1,1,0)
\\
h_3, h_2 & 
& & \vec{0} & (0,1,0)
& \muc{2}{|c|}{}
\\
h_1 & 
& & \vec{0} & (1,0,0)
& \muc{2}{|c|}{}
\\\hline
b & (\frac{1}{6},-\frac{1}{3},\frac{1}{6})
& (1,1;2,-1;1,\omega) 
& (1,0;0,0) & (1,1,0)
& (1,0;0,0) & (1,1,0) 
\\\hline
c & (-\frac{1}{3},\frac{1}{3},0)
& (1,-1;-1,2;1,0)
& (1,0;0,1) & (0,1,1) 
& (1,0;0,0) & (1,1,0) 
\\\hline
d & (\frac{1}{6},\frac{1}{3},-\frac{1}{2})
& (1,1;1,-2;0,1) 
& (0,0;1,0) & (0,1,1)
& (0,0;1,0) & (0,0,0)
\\
\hat{h}_1 & 
& & (1,0;0,0) & (0,1,0)
& \muc{2}{|c|}{}
\\\hline
\end{array}
}{smmodez6p}{Geometric setup of the Standard Model examples on the {\bf ABa} lattice in the 
$T^6/\mathbb{Z}_6'$ background with complex structure parameter $\varrho \equiv \frac{\sqrt{3}}{2} \frac{R_2}{R_1} = \frac{1}{2 \omega} $ on $T^2_3$ 
and  $\omega=1$ for the models with hidden sectors and $\omega=2$ for the model without hidden sector. In the fourth and fifth column, the values of 
the discrete displacements $(\sigma^1,\sigma^3)$, Wilson lines $(\tau^1,\tau^3)$ and $\Z_2$ eigenvalues $\tau^0$
for the models with hidden sectors are listed, and the last two columns contain the same data for the model without
hidden sector.}

\mathtabfix{
\begin{array}{|c||c|c|c||c|c|c||c|c|c||c|c|c|}\hline
x & I_{ \cdot a} & I_{ \cdot (\theta a)}  & I_{ \cdot (\theta^2 a)} 
 & I_{ \cdot b} & I_{ \cdot (\theta b)}  & I_{ \cdot (\theta^2 b)} 
 & I_{ \cdot c} & I_{ \cdot (\theta c)}  & I_{ \cdot (\theta^2 c)} 
 & I_{ \cdot d} & I_{ \cdot (\theta d)}  & I_{ \cdot (\theta^2 d)} 
 \\\hline\hline
a &0_{123} &  (-1) \cdot 0_{3}   & (-1) \cdot 0_{3}  
& 2 & 2  & (-1)
& 0_{1} \cdot (-2) &(-1)  & 1 
& (-4) \cdot 0_3 & (-1) \cdot 0_3 & (-1) \cdot  0_3
\\\hline
b & \muc{3}{|c|}{} & 0_{123} & (-9) \cdot 0_3 & (-9)\cdot 0_3
& 6 \, \omega  & 3 \, \omega & (- \omega ) \cdot 0_2
& 0_1 \cdot  (-3) & (-9) & (-3) \cdot 0_2
\\\hline
c &  \muc{6}{|c|}{} & 0_{123} &(-3) \cdot 0_3 &(-3) \cdot 0_3
& (-2) \cdot  0_2 & 3 & -3
\\\hline 
d &  \muc{9}{|c|}{} & 0_{123} & (-9) \cdot 0_3 & (-9)\cdot 0_3
 \\\hline
\end{array}
}{smz6pinter1}{Toroidal intersection numbers of the observable sector of the 
supersymmetric Standard Model examples on $T^6/\mathbb{Z}_6'$, Part I. $0_3$ denotes that the D6-branes are parallel along $T^2_3$, 
branes parallel everywhere have intersection number $0_{123}$.}

\mathtabfix{
\begin{array}{|c||c|c|c||c|c|c||c|c|c||c|c|c|}\hline
x & I_{ \cdot a}^{\bZ_2} & I_{ \cdot (\theta a)}^{\bZ_2}  & I_{ \cdot (\theta^2 a)}^{\bZ_2}
 & I_{ \cdot b}^{\bZ_2} & I_{ \cdot (\theta b)}^{\bZ_2}  & I_{ \cdot (\theta^2 b)}^{\bZ_2}
 & I_{ \cdot c}^{\bZ_2} & I_{ \cdot (\theta c)}^{\bZ_2}  & I_{ \cdot (\theta^2 c)}^{\bZ_2}
 & I_{ \cdot d}^{\bZ_2} & I_{ \cdot (\theta d)}^{\bZ_2}  & I_{ \cdot (\theta^2 d)}^{\bZ_2}
 \\\hline\hline
a & 4 \cdot 0_2 & -2 & 2  
& (-2) & (-2) & 1
& 4 & 1  & 1
& \emptyset  & (-2) & 2 
\\\hline
b & \muc{3}{|c|}{} &4 \cdot 0_2 & -6 & 6
& 6 \, \omega & 3 \, \omega & (- \omega ) \cdot 0_{2}
& \emptyset  & (-3) & (-1) \cdot 0_{2}
\\\hline
c &  \muc{6}{|c|}{} & 4 \cdot 0_2 &  -6 &  6 
&\emptyset  & 3  &  3
\\\hline 
d &  \muc{9}{|c|}{} & 4 \cdot 0_2 & 6 & -6 
\\\hline
\end{array}
}{smz6pinter2}{$\Z_2$ invariant intersection numbers of the observable sector of the 
supersymmetric Standard Model examples on $T^6/\mathbb{Z}_6'$, Part I.  $\emptyset$ 
occurs when D6-branes carry a relative Wilson line or are at a distance along a two-torus
with wrapping numbers of the same type, e.g. $(n_1,m_1)=$(odd,odd). Whenever the D6-branes
are parallel along $T^2_2$ labeled by $0_2$, non-chiral pairs of bifundamental representations occur.}

%%%%%%%%%%%%%%%%%%%%%%%%%%%%%%%%%%%

%%%%%%%%%%%%%%%%%%%%%%%%%%%%%%%%%%%
\mathtabfix{
\begin{array}{|c||c|c|c||c|c|c||c|c|c|}\hline
x & I_{ \cdot a'} & I_{ \cdot (\theta a')}  & I_{ \cdot (\theta^2 a')} 
 & I_{ \cdot b'} & I_{ \cdot (\theta b')}  & I_{ \cdot (\theta^2 b')} 
 & I_{ \cdot d'} & I_{ \cdot (\theta d')}  & I_{ \cdot (\theta^2 d')} 
 \\\hline\hline
a & (-1) \cdot 0_3  & 0_{123} & (-1) \cdot 0_3 
& (-2)  & (-2) & 1 
& (-1) \cdot 0_3 & (-4) \cdot 0_3 & (-1) \cdot 0_3
\\\hline
b &  \muc{3}{|c|}{} & 18 \, \omega  & 0_1 \cdot (-6 \, \omega )  & (-6 \, \omega ) \cdot 0_2 
&  (-3) \cdot 0_2  & 0_1 \cdot (-3) & (-9)
\\\hline
d &  \muc{6}{|c|}{} & (-9) \cdot 0_3  & 0_{123}   & (-9) \cdot 0_3
\\\hline
\end{array}
}{smz6pinter3}{Toroidal intersection numbers of the observable sector of the 
supersymmetric Standard Model examples on $T^6/\mathbb{Z}_6'$, Part II. Since brane $c$ is orientifold invariant
for no continuous displacement or Wilson line on $T^2_2$, the corresponding intersection numbers for $(\theta^k c')$ 
can be read off from table~\protect\ref{tab:smz6pinter1}.}

\mathtab{
\begin{array}{|c||c|c|c||c|c|c||c|c|c|}\hline
x & I_{ \cdot a'}^{\bZ_2} & I_{ \cdot (\theta a')}^{\bZ_2}  & I_{ \cdot (\theta^2 a')}^{\bZ_2}
 & I_{ \cdot b'}^{\bZ_2} & I_{ \cdot (\theta b')}^{\bZ_2}  & I_{ \cdot (\theta^2 b')}^{\bZ_2}
 & I_{ \cdot d'}^{\bZ_2} & I_{ \cdot (\theta d')}^{\bZ_2}  & I_{ \cdot (\theta^2 d')}^{\bZ_2}
 \\\hline\hline
a & (-2)   & 4 \cdot 0_2 & 2
& (-2) & (-2) & 1
& (-2) & \emptyset &  2
\\\hline
b & \muc{3}{|c|}{} & (-6) & (-12) & (-2) \cdot 0_2
& 1 \cdot 0_2 & \emptyset  &  3 
\\\hline
d &  \muc{6}{|c|}{} & 6 & (-4) \cdot 0_2 & (-6)
\\\hline
\end{array}
}{smz6pinter4}{$\Z_2$ invariant intersection numbers of the observable sector of the 
supersymmetric Standard Model examples on $T^6/\mathbb{Z}_6'$, Part II. Intersection numbers involving $(\theta^k c')$ 
can be read off from table~\protect\ref{tab:smz6pinter2}.}

\mathtab{
\begin{array}{|c||c|c||c|c||c|c|}\hline
x & \tilde{I}_x^{\OR} & \tilde{I}_x^{\OR\theta^{-3}} & \tilde{I}_x^{\OR\theta^{-1}} & \tilde{I}_x^{\OR\theta^{-4}}  & \tilde{I}_x
^{\OR\theta^{-2}} & \tilde{I}_x^{\OR\theta^{-5}} 
\\\hline\hline
a, h_3,h_2,h_1 &  (-2) &   (-1) \cdot 0_3
& (-4) \cdot 0_3 & 0_1 \cdot (-4)
& 2 &  (-1) \cdot 0_3
\\\hline
b & 6 \, \omega  & -18 
& 0_1 \cdot (-6) & 12 \,  \omega 
& (-2 \, \omega ) \cdot 0_2 & (-6) \cdot 0_2
\\\hline
c &  (-3) \cdot 0_3 & 6
& (-4) \cdot 0_2 & 0_{123}
& (-3) \cdot 0_3 & (-6)
\\\hline 
d,\hat{h}_1 & 6 & (-9) \cdot 0_3
& 0_{123} & (-4) \cdot 0_2
& (-6) & (-9) \cdot 0_3
\\\hline
\end{array}
}{smz6pinter5}{Intersection numbers with O6-planes
of the supersymmetric Standard Model examples on $T^6/\mathbb{Z}_6'$.}
%%%%%%%%%%%%%%%%%%%%%%%%%%%%%%%%%%%%%%%%%%%%%%%%%%%%%%%%%%

\mathtabfix{
\begin{array}{|c||c|c|c||c|c|c||c|c|c||c|c|c|}\hline
x & I_{ \cdot h_3} & I_{ \cdot (\theta h_3)}  & I_{ \cdot (\theta^2 h_3)} 
 & I_{ \cdot \hat{h}_1} & I_{ \cdot (\theta \hat{h}_1)}  & I_{ \cdot (\theta^2 \hat{h}_1)
} 
%%%%%
 & I_{ \cdot h_3}^{\bZ_2} & I_{ \cdot (\theta h_3)}^{\bZ_2}  & I_{ \cdot (\theta^2 h_3)}^
{\bZ_2} 
 & I_{ \cdot \hat{h}_1}^{\bZ_2} & I_{ \cdot (\theta \hat{h}_1)}^{\bZ_2}  & I_{ \cdot (\theta^2 \hat{h}_1)}^{\bZ_2} 
 \\\hline\hline
a & 0_{123} &  (-1)\cdot 0_3  & (-1) \cdot 0_3  
& (-4) \cdot 0_3 & (-1)  \cdot 0_3 & (-1) \cdot 0_3 
& \emptyset  & \emptyset  & \emptyset 
& \emptyset  & \emptyset  & \emptyset 
\\\hline
b & (-2) & 1 & (-2)
& 0_1 \cdot (-3) & (-9) & (-3) \cdot 0_2
& \emptyset  & 1 & 2
& 6 & 3   & (-1) \cdot  0_2
\\\hline
c & 0_1 \cdot (-2)  & (-1)  & 1
& (-2) \cdot 0_2 & 3  & (-3)
& \emptyset & (-1)  & 1 
& (-2) \cdot 0_2 & 3  & (-3)
\\\hline 
d & (-4) \cdot 0_3 & (-1) \cdot 0_3 & (-1) \cdot 0_3
&  0_{123} & (-9) \cdot 0_3  &  (-9)\cdot 0_3  
& \emptyset & \emptyset & \emptyset 
& \emptyset  & \emptyset  & \emptyset 
\\\hline\hline
h_3 &  0_{123} & (-1) \cdot 0_3 & (-1) \cdot 0_3 
& \muc{3}{|c|}{}
& 4 \cdot 0_2 & 2  & (-2) 
& \muc{3}{|c|}{}
\\\hline\hline
\hat{h}_1 & \muc{3}{|c|}{}
 &  0_{123} & (- 9) \cdot 0_3 &  (-9) \cdot 0_3
& \muc{3}{|c|}{}
& 4 \cdot 0 & (-6) & 6
\\\hline
\end{array}
}{smz6pinter6}{Intersection numbers involving the hidden sector branes $h_3$ or $\hat{h}_1$ 
of the supersymmetric Standard Model examples on $T^6/\mathbb{Z}_6'$. All intersection numbers 
for $h_2$ agree with those of $h_3$, and so do the toroidal intersections for $h_1$. Due to the different
$\Z_2$ eigenvalue and Wilson line assignments, some of the $I^{\Z_2}$ differ for $h_1$ and can be easily computed
from the $\Z_2$ eigenvalue and Wilson line assignments in table~\protect\ref{tab:smmodez6p}. }

%%%%%%%%%%%%%%%%%%%%%%%%%%%%%%%%%%%%%%%%%%
\subsubsection{Spectrum}

The complete massless matter spectra for both kinds models, with and without a hidden
sector, are given below in a compact form. The notation is the same as in
the previous example, $\left((S)U(3)_a,(S)U(2)_b \right)_{Y,B-L}^{(U(1)_c,U(1)_d)}$ in the observable sector and a third non-Abelian entry 
for the hidden sector.

The massless spectrum for each model consists of three (with hidden sector) or two (no hidden sector) components:
\begin{itemize}
\item
The `chiral' spectrum due to non-vanishing intersection numbers
\begin{equation}\label{Eq:ex2-explicitspectrum-chiral}
\begin{aligned}
{}
[C] &=3\times\bigg[
    \left(\3,\2\right)_{\bf 1/6, 1/3}^{(0,0)}
  + \left(\ov{\3},\1\right)_{\bf 1/3,-1/3}^{(1,0)}
  + \left(\ov{\3},\1\right)_{\bf -2/3, -1/3}^{(-1,0)}
\\
&\qquad\qquad
  + \left(\1,\1\right)_{\bf 1,1}^{(1,1)}
  +  \left(\1,\1\right)_{\bf 0,1}^{(-1,1)}
  + 2 \times \left(\1,\2\right)_{\bf -1/2,-1}^{(0,-1)}
  + \left(\1,\2\right)_{\bf 1/2,1}^{(0,1)}
\\
&\qquad\qquad
  + 3 \, \omega \times  \left(\1,\ov{\2}\right)_{\bf -1/2,0}^{(-1,0)}
  + 3 \, \omega \times  \left(\1,\ov{\2}\right)_{\bf 1/2,0}^{(1,0)}
  + 3 \, (\omega -1 )\times \left(\1,\1_{\ov{A}}\right)_{\bf 0,0}^{(0,0)}
\bigg]
\\
& \equiv 3 \times \bigg[Q_L +d_R + u_R + e_R +\nu_R + 2\times L + \ov{L}  \bigg]
\\
&\quad
+ 9 \, \omega \times \bigg[ H_d + H_u\bigg]
+ 9 \, (\omega -1) \times S 
.
\end{aligned}
\end{equation}
As for the previously discussed $T^6/\Z_6$ example, the Higgses $H_u+H_d$ group into non-chiral ${\cal N}=2$ supersymmetric
pairs when $U(1)_b \subset U(2)_b$ acquires a mass through the Green-Schwarz mechanism. In the same way, three non-chiral
lepton pairs and $9(\omega-1)$ gauge singlets under the unbroken gauge group arise.
The truly chiral spectrum after the Green Schwarz mechanism consists therefore of exactly three quark-lepton generations.
%%%%%%%
\item
The `universal non-chiral' spectrum coming only from branes $a,b,c,d$ supporting the
Standard Model gauge group~\footnote{The number of symmetric and antisymmetric non-chiral pairs of $U(2)_b$ 
has been corrected as well as the multiplicities of non-chiral $bd$ and $bd'$ pairs exchanged w.r.t.~\cite{Gmeiner:2008xq}.
Similarly the number of $b\hat{h}_1$ pairs is augmented by 1.
This slightly changes the values of the beta function coefficients of the $SU(2)_b$ and for the models with hidden sector also the $U(1)_Y$ factor
as listed in section~\protect\ref{Sss:Z6pBeta}. } 
\begin{equation}\label{Eq:ex2-explicitspectrum-non-chiral}
\begin{aligned}
{}
[V_U] &= 2 \times \left(\bf{8},\1\right)_{\bf 0,0}^{(0,0)}
  + 10 \times \left(\1,\3\right)_{\bf 0,0}^{(0,0)}
  + 26  \times \left(\1,\1\right)_{\bf 0,0}^{(0,0)}
\\
&\quad + \bigg[
     \left(\3,\2\right)_{\bf 1/6,1/3}^{(0,0)}
  + 3\times \left(\ov{\3},\1\right)_{\bf 1/3,2/3}^{(0,1)}
  + 3 \times \left(\ov{\3},\1\right)_{\bf -2/3,-4/3}^{(0,-1)}
\\
&\qquad\quad
  + (3-x+1_m) \times \left(\1,\1 \right)_{\bf 1,0}^{(2,0)}
  + \left(1+2_m  \right) \times (\ov{\3}_A,\1)_{\bf 1/3,2/3}^{(0,0)}
\\
&\qquad\quad
  + \left( 3 \, \omega + (\omega -1)_m  \right) \times (\1,\3_S)_{\bf 0,0}^{(0,0)}
  + \omega_m \times  \left(\1,\ov{\2}\right)_{\bf -1/2,0}^{(-1,0)}
\\
&\qquad\quad
  + \omega_m \times \left(\1,\ov{\2}\right)_{\bf 1/2,0}^{(1,0)}
  + 2_m \times  \left(\1,\2\right)_{\bf -1/2,-1}^{(0,-1)}
  + 1_m \times  \left(\1,\2\right)_{\bf 1/2,1}^{(0,1)}
\\
&\qquad\quad
  + \left(5+2 \, \omega \right)_m \times (\1,\1_A)_{\bf 0,0}^{(0,0)}
  + 1_m \times   (\1,\1)_{\bf 0,-1}^{(1,-1)}
\\
&\qquad\quad
  + 1_m \times (\1,\1)_{\bf 1,1}^{(1,1)}
  + \;c.c.\;\bigg].
\end{aligned}
\end{equation}
It consists of adjoints, symmetric and antisymmetric representations of the broken $SO(2)_c$ or $Sp(2)_c$ gauge group and 
bifundamentals, symmetrics and antisymmetrics in ${\cal N}=2$ hyper multiplets of the unitary and Abelian groups.
As for the example on $T^6\Z_6$ presented in section~\ref{Ss:Ex1}, a index $m$ denotes that these state acquire a mass if the D6-branes
supporting them carry a relative Wilson line or are spatially separated along $T^2_2$.
In appendix~\ref{App:SO-Sp}, we argue that $U(1)_c$ derives from a broken $Sp(2)_c$ gauge group with one chiral multiplet
in the symmetric and three in the antisymmetric representation corresponding to $x=3$.
\\
For the model without hidden sector, the massless matter spectrum consists of 
\begin{equation}
\left( [C] + [V_U] \right)_{\omega = 2}.
\end{equation}
The models with hidden sector have an additional contribution:
%%%%%%%%%%%%%%%%%
\item
The `non-universal non-chiral' spectrum involving hidden branes $h_3$ or $\hat{h}_1$:
\begin{itemize}
\item
For the hidden gauge group $SO/Sp(6)_{h_3}$
\begin{equation}
\begin{aligned}
{}
[V_{h_3}]=\;&
z \times (\1,\1;{\bf 15})^{(0,0)}_{\bf 0,0} +
(2-z) \times  (\1,\1;{\bf 21})^{(0,0)}_{\bf 0,0}
\\
&
+ \bigg[ (\1,\2;\6)^{(0,0)}_{\bf 0,0} + (\1,\1;{\bf 6})_{1/2,0}^{(1,0)}
 \; + \cc \; \bigg] .
\end{aligned}
\end{equation}
Since brane $h_3$ is orthogonal to the $\OR\theta^{-1}$ and $\OR\theta^{-4}$ planes along $T^2_2$, 
there is no obvious way of breaking this group down to a unitary subgroup by parallel displacement.
According to the computation in appendix~\ref{App:SO-Sp}, the hidden gauge group is $Sp(6)_{h_3}$,
and the two chiral multiplets are in the antisymmetric representation, i.e. $z=2$.
%%%%
\item
Or for a hidden $\widehat{SO/Sp(2)}_{\hat{h}_1}$
\begin{equation}
\begin{aligned}
{}
[V_{\hat{h}_1}]=\;&
y \times (\1,\1;\1_A)^{(0,0)}_{\bf 0,0}
+ (10-y) \times (\1,\1;\3_S)^{(0,0)}_{\bf 0,0}
\\
&
+\bigg[ (3+2_m) \times (\1,\2;\2)^{(0,0)}_{\bf 0,0}  + (3+2_m) \times (\1,\1;\2)^{(1,0)}_{\bf 1/2,0}
 \; + \cc \; \bigg].
\end{aligned}
\end{equation}
Brane $\hat{h}_1$ is parallel to the $\OR\theta^{-1}$ and $\OR\theta^{-4}$ planes on $T^2_2$, 
and the non-Abelian hidden group can be broken down to $U(1)_{\hat{h}_1}$ by continuously displacing the brane and its image
from the orientifold planes on this two-torus. 
Before the gauge symmetry breaking, the analysis in appendix~\ref{App:SO-Sp} leads to an $\widehat{Sp(2)}_{\hat{h}_1}$ gauge group
with one chiral multiplet in the symmetric and nine in the antisymmetric representation, i.e. $y=9$.
\end{itemize}
\end{itemize}

%%%%%%%%%%%%%%%%%%%%%%%%%%%%%%%
\subsubsection{Beta functions}\label{Sss:Z6pBeta}

For the coefficients in front of the $\ln(M_s^2/\mu^2)$-term in~\eqref{Eq:Def-gauge}, the field theoretical formulae~\eqref{Eq:beta-SU(N)} and~\eqref{Eq:Def-U(1)-massless} give
\begin{equation}
\begin{aligned}
&b_{SU(3)_a}=12+2_m,
&&b_{U(1)_{B-L}} = 72 + \frac{64}{3}_m
\\
&b_{SU(2)_b}=\left\{\begin{array}{cc} 68+11_m&\mbox{no hidden}\\  53+9_m &\hat{h}_1\\ 59+5_m &h_3,h_1+h_2\end{array}\right.,
&&b_{U(1)_Y}=
\left\{\begin{array}{cc} 48-2x+\frac{37}{3}_m &\mbox{no hidden}\\42-2x+ \frac{37}{3}_m&\hat{h}_1\\ 42-2x+\frac{31}{3}_m&h_3,h_1+h_2\end{array}\right.
,
\end{aligned}
\end{equation}
and from the analysis in appendix~\ref{App:SO-Sp}, the one-loop running of the hidden sector branes is computed according to~\eqref{Eq:beta-Sp(2M)},
\begin{equation}
b_{Sp(6)_{h_3}} = -5,
\qquad\quad
b_{\widehat{Sp(2)}_{\hat{h}_1} }= 5+6_m.
\end{equation}
The hidden gauge group $Sp(6)_{h_3}$ has very little massless matter and can accordingly run to strong coupling at low energies, whereas the other choice of hidden sector 
$\widehat{Sp(2)}_{\hat{h}_1} $  contains a larger amount of massless vector like matter and cannot become strongly coupled, unless so far undetermined field theory couplings render some of the vector-like states massive. It remains to be seen if the hidden gauge group   $Sp(6)_{h_3}$  can provide for supersymmetry breaking via a gaugino condensate.

%%%%%%%%%%%%%%%%%%%%%%%%%%%%%%%%%%%%%
\subsubsection{Threshold corrections}

The individual threshold contributions are computed analogously to the previous example using the intersection numbers in tables~\ref{tab:smz6pinter1}
to~\ref{tab:smz6pinter6}, relative angles among D6-branes and O6-planes and inserting them in the general formulae in table~\ref{tab:Amplitudes-thresholds}. 
In contrast to the $T^6/\Z_6$ example presented in section~\ref{Ss:Ex1}, here 
all Standard Model branes $a\ldots d$ wrap different torus cycles.

The generic result for the $SU(3)_a \times SU(2)_b \times U(1)_{Y} \times U(1)_{B-L}$ gauge symmetry is 
(with the three entries of a column denoting the three possible hidden sector configurations $(h_3,\hat{h}_1,\emptyset)$):
\begin{equation}
 \begin{aligned}
 \Delta_{SU(3)_a} 
=\;& - 7 \ln 5 
- 2 \Lambda(0,0,v_1;\frac{2}{\sqrt{3}})+ 4 \Lambda(0,0,2v_1;\frac{4}{\sqrt{3}})
 + 6 \Lambda(0,0,v_2;\frac{2}{\sqrt{3}})
\\ &
+ 6 \Lambda(\sigma^2_{aa'},\tau^2_{aa'},v_2;\frac{2}{\sqrt{3}})
-12 \Lambda(0,0,v_3;\frac{1}{\omega\sqrt{3}}) + 6 \Lambda(0,0,v_3; \frac{4}{\omega\sqrt{3}})
\\ &
-6 (2-\omega)\Lambda(1,1,v_3;\frac{1}{\sqrt{3}\omega})
,
\end{aligned}
\end{equation}
\begin{equation}
\begin{aligned}
\Delta_{SU(2)_b} 
=\;& \left(41 - 16 \omega +  \left\{\begin{array}{c} -5 \\ -13 \\\emptyset\end{array}\right\}   \right) \ln 2 - \left( 36 + 3 \omega + \left\{\begin{array}{c} 9 \\ 0 \\\emptyset\end{array}\right\}  \right) \ln 5
\\
&
 -\left( 6 \omega +\left\{\begin{array}{c} 0 \\ 3 \\\emptyset\end{array}\right\}  \right) \Lambda(0,0,v_1;2\sqrt{3}) 
- 3 \Lambda(1,\omega-1,v_1; 2\sqrt{3})
\\
&
  + 6 \Lambda(0,0,2v_1;4\sqrt{3})
 + 4 \, \Lambda(0,0,v_2;2\sqrt{3}) - (2+ 6 \omega) \Lambda(\sigma^2_{bb'},\tau^2_{bb'},v_2;2\sqrt{3})
\\
&
-\omega \Lambda(\sigma^2_{bc},\tau^2_{bc},v_2;2\sqrt{3})
-\omega \Lambda(\sigma^2_{bc'},\tau^2_{bc'},v_2;2\sqrt{3})
-2 \Lambda(\sigma^2_{bd},\tau^2_{bd},v_2;2\sqrt{3})
\\
&
 - \Lambda(\sigma^2_{bd'},\tau^2_{bd'},v_2;2\sqrt{3})
+(2\omega + 6) \Lambda(\sigma^2_{bb'},\tau^2_{bb'},2v_2;4\sqrt{3})
\\
&
-2 \left\{\begin{array}{c} 0 \\  \Lambda(\sigma^2_{b\hat{h}_1},\tau^2_{b\hat{h}_1},v_2;2\sqrt{3})+ \Lambda(\sigma^2_{b\hat{h}_1'},\tau^2_{b\hat{h}_1'},v_2;2\sqrt{3})\\\emptyset\end{array}\right\} 
\\
&
-18 \, \Lambda(0,0,v_3;\frac{4\omega}{\sqrt{3}})
,
\end{aligned}
\end{equation}
\begin{equation}
\begin{aligned}
 \Delta_{U(1)_{B-L}} 
=\;& - 20 \, \ln 2 -\frac{32}{3} \, \ln 5 -\frac{4}{3}  \Lambda(0,0,v_1;\frac{2}{\sqrt{3}})+ \frac{8}{3} \Lambda(0,0,2v_1;\frac{4}{\sqrt{3}})
\\
&
- 12 \,  \Lambda(1,\omega-1,v_1; 2\sqrt{3})
+ 8 \, \Lambda(\sigma^2_{aa'},\tau^2_{aa'},v_2;\frac{2}{\sqrt{3}})-8 \,  \Lambda(\sigma^2_{bd},\tau^2_{bd},v_2;2\sqrt{3})
\\
&
 -4 \,  \Lambda(\sigma^2_{bd'},\tau^2_{bd'},v_2;2\sqrt{3})
-2 \, \Lambda(\sigma^2_{cd},\tau^2_{cd},v_2;2\sqrt{3}) -2 \, \Lambda(\sigma^2_{cd'},\tau^2_{cd'},v_2;2\sqrt{3})
\\
&
 - 8 \, \Lambda(\sigma^2_{dd'},\tau^2_{dd'},v_2;2\sqrt{3})
+ 8 \, \Lambda(\sigma^2_{dd'},\tau^2_{dd'},2v_2;4\sqrt{3})
- 80 \, \Lambda(0,0,v_3;\frac{1}{\omega\sqrt{3}})
\\
&
 + 40 \, \Lambda(0,0,v_3; \frac{4}{\omega\sqrt{3}})
-40 \, (2 - \omega ) \Lambda(1,1,v_3; \frac{1}{\sqrt{3}\omega})
,
\end{aligned}
\end{equation}
\begin{equation}
\begin{aligned}
\Delta_{U(1)_Y} 
=\;&\left( 5\omega  +\left\{\begin{array}{c} -2 \\ -1 \\ \emptyset  \end{array}\right\} \right) \ln 2 
-  \left(\frac{17}{3} + 12 \omega  +\left\{\begin{array}{c} -3 \\ 3\\ \emptyset \end{array}\right\} \right) \ln 5 
\\
&
- \frac{10}{3} \, \Lambda(0,0,v_1;\frac{2}{\sqrt{3}})+ \frac{2}{3} \Lambda(0,0,2v_1;\frac{4}{\sqrt{3}})
\\
&
  - 3 \left\{\begin{array}{c} \Lambda(1,0,v_1; \frac{2}{\sqrt{3}} ) +\Lambda(1,0,v_1;2\sqrt{3})  \\ \Lambda(1,0,v_1;2\sqrt{3})  \\\Lambda(1,1,v_1;2\sqrt{3})  \end{array}\right\}
  + 2 \, \Lambda(\sigma^2_{aa'},\tau^2_{aa'},v_2;\frac{2}{\sqrt{3}})
\\
&
-2 \,  \Lambda(\sigma^2_{bd},\tau^2_{bd},v_2;2\sqrt{3})
 -  \Lambda(\sigma^2_{bd'},\tau^2_{bd'},v_2;2\sqrt{3})
-2\, \Lambda(\sigma^2_{cd'},\tau^2_{cd'},v_2;2\sqrt{3})
\\
&
- 2 \, \Lambda(\sigma^2_{dd'},\tau^2_{dd'},v_2;2\sqrt{3})
+ 2 \, \Lambda(\sigma^2_{dd'},\tau^2_{dd'},2v_2;4\sqrt{3})
 -\omega \Lambda(\sigma^2_{bc},\tau^2_{bc},v_2;2\sqrt{3})
\\
&
- \omega \Lambda(\sigma^2_{bc'},\tau^2_{bc'},v_2;2\sqrt{3})
 + 2 \, \Lambda(\sigma^2_{cc'},\tau^2_{cc'},v_2;2\sqrt{3}) + 2 \, \Lambda(\sigma^2_{cc'},\tau^2_{cc'}, 2 v_2; 4\sqrt{3})
\\
&
-  \left\{\begin{array}{c} 0 \\\Lambda(\sigma^2_{c\hat{h}_1},\tau^2_{c\hat{h}_1},v_2;2\sqrt{3})
 +  \Lambda(\sigma^2_{c\hat{h}_1'},\tau^2_{c\hat{h}_1'},v_2;2\sqrt{3})
\\ \emptyset \end{array}\right\}
\\
&
 - 20 \, \Lambda(0,0,v_3;\frac{1}{\omega\sqrt{3}}) + 10 \, \Lambda(0,0,v_3; \frac{4}{\omega\sqrt{3}})
- 3 \, \Lambda(0,0,v_3;\sqrt{3}\omega)
\\
&
 + 3 \,\Lambda(0,0,v_3;4\omega\sqrt{3}) 
-10 \, (2 - \omega ) \Lambda(1,1,v_3; \frac{1}{\sqrt{3}\omega})
,
\end{aligned}
\end{equation}
and for the two choices of hidden gauge groups $Sp(6)_{h_3}$ or $Sp(2)_{\hat{h}_1}$
\begin{equation}
\begin{aligned}
%%%%%%%%%%%%%%%%%%%%%%%%%%%
\Delta_{Sp(6)_{h_3}} =\;&  -4 \, \ln 5 + \frac{11}{3} \, \ln 2 - \Lambda(1,0,v_1;\frac{2}{\sqrt{3}}) +  2 \Lambda(0,0,2v_1;\frac{4}{\sqrt{3}})
\\ &
+ 6 \, \Lambda(0,0,v_2;\frac{2}{\sqrt{3}})
- 6 \, \Lambda(1,1,v_3;\frac{1}{\omega\sqrt{3}})
,
\\
%%%%%%%%%%%%%%%%%%%%%%%%%%%
\Delta_{Sp(2)_{\hat{h}_1}} 
=\;& -3 \, \ln 5 + 12 \, \ln 2 -3 \,  \Lambda(0,0,v_1; 2 \sqrt{3}) + 2 \,  \Lambda(0,0,v_2;2\sqrt{3})
\\ &
 + 2 \Lambda(0,0,2v_2;4\sqrt{3})
 -4 \, \Lambda(\sigma^2_{b\hat{h}_1},\tau^2_{b\hat{h}_1},v_2;2\sqrt{3})
\\ &
 -2 \, \Lambda(\sigma^2_{c\hat{h}_1},\tau^2_{c\hat{h}_1},v_2;2\sqrt{3})
 -18 \,  \Lambda(1,1,v_3; \frac{1}{\sqrt{3}\omega})
.
\end{aligned}
\end{equation}

As in the previous example, the generic formulae for arbitrary displacements and Wilson lines on the second two-torus
simplify significantly for choosing identical two-torus volume moduli $v_1=v_2=v_3=v$ and only $(\sigma^2_c,\tau^2_c) \neq (0,0)$ in order to 
break $SO/Sp(2)_c \rightarrow U(1)_c$, whereas all other branes $x \in \{a,b,d,h_3,\hat{h}_1\}$ have $(\sigma^2_x,\tau^2_x)=(0,0)$.
The gauge threshold corrections  in this case are given by 
\begin{equation}\label{eq:tcex1}
\begin{aligned}
\Delta_{SU(3)_a} =\;& 4 \, \tilde{\Lambda}(0,0,v) + 4 \, \tilde{\Lambda}(0,0,2v) -6(2-\omega) \tilde{\Lambda}(1,1,v)
\\
&
 -7 \, \ln 5 + 30 \, \ln 2 + 6 \, \ln \omega
-4 \, \ln 3 
,
\end{aligned}
\end{equation}

\begin{equation}
\begin{aligned}
\Delta_{SU(2)_b} =\;&\left(-12 \omega -19 + \left\{\begin{array}{c} 0 \\ -7 \\ 0 \end{array}\right\}  \right) \, \tilde{\Lambda}(0,0,v)
+ (2 \omega + 12) \, \tilde{\Lambda}(0,0,2v)
\\
&
 - 3 \, \tilde{\Lambda}(1,\omega-1,v) -2 \, \tilde{\Lambda}(\sigma^2_{c},\tau^2_{c},v)
\\
&
 +\left( - 24\omega + 28 + \left\{\begin{array}{c} - 5  \\ -20 \\ 0 \end{array}\right\}  \right) \, \ln 2 
- \left(36 + 3 \omega + \left\{\begin{array}{c} 9 \\ 0 \\ 0 \end{array}\right\}  \right) \, \ln 5
\\
&
 - 18 \, \ln \omega 
+ \left( \frac{29}{2} - 5 \omega + \left\{\begin{array}{c} 0 \\ -\frac{7}{2} \\ 0 \end{array}\right\}  \right) \, \ln 3
,
\end{aligned}
\end{equation}

\begin{equation}
\begin{aligned}
\Delta_{U(1)_{B-L}} =\;& - \frac{160}{3} \tilde{\Lambda}(0,0,v) + \frac{32}{3} \tilde{\Lambda}(0,0,2v) -12 \tilde{\Lambda}(1,\omega-1,v) 
-40(2-\omega) \tilde{\Lambda}(1,1,v)
\\
&
-4 \, \tilde{\Lambda}(\sigma^2_{c},\tau^2_{c},v)
-\frac{32}{2} \, \ln 5 + 36 \, \ln 2  + \frac{28}{3} \, \ln 3 + 40 \, \ln \omega
,
\end{aligned}
\end{equation}

\begin{equation}
\begin{aligned}
\Delta_{U(1)_Y} =\;& -\frac{49}{3} \tilde{\Lambda}(0,0,v) + \frac{8}{3} \tilde{\Lambda}(0,0,2v) -2(1+\omega)\tilde{\Lambda}(\sigma^2_{c},\tau^2_{c},v)
\\
&
+ 2 \tilde{\Lambda}(2\sigma^2_{c},\tau^2_{2c},v)+ 2 \tilde{\Lambda}(2\sigma^2_{c},2\tau^2_{c},2v) 
+ 10(\omega-2) \tilde{\Lambda}(1,1,v)
\\
&
+ \left\{\begin{array}{c} -6 \tilde{\Lambda}(1,0,v) \\ -3 \tilde{\Lambda}(1,0,v) -2 \tilde{\Lambda}(\sigma^2_{c},\tau^2_{c},v) \\ -3 \tilde{\Lambda}(1,1,v)\end{array}\right\}
+ \left(\frac{59}{3} +  5 \omega  + \left\{\begin{array}{c} -2 \\ -1 \\ 0 \end{array}\right\}\right) \ln 2 
\\
&
- \left(\frac{17}{3} + 12 \omega + \left\{\begin{array}{c} -3 \\ 3 \\  0\end{array}\right\} \right) \ln 5 
+ \frac{19}{6} \ln 3 
+ 10 \ln \omega 
,
\end{aligned}
\end{equation}

\begin{equation}\label{eq:tcex2}
\begin{aligned}
\Delta_{Sp(6)_{h_3}} =\;& -4 \, \ln 5 -4 \, \ln 3 + \frac{41}{3} \, \ln 2 + 6 \, \tilde{\Lambda}(0,0,v) + 2 \, \tilde{\Lambda}(0,0,2v)
\\ &
-\tilde{\Lambda}(1,0,v)- 6 \, \tilde{\Lambda}(1,1,v)
,
\\
\Delta_{Sp(2)_{\hat{h}_1}} =\;& -3 \, \ln 5 -\frac{3}{2} \, \ln 3 + 11 \, \ln 2 - 5 \, \tilde{\Lambda}(0,0,v)+ 2 \, \tilde{\Lambda}(0,0,2v)
\\ &
-2 \, \Lambda(\sigma^2_{c},\tau^2_{c},v)
 -18 \,  \tilde{\Lambda}(1,1,v)
.
\end{aligned}
\end{equation}

The lattice sums are abbreviated as follows. For vanishing relative displacements and Wilson lines, the dependence on the (length)$\!{}^2$
$V$ of the two-cycle has been split off, and since for $(\sigma,\tau) \neq (0,0)$ the lattice sum does not depend on $V$, the fourth argument has been dropped,
\begin{equation}
\begin{aligned}
\tilde{\Lambda}(0,0,v) &\equiv \Lambda(0,0,v;V) - \ln V = \ln \left(2 \pi v \, \eta^4 (iv) \right)
,
\\
\tilde{\Lambda}(\sigma,\tau,v)_{(\sigma,\tau)\neq(0,0)} &\equiv \Lambda(\sigma,\tau,v;V)_{(\sigma,\tau)\neq(0,0)} 
=  \ln\left|e^{-\pi \sigma^2 v/4}\frac{\vartheta_1 (\frac{\tau}{2}-i\frac{\sigma}{2} \, v,i v)}{\eta (i v)}\right|^2
.
\end{aligned}
\end{equation}

The two-torus volume dependence of the simplified threshold contributions~\eqref{eq:tcex1}~--~\eqref{eq:tcex2} is shown
in figures~\ref{fig:thresholds_ex_h} and~\ref{fig:thresholds_ex_nh} for the three cases of hidden sector configurations $(h_3,\hat{h}_1,\emptyset)$
and the choice $(\sigma^2_c,\tau^2_c)=(\frac{1}{2},0)$ and $(\sigma^2_x,\tau^2_x)=(0,0)$ for $x \neq c$.
As can be seen in the plots, the $SU(3)_a$ and the hidden $Sp(6)_{h_3}$ or $Sp(2)_{\hat{h}_1}$ branes obtain in the whole geometric regime $v>1$ a negative threshold correction
for the given choice of continuous Wilson lines and displacements on $T^2_3$,
which means that the respective gauge couplings are enhanced.

\twofig{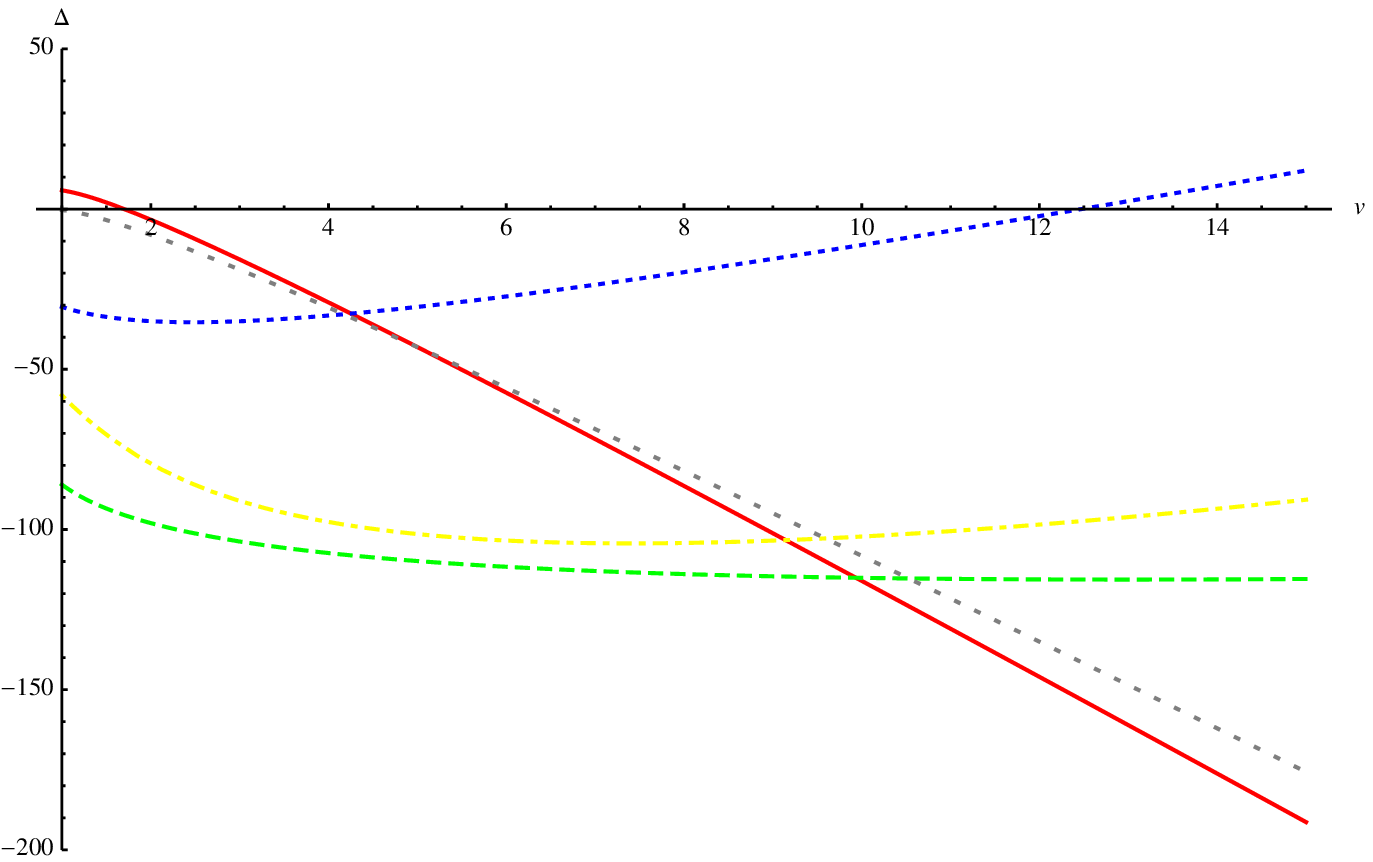}{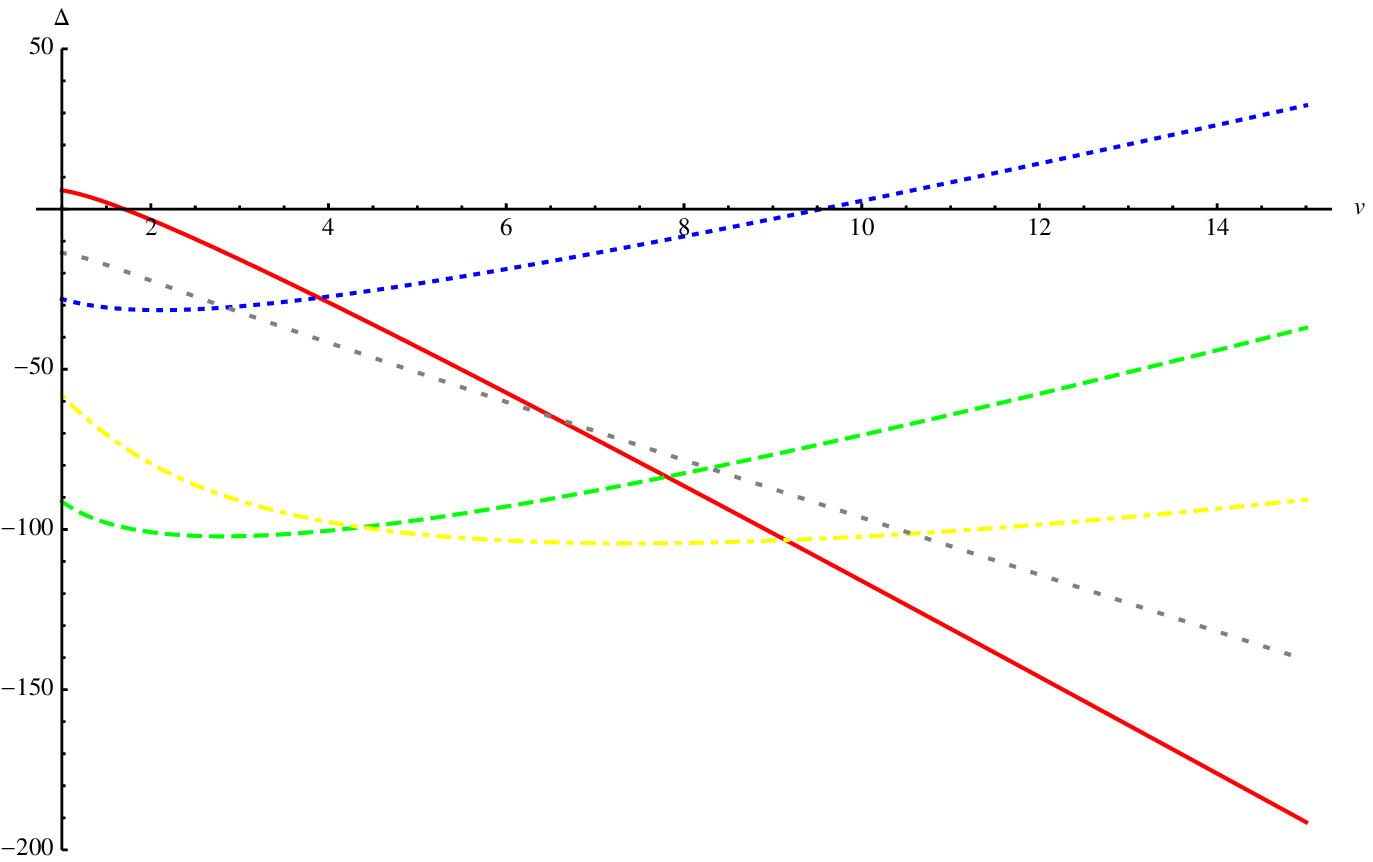}{Gauge threshold contributions in dependence on the two-torus volume $v$ for the $T^6/\bZ_6'$ examples with hidden sector branes (a) $Sp(6)_{h_3}$ and (b) $Sp(2)_{\hat{h}_1}$. The contribution to $SU(3)_a$ is shown in solid red, $SU(2)_b$ in dashed green, $U(1)_Y$ in dotted blue, $U(1)_{B-L}$ in dot-dashed yellow and for the hidden sector as sparse dashed grey. The geometric regime requires $v>1$, and the gauge thresholds are plotted for the choice of continuous displacements and Wilson lines  $(\sigma^2_c,\tau^2_c)=(\frac{1}{2},0)$ and $(\sigma^2_x,\tau^2_x)=(0,0)$ for $x \neq c$.}{fig:thresholds_ex_h}
\fig{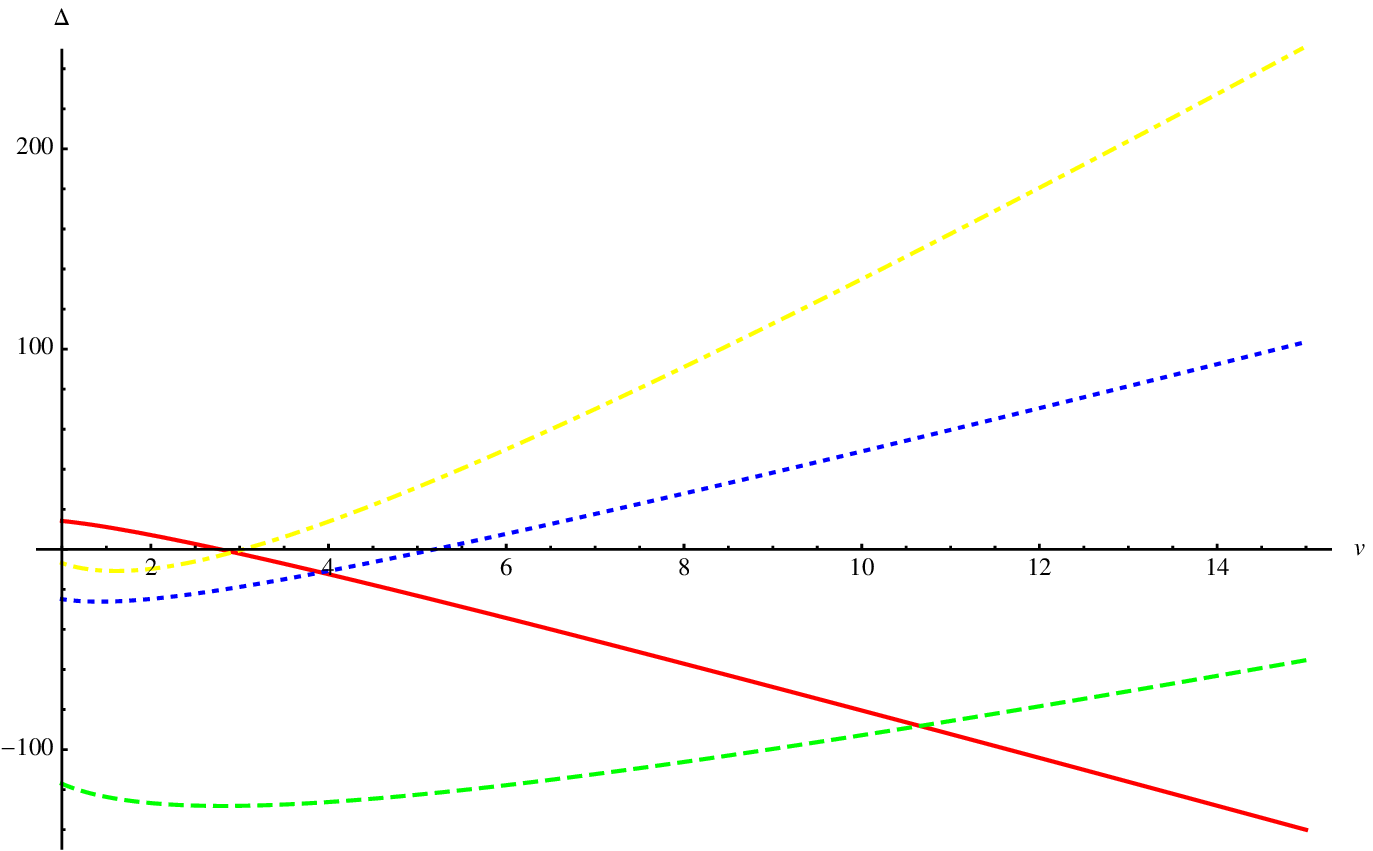}{Gauge threshold contributions in dependence of the two-torus volume $v$ for the $T^6/\bZ_6'$ example without a hidden sector. The color coding for the visible sector is the same as in figures~\protect\ref{fig:thresholds_ex_h}, the continuous open string moduli are again chosen as $(\sigma^2_c,\tau^2_c)=(\frac{1}{2},0)$ and $(\sigma^2_x,\tau^2_x)=(0,0)$ for $x \neq c$.}

%%%%%%%%%%%%%%%%%%%%%%%%%%%%%%%%%%%%%%%%%%%%%%%%%%%%%%%%%%%%%%%%

\subsection{Comments on gauge coupling unification}

The $T^6/\mathbb{Z}_6$ example in section~\ref{Ss:Ex1} fulfills GUT relations for the gauge couplings at string tree level, $\frac{\alpha_{s, {\rm tree}}}{\alpha_{w, {\rm tree}}}=1$ and $\sin^2 \theta_{W, {\rm tree}} = 0.375$, provided that the D6-brane $e$ is included in the definition of the hyper charge. 

The $T^6/\mathbb{Z}_6'$ examples presented in section~\ref{Ss:Ex2} on the other hand, deviate at string tree level considerably from a unification of the gauge couplings, $\frac{\alpha_{s, {\rm tree}}}{\alpha_{w, {\rm tree}}}=6 \omega$ and $\sin^2 \theta_{W, {\rm tree}} = 0.72$ in the case without hidden sector and $0.65$ in the case with hidden sectors. 

In order to see how these relations change at 1-loop in the string coupling, we study here some numerical values at the string scale $M_{\rm string}$ for the tree level gauge couplings and threshold corrections. For a $SU(N_x)$ gauge factor on a {\it fractional} D6-brane $x$, one has
\begin{equation}\label{Eq:gauge-tree-frac}
\frac{1}{\alpha_{x,{\rm tree}}^{\rm frac}} = \frac{M_{\rm Planck}}{M_{\rm string}} \frac{ \prod_{i=1}^3 \sqrt{V^{(i)}_{xx}}}{4 \sqrt{2}}
\end{equation}
with $\frac{ \prod_{i=1}^3 \sqrt{V^{(i)}_{xx}}}{4 \sqrt{2}} = 1/(2 \cdot 3^{1/4}) \sim 0.38$ for all D6-branes $a,b,c,d,e$ in the $T^6/\mathbb{Z}_6$ example and the values listed in table~\ref{tab:GG-numerics} in the $T^6/\mathbb{Z}_6'$ example.
\mathtab{
\begin{array}{|c||c|c|c|c|}\hline
{\rm D6-brane } x & a & b & c & d 
\\\hline\hline
\frac{\prod_{i=1}^3 \sqrt{V^{(i)}_{xx}}}{4 \sqrt{2}}  & \frac{1}{2(6\sqrt{3}\omega)^{1/2}} & \left(\frac{\sqrt{3}\omega}{2}\right)^{1/2} & \frac{1}{2}  \left(\frac{\sqrt{3}\omega}{2}\right)^{1/2} & \frac{1}{2}  \left(\frac{\sqrt{3}}{2 \omega}\right)^{1/2} 
\\\hline
\omega =1 & 0.16 & 0.93 & 0.47 & 0.47
\\\hline
\omega =2 & 0.11 & 1.3 & 0.66 & 0.33
\\\hline
\end{array}
}{GG-numerics}{Numerical values of the 3-cycle volumes of the Standard Model branes in the $T^6/\Z_6'$ examples. $\omega=1$ corresponds to the examples with hidden sectors $Sp(6)_{h_3}$ and $Sp(2)_{\hat{h}_1}$, whereas $\omega=2$ pertains to the example without hidden sector.
}

Secondly, the values of the gauge threshold corrections depend on the size of the compact six-dimensional volume $V_6 \equiv \prod_{i=1}^3 v_i$, where $v_i$ are the two-cycle volumes in units of $\alpha'$ defined in~(\ref{Eq:Def-Vaa}). 
For isotropic two-torus volumes $v_i \equiv v$, rewriting the dimensionally reduced gravitational action as done e.g. in~\cite{Blumenhagen:2003jy} leads to the relation
\begin{equation}\label{Eq:volume-scales}
v = \frac{1}{2} \left( g_{\rm string} \frac{M_{\rm Planck}}{M_{\rm string}} \right)^{2/3}.
\end{equation}
Let us now assume that $M_{\rm string}=M_{GUT} \sim 2 \cdot 10^{16} GeV$ leading to the ratio of mass scales $M_{\rm Planck}/M_{\rm string} \sim 600$, on which both the tree level values and the 1-loop corrections to the gauge couplings depend, and $g_{\rm string}=g_{GUT} \sim \sqrt{\pi/6 } $, which gives the size of isotropic two-tori $v \sim 30$. For these values and the most simple choice of continuous displacements and Wilson lines, $(\sigma_c^i,\tau_c^i)=(\frac{1}{2},0)$ and all other $\{\sigma_x^i,\tau_x^i\}$ vanishing with $i=3$ for the $T^6/\mathbb{Z}_6$ and $i=2$ for the $T^6/\mathbb{Z}_6'$ examples, one arrives for the $T^6/\mathbb{Z}_6$ example at 
\begin{equation}
\begin{aligned}
\frac{1}{\alpha_{SU(3)_a,{\rm tree}}} + \frac{\Delta_{SU(3)_a}}{4 \pi} & \sim 228 + 16 = 244
, \\
\frac{1}{\alpha_{SU(2)_b,{\rm tree}}} + \frac{\Delta_{SU(2)_b}}{4 \pi} & \sim 228 + 15 = 243
,
\end{aligned}
\end{equation}
and for the $T^6/\mathbb{Z}_6'$ examples at 
\begin{equation}\label{eq:gauge-values-at-string}
\begin{aligned}
\frac{1}{\alpha_{SU(3)_a,{\rm tree}}} + \frac{\Delta_{SU(3)_a}}{4 \pi}  & \sim \left\{\begin{array}{c}  
93 - 34
\\
93 - 34
\\
66 - 26
\end{array}\right\}
= \left\{\begin{array}{c} 59 \\ 59 \\ 40
\end{array}\right\} , 
\\
\frac{1}{\alpha_{SU(2)_b,{\rm tree}}} + \frac{\Delta_{SU(2)_b}}{4 \pi}  & \sim \left\{\begin{array}{c}  
558 - 9 
\\
558 + 6 
\\
790 + 7 
\end{array}\right\}
= \left\{\begin{array}{c} 549 \\ 564 \\ 797
\end{array}\right\} , 
\end{aligned}
\end{equation}
where the entries in the columns correspond to the configurations with hidden sectors of rank three or one or no hidden sector, $(h_3,\hat{h}_1,\emptyset)$, as before. The running of the $SU(3)_a$ and $SU(2)_b$ couplings for the example without a hidden sector and the same simple choice of open string moduli 
is shown graphically in figure~\ref{fig:runningz6pnh}. 

\fig{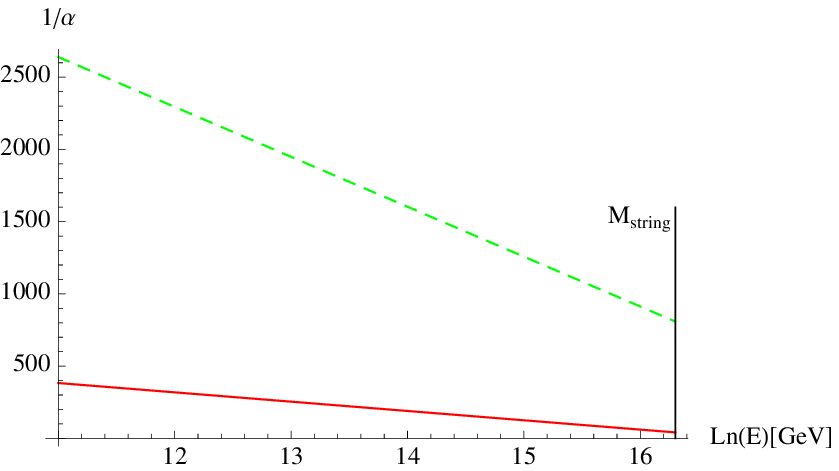}{Running of the gauge couplings in the $T^6/\mathbb{Z}_6'$ example without a hidden sector for the choice $(\sigma_c^2,\tau_c^2)=(\frac{1}{2},0)$ and all others vanishing, $\sigma_x^2=\tau_x^2=0$ for $x \in \{a,b,d\}$. For these values of the moduli, some non-chiral particles $\omega \times [(\1,\2)_{\bf 1/2,0}^{(1,0)} + (\1,\2)_{\bf -1/2,0}^{(-1,0)} + c.c.]$ acquire a mass which is according to~(\protect\ref{Eq:Open-Massformula}) proportional to $M^2_{0,0}(b,c) = v_2/(16 \cdot V^{(2)}_{bc}) \sim 30/(32\sqrt{3})$. Assuming that these masses are exactly at the string scale gives $b_{SU(3)_a}=14$ and $b_{SU(2)_b}=75$ for the model without hidden sector. The 1-loop corrected values of the gauge couplings at $M_{\rm string}$ are given in~(\protect\ref{eq:gauge-values-at-string}).  
The red, solid line shows the running of the $SU(3)_a$ coupling down to the electro-weak scale, the green, dashed line the $SU(2)_b$ coupling.}

All numerical values of the 1-loop corrected gauge couplings for $SU(3)_a$ and $SU(2)_b$ of the examples with the given choices of string scale $M_{\rm string}$ and string coupling $g_{\rm string}$ to match the GUT-scale, and with the most simple choice of open string moduli $\{\sigma^i_x,\tau^i_x\}$ obviously deviate from the value $\alpha_{GUT}^{-1} \sim 24$.  In addition, since most non-chiral particles are treated as massless, the couplings at the 
electro-weak scale $M_Z \approx 91 GeV$ are too weak compared to the experimental values $\alpha_s^{-1}(M_Z)\sim 9$ and $\alpha_w^{-1}(M_Z) \sim 29$.   
The fine structure constant $\alpha_{SU(3)_a}$ in the $T^6/\mathbb{Z}_6'$ examples, however, clearly shows that 1-loop corrections can change the numerical value at the string scale significantly.

One should keep in mind that $M_{\rm string}$ is in type II string constructions not naturally related to the GUT-scale, and that gauge coupling unification is not expected to occur in general. In order to study the running down to the electro-weak in detail and match with experimental data, it is furthermore essential to know the scale at which non-chiral particles acquire masses. The full study of the gauge coupling dependence on these masses as well as generic choices of open string moduli $\{\sigma^i_x,\tau^i_x\}$ for all D6-branes $x$ and string coupling constant $g_{\rm string}$ and string scale $M_{\rm string}$ goes beyond the scope of the present article.

%%%%%%%%%%%%%%%%%%%%%%%%%%%%%%%%%%%%%%%%%%%%%%%%%%%%%%%%%%%%%%%%%%%%%%%%5
\section{Statistics for \texorpdfstring{$T^6/\Z_6'$}{T6/Z6'}}\label{S:Statistics}

In this section, a systematic computer analysis of the diversity at one-loop of
three generation Standard Model spectra without chiral exotics on the {\bf ABa}
lattice  on $T^6/\Z_6'$ is considered. 
In~\cite{Gmeiner:2008xq} we found that a huge number of different configurations
of exceptional three-cycles only gave three different values of {\it chiral}
observable sector spectra and tree level gauge couplings of
$SU(3) \times SU(2) \times U(1)_Y$.
Here we analyze the {\it complete}, i.e. chiral plus non-chiral, massless matter
spectrum, the respective one-loop beta function coefficients and gauge threshold
corrections. Not surprisingly, the number of models with distinct features is 
enhanced from three to 196, which is about one quarter of the number of
configurations of fractional three-cycles on which these are realized.

%%%%%%%%%%%%%%%%%%%%%
\subsection{Computation}\label{S:Computation}

The setup used for a statistical evaluation of the one-loop threshold
corrections is the ensemble of models on $T^6/\mathbb{Z}_6'$ considered
in~\cite{Gmeiner:2007zz,Gmeiner:2008xq}.
The total number of supersymmetric solutions to the RR tadpole cancellation
conditions found in this background was $\mathcal{O}(10^{23})$.
In order to compute the threshold corrections for this set of models, we
have to obtain the relevant intersection numbers and angles in the toroidal
and $\mathbb{Z}_2$ twisted sectors for all of them. At one-loop in the gauge couplings, the models are completely characterized by these three quantities.

The $\Z_2$ invariant intersection numbers in appendix~\ref{AppSs:Z2-numbers} turn out to be identical for configurations
where some brane displacement parameters $(\tilde{\sigma}^2,\tilde{\sigma}^2;\tilde{\sigma}^5,\tilde{\sigma}^6) $ and $\Z_2$ eigenvalues $\tau^0$ 
are exchanged. With the shortened notation of $\sigma^1,\sigma^3 \in \{0,1\}$ if branes pass through the origin or not on the the corresponding two-torus, 
the number of (at one-loop in the gauge couplings) nonequivalent
supersymmetric solutions to the RR tadpole cancellation conditions is by a
factor of $2 \cdot 4^k$ reduced compared to the values obtained
in~\cite{Gmeiner:2008xq}, where $k$ is the total number of (visible and hidden)
stacks of D6-branes. The factor arises from an exchange symmetry of the
`right-brane' $c$ and the `leptonic brane' $d$. This reduces the number of $\mathcal{O}(7 \cdot 10^6)$ Standard Models without chiral exotics of~\cite{Gmeiner:2008xq}
to 768 distinct models, for which the massless spectra, beta function coefficients and gauge thresholds can be computed explicitly.

The values of the gauge thresholds depend on unfixed moduli, in
particular the overall volume of the three two-tori, continuous Wilson lines
and brane displacements. A full comparison of the one-loop gauge thresholds for all continuous parameters does, however, overcharge the computational effort. 
For practical reasons, we will therefore fix the two-tori volumes $v$ at specific values in the
following and set continuous Wilson lines $\tau^2$ and displacements $\sigma^2$
to zero, except for the `right-brane' $c$ (cf. the explicit examples in
section~\ref{Ss:Ex2}).

%%%%%%%%%%%%%%%%%%%%%%%%%%%%%%%%%%%%%%%%%
\subsection{Non-chiral matter content}\label{S:nonchiral}

In total we found  768 different models that have a visible sector with the
chiral spectrum of the MSSM (except for the Higgs sector), a massless
hypercharge and no chiral exotics.Out of these there are actually only 16 distinct classes of models, listed in
table~\ref{tab:nonchiralmatter},  which differ in their non-chiral matter content.
To keep this table readable we list the sum of the total massless matter content
for the various visible and hidden sectors.
The examples in section~\ref{Ss:Ex2} belong to  class (viii) for no hidden
sector, (x) for a hidden $Sp(2)_{\hat{h}_1}$ and (xii) for a hidden
$Sp(6)_{h_3}$ gauge group.

The numbers of bifundamentals and symmetric and antisymmetric 
representations on branes $a,b,d$ matches those computed in section 5, 
whereas the number of antisymmetric and symmetric representations of 
brane $c$ does not match exactly. This is on the one hand due to the 
fact that the computer algorithm for this matter on $U(N)$ stacks does 
not apply to orientifold invariant branes such as $c$, for which we 
discussed the matter content in appendix A.3. On the other hand, in 
section 5, we used that upon decomposing $Sp(2N) \rightarrow U(N)$ by 
separating branes and their orientifold images the vector and one chiral 
multiplet in the representation $[\Sym_{U(N)} + c.c.]$ acquire masses.

As can be seen from  table~\ref{tab:nonchiralmatter} there are also only three 
classes of models with different ratios of the tree level gauge couplings at the
string scale and Higgs candidates, for more details see~\cite{Gmeiner:2008xq}.
However, due to the different non-chiral matter sector one expects the one-loop 
running of the gauge couplings and the threshold corrections to be more diverse.
We will see in the next section that this is indeed the case.

\begin{sidewaystable}[ht]
\centering
\resizebox{\linewidth}{!}{%
\begin{tabular}{|c|rrrrr|rrrrrrrrr|r|}\hline
no. & $s$ & $rk_h$ & $h$ & $\frac{\alpha_s}{\alpha_w}$ & $\sin^2\theta_W$
& $\sum\limits_{x}\varphi^{ax}$ & $\sum\limits_{x}\varphi^{ax'}$
& $\sum\limits_{x}\varphi^{bx}$ & $\sum\limits_{x}\varphi^{bx'}$
& $\sum\limits_{x,h}\varphi^{xh}$ & $\sum\limits_{x,h}\varphi^{xh'}$
& $\sum\limits_{x}\varphi^{\Adj_x}$ & $\sum\limits_{x}\varphi^{\Anti_x}$
& $\sum\limits_{x}\varphi^{\Sym_x}$ & \#\\\hline
i   & 0 & 0 & 18 & 4  & 0.67 & 12 & 15 & 30 & 32 &  0 &  0 & 20 & 52 & 53 &  12\\
ii  & 0 & 0 & 18 & 4  & 0.67 & 12 & 15 & 30 & 32 &  0 &  0 & 20 & 64 & 39 &  12\\
iii & 0 & 0 & 18 & 4  & 0.67 & 12 & 15 & 32 & 30 &  0 &  0 & 20 & 52 & 53 &  24\\
iv  & 0 & 0 & 18 & 4  & 0.67 & 12 & 15 & 32 & 30 &  0 &  0 & 20 & 64 & 39 &  24\\
v   & 0 & 0 & 18 & 4  & 0.67 & 14 & 19 & 30 & 32 &  0 &  0 & 20 & 50 & 55 &  12\\
vi  & 0 & 0 & 18 & 4  & 0.67 & 14 & 19 & 30 & 32 &  0 &  0 & 20 & 66 & 37 &  12\\
vii & 0 & 0 & 21 & 12 & 0.72 & 9  & 14 & 32 & 32 &  0 &  0 & 26 & 44 & 37 &  12\\
viii& 0 & 0 & 21 & 12 & 0.72 & 9  & 14 & 32 & 32 &  0 &  0 & 26 & 48 & 31 &  12\\
ix  & 1 & 1 & 12 & 6  & 0.65 & 9  & 14 & 21 & 21 & 20 & 20 & 26 & 34 & 26 &  36\\
x   & 1 & 1 & 12 & 6  & 0.65 & 9  & 14 & 21 & 21 & 20 & 20 & 26 & 37 & 21 &  36\\
xi  & 1 & 3 & 12 & 6  & 0.65 & 9  & 14 & 21 & 21 &  4 &  4 & 26 & 34 & 26 & 108\\
xii & 1 & 3 & 12 & 6  & 0.65 & 9  & 14 & 21 & 21 &  4 &  4 & 26 & 37 & 21 & 108\\
xiii& 2 & 3 & 12 & 6  & 0.65 & 9  & 14 & 21 & 21 &  8 &  8 & 26 & 34 & 26 & 108\\
xiv & 2 & 3 & 12 & 6  & 0.65 & 9  & 14 & 21 & 21 &  8 &  8 & 26 & 37 & 21 & 108\\
xv  & 3 & 3 & 12 & 6  & 0.65 & 9  & 14 & 21 & 21 & 12 & 12 & 26 & 34 & 26 &  72\\
xvi & 3 & 3 & 12 & 6  & 0.65 & 9  & 14 & 21 & 21 & 12 & 12 & 26 & 37 & 21 &  72\\\hline
\end{tabular}}
\caption{Classes of models with a visible sector resembling the Standard Model. The number of hidden sector stacks is given by $s$, the total rank of the hidden sector gauge group by $rk_h$ and the number of Higgs pair candidates by $h$ (which can be either Higgs pair $H_u+H_d$ multiplets or lepton/anti-lepton pairs $L + \bar{L}$). Furthermore the ratios of values for the tree-level gauge couplings at the string scale and the total amount of chiral and non-chiral matter in the various sectors is given. The variable $x$ in the sums runs over the Standard Model D6-branes $a,b,c,d$, the variable $h$ over all stacks in the hidden sector. The last column gives the number of different geometrical realizations of each class.}
\label{tab:nonchiralmatter}
\end{sidewaystable}

%%%%%%%%%%%%%%%%%%%%%
\subsection{Field theoretic running of gauge couplings at one-loop}\label{Ss:S-beta-function}

The one-loop running of the gauge couplings due to massless charged string states is determined by the beta-function 
coefficient $b_a$ in~\eqref{Eq:Def-gauge}. It can be computed as outlined in
section~\ref{sec:Thresholds}. The relevant expressions in terms of
intersection numbers and angles are summarized in the appendix~\ref{AppS:Tables} in the fourth column of
Table~\ref{tab:Amplitudes-thresholds}.
The running of the couplings does not depend on the volume of the compactification space or the exact values of 
the continuous position and Wilson line moduli, 
in contrast to the threshold corrections $\Delta$ which we will consider in the next section.

\wfig{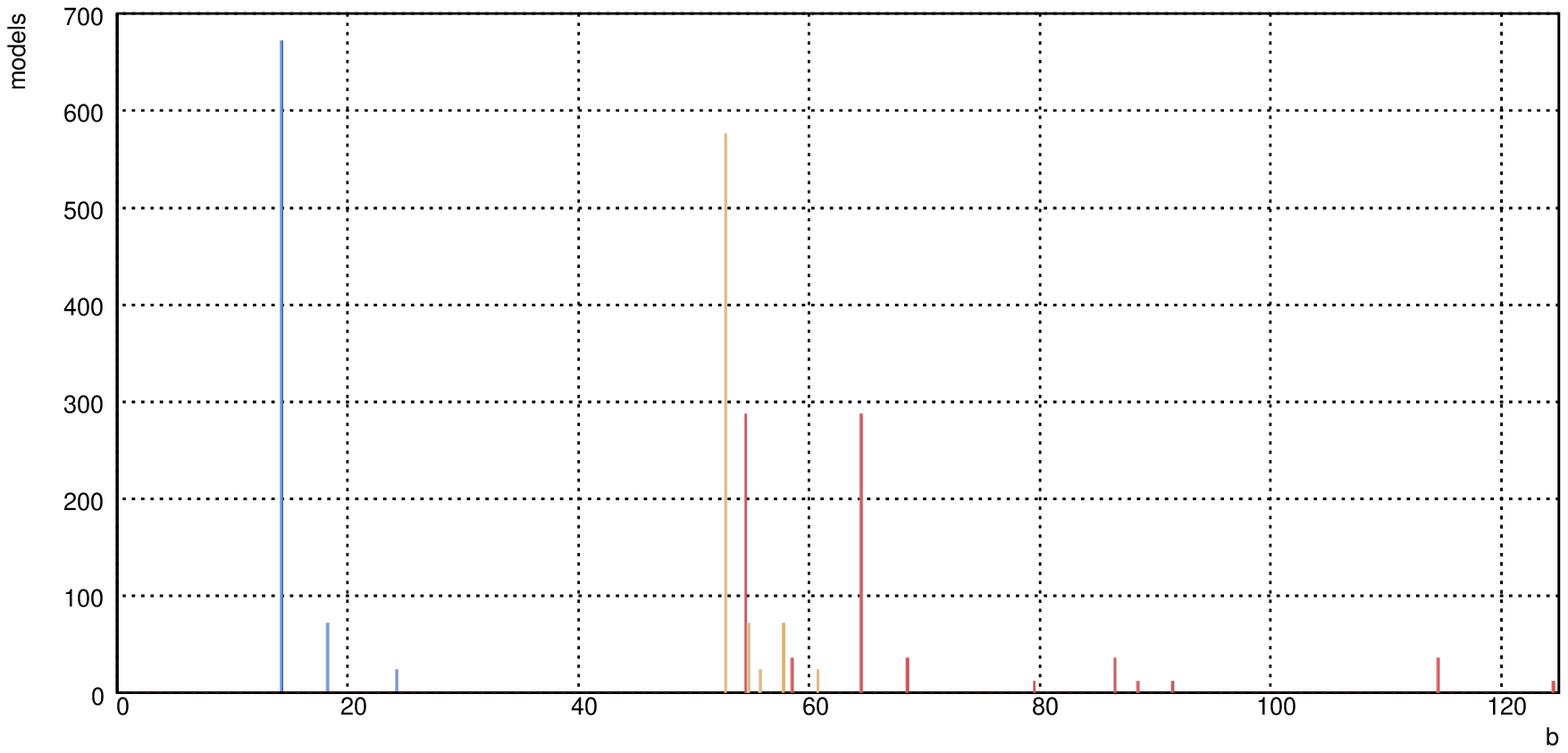}{Number of models with different values of $b_G$ for $G=SU(3)_a$ (blue), $SU(2)_b$ (red) and $U(1)_Y$ (yellow). Note that the color coding differs from the plots of the examples in section~\protect\ref{Ss:Ex2}}

We have computed the distribution of the beta function coefficients for the ensemble
of Standard Models on $T^6/\mathbb{Z}_6'$.
The resulting frequency distribution for the running of $SU(3)$, $SU(2)$ and $U(1)_Y$
is shown in figure~\ref{fig:z6p_beta}, 
and in table~\ref{tab:beta-correlations}, the correlations among different beta function coefficients are listed.
On the {\bf ABa} lattice on $T^6/\Z_6'$, a total of ten different combinations of one-loop beta function coefficients for the 
Standard Model gauge group occur.

\mathtab{
\begin{array}{|c||c|c|c|c||c|}\hline
{\rm no.} & b_{SU(3)_a} & b_{SU(2)_b} & b_{U(1)_Y} & b_{U(1)_{B-L}} & \# \\\hline\hline
{\rm xii, xiv, xvi} & 14 &  54 & 157/3 &  280/3 & 288 \\\hline
{\rm x}             & 14 &  62 & 163/3 &  280/3 &  36 \\\hline
{\rm xi, xiii, xv}  & 14 &  64 & 157/3 &  280/3 & 288 \\\hline
{\rm ix}            & 14 &  68 & 163/3 &  280/3 &  36 \\\hline
{\rm viii}          & 14 &  79 & 181/3 &  280/3 &  12 \\\hline
{\rm vii}           & 14 &  91 & 181/3 &  280/3 &  12 \\\hline
{\rm ii, iv}        & 18 &  86 & 58    &  132   &  36 \\\hline
{\rm i, iii}        & 18 & 114 & 58    &  132   &  36 \\\hline
{\rm vi}            & 24 &  88 & 56    &  124   &  12 \\\hline
{\rm v}             & 24 & 124 & 56    &  124   &  12 \\\hline
\end{array}
}{beta-correlations}{Combinations of beta function coefficients for $SU(3)_a$, $SU(2)_b$, $U(1)_Y$ and $U(1)_{B-L}$ for the classes of spectra in table~\protect\ref{tab:nonchiralmatter}.}

The beta function coefficients for $SU(3)$, $SU(2)$ and $U(1)_{B-L}$ 
match the values in section 5, whereas for $U(1)_Y$ the values are 
augmented by 6 compared to section 5. The reason is, as explained in 
section 6.2, the mismatch in the counting of symmetric representations 
on brane $c$.

%%%%%%%%%%%%%%%%%%%%%
\subsection{Threshold corrections}\label{Ss:S-thresholds}
 In contrast to the beta-function coefficients considered above, there is
a dependency on the two-torus volumes $v$ present in the gauge  thresholds $\Delta$ coming from sectors where some D6-branes are 
parallel.\footnote{The threshold contributions from non-trivially intersecting D6-branes depend on the complex structure moduli through
the angles. These complex structure moduli take discrete values for supersymmetric models and thereby the threshold contributions
are just some constants, whereas the K\"ahler moduli are unconstrained.}
Since these correspond to unfixed moduli, their values are a priori arbitrary.
On physical grounds however, one has to put lower and upper bounds. In order to be in the geometric regime where $\alpha'$ corrections
are sub-leading, $v>1$ is necessary, 
and an upper bound comes from the fact that large extra dimensions have not been observed so far.
In order to do a numerical comparison of the contributions in our ensemble of models,
we will choose different values for $v$ between $1$ and $10$.

\wfig{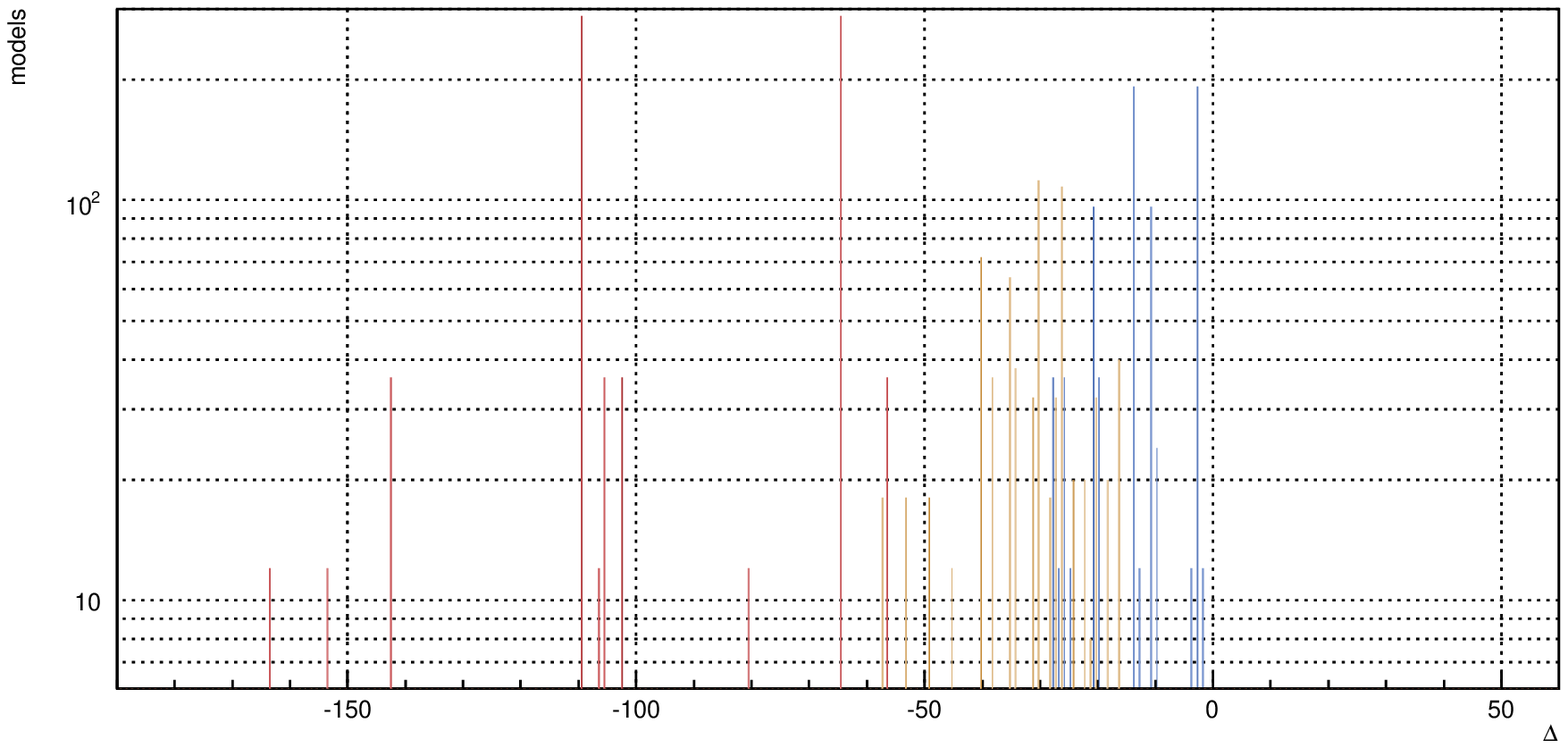}{Number of models with different values of the one-loop gauge threshold corrections $\Delta$ at common volume $v=1$ for $G=SU(3)_a$ (blue), $SU(2)_b$ (red) and $U(1)_Y$ (yellow).}
\wfig{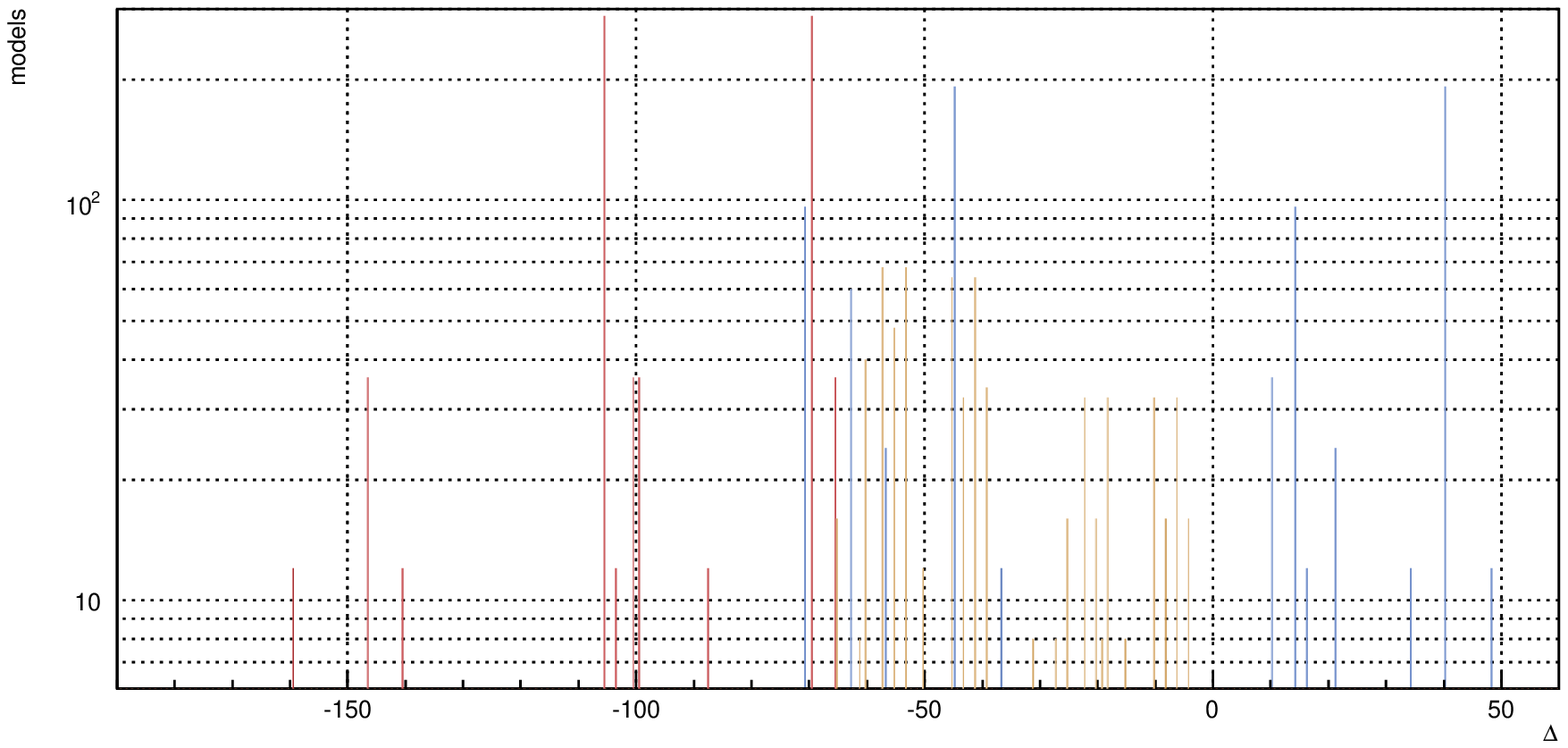}{Number of models with different values of the one-loop gauge threshold corrections $\Delta$ at common volume $v=5$ for $G=SU(3)_a$ (blue), $SU(2)_b$ (red) and $U(1)_Y$ (yellow).}
\wfig{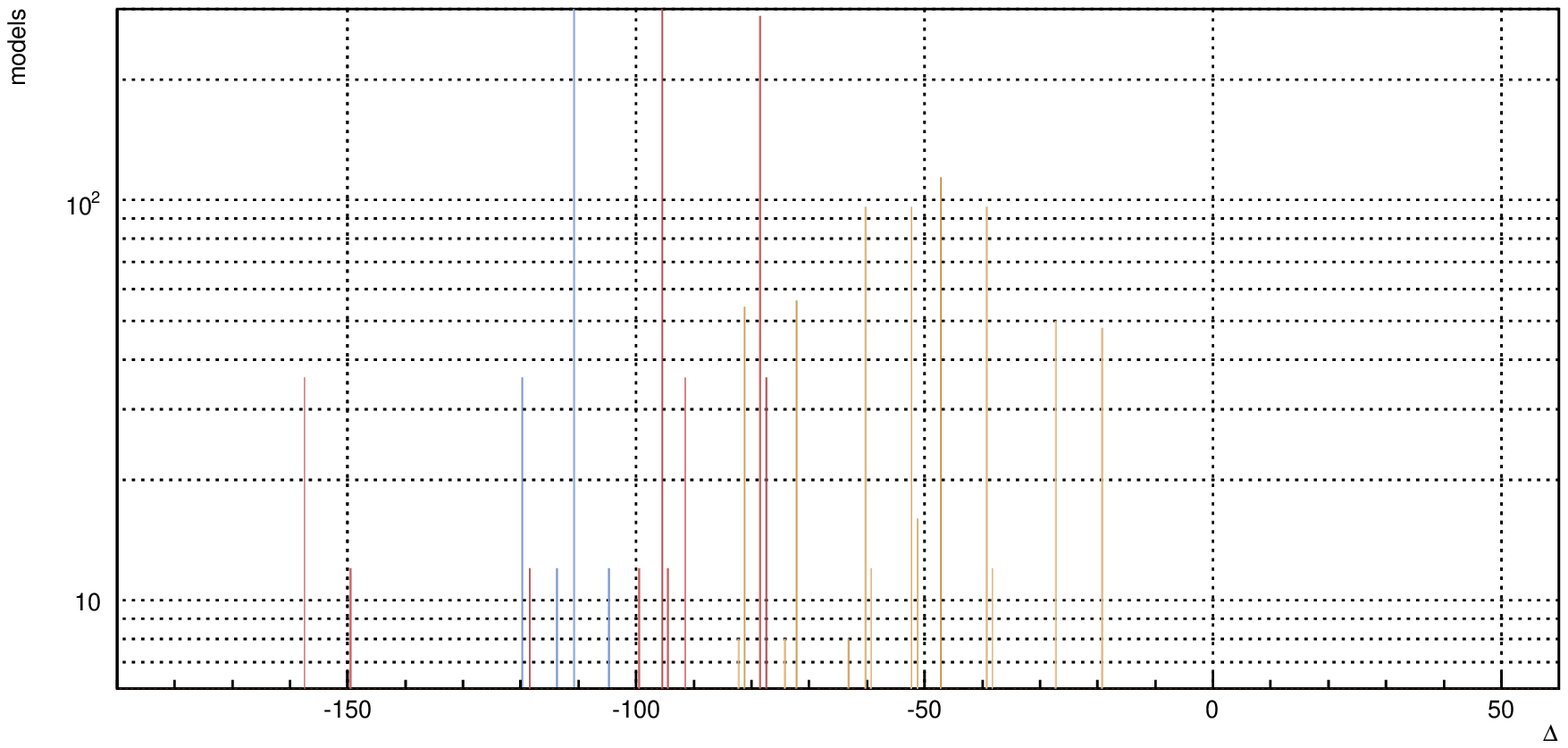}{Number of models with different values of the one-loop gauge threshold corrections $\Delta$ at common volume $v=10$ for $G=SU(3)_a$ (blue), $SU(2)_b$ (red) and $U(1)_Y$ (yellow).}

The resulting frequency distributions are shown in figures~\ref{fig:z6p_thresh_v1} to~\ref{fig:z6p_thresh_v10}. In table~\ref{tab:betathresh} we show the different combinations of beta functions and threshold corrections that occur for $v=5$.
It is interesting to note that the threshold corrections are considerably more diverse than the values for the running of the gauge coupling.
For $v=5$, a total of 84 combinations of beta function coefficients and values of gauge thresholds occurs. At the limiting value for a reliable supergravity description at  $v=1$ we find 112 different combinations of threshold corrections for the $SU(3)$, $SU(2)$ and $U(1)_Y$ gauge groups.
Including the different values for the beta functions we end up with 196 classes of models.
This shows that although there seemed to be not much variation in the different Standard Model-like constructions from the point of view of the chiral matter content, they actually do behave quite differently as soon as one includes one loop corrections.
This result does not come as a total surprise though, since we know that the non-chiral matter spectrum differs, which in turn gives contributions to the corrections of the gauge couplings, and identical multiplicities of massless states can differ in terms of their individual composition of intersection numbers.
Another characteristic feature of the distribution of $\Delta$ is the behavior of the thresholds corrections for different values of $v$. As we have already seen in the explicit examples in section~\ref{Ss:Ex2} there is a difference between the $v$-dependence of the $SU(3)$ and $SU(2)$ corrections. While the $SU(3)$ corrections tend to large negative values for larger $v$, the $SU(2)$ part remains more or less constant in the negative regime. This behavior can also be seen in the larger class of models under consideration here.

\mathtabfix{
\begin{array}{|rrrrrr|r||rrrrrr|r|}\hline
b_{SU(3)_a} & b_{SU(2)_b} & b_{U(1)_Y} & \Delta_{SU(3)_a} & \Delta_{SU(2)_b} & \Delta_{U(1)_Y} & \# &
b_{SU(3)_a} & b_{SU(2)_b} & b_{U(1)_Y} & \Delta_{SU(3)_a} & \Delta_{SU(2)_b} & \Delta_{U(1)_Y} & \# \\\hline\hline
14 & 54 & 52.37 & -44.12 & -69.92 & -53.64 & 32 &
14 & 54 & 52.37 & -44.12 & -69.92 & -55.82 & 16\\[-1ex]
14 & 54 & 52.37 & -44.12 & -69.92 & -57.89 & 32 &
14 & 54 & 52.37 & -44.12 & -69.92 & -60.07 & 16\\[-1ex]
14 & 54 & 52.37 & -70.32 & -69.92 & -18.71 & 32 &
14 & 54 & 52.37 & -70.32 & -69.92 & -20.90 & 16\\[-1ex]
14 & 54 & 52.37 & 14.52 & -69.92 & -4.57 & 16 &
14 & 54 & 52.37 & 14.52 & -69.92 & -6.76 & 32\\[-1ex]
14 & 54 & 52.37 & 40.72 & -69.92 & -39.50 & 16 &
14 & 54 & 52.37 & 40.72 & -69.92 & -41.68 & 32\\[-1ex]
14 & 54 & 52.37 & 40.72 & -69.92 & -43.75 & 16 &
14 & 54 & 52.37 & 40.72 & -69.92 & -45.93 & 32\\[-1ex]
14 & 62 & 54.37 & -36.78 & -65.02 & -61.95 & 4 &
14 & 62 & 54.37 & -36.78 & -65.02 & -64.14 & 2\\[-1ex]
14 & 62 & 54.37 & -62.97 & -65.02 & -27.03 & 4 &
14 & 62 & 54.37 & -62.97 & -65.02 & -29.21 & 2\\[-1ex]
14 & 62 & 54.37 & -62.97 & -65.02 & -31.27 & 4 &
14 & 62 & 54.37 & -62.97 & -65.02 & -33.46 & 2\\[-1ex]
14 & 62 & 54.37 & 21.87 & -65.02 & -12.89 & 2 &
14 & 62 & 54.37 & 21.87 & -65.02 & -15.07 & 4\\[-1ex]
14 & 62 & 54.37 & 21.87 & -65.02 & -17.13 & 2 &
14 & 62 & 54.37 & 21.87 & -65.02 & -19.32 & 4\\[-1ex]
14 & 62 & 54.37 & 48.06 & -65.02 & -47.81 & 2 &
14 & 62 & 54.37 & 48.06 & -65.02 & -50 & 4\\[-1ex]
14 & 64 & 52.37 & -44.12 & -105.36 & -53.64 & 32 &
14 & 64 & 52.37 & -44.12 & -105.36 & -55.82 & 16\\[-1ex]
14 & 64 & 52.37 & -44.12 & -105.36 & -57.89 & 32 &
14 & 64 & 52.37 & -44.12 & -105.36 & -60.07 & 16\\[-1ex]
14 & 64 & 52.37 & -70.32 & -105.36 & -22.96 & 32 &
14 & 64 & 52.37 & -70.32 & -105.36 & -25.14 & 16\\[-1ex]
14 & 64 & 52.37 & 14.52 & -105.36 & -11 & 32 &
14 & 64 & 52.37 & 14.52 & -105.36 & -8.82 & 16\\[-1ex]
14 & 64 & 52.37 & 40.72 & -105.36 & -39.50 & 16 &
14 & 64 & 52.37 & 40.72 & -105.36 & -41.68 & 32\\[-1ex]
14 & 64 & 52.37 & 40.72 & -105.36 & -43.75 & 16 &
14 & 64 & 52.37 & 40.72 & -105.36 & -45.93 & 32\\[-1ex]
14 & 68 & 54.37 & -36.78 & -100.46 & -66.20 & 4 &
14 & 68 & 54.37 & -36.78 & -100.46 & -68.39 & 2\\[-1ex]
14 & 68 & 54.37 & -62.97 & -100.46 & -27.03 & 4 &
14 & 68 & 54.37 & -62.97 & -100.46 & -29.21 & 2\\[-1ex]
14 & 68 & 54.37 & -62.97 & -100.46 & -31.27 & 4 &
14 & 68 & 54.37 & -62.97 & -100.46 & -33.46 & 2\\[-1ex]
14 & 68 & 54.37 & 21.87 & -100.46 & -12.89 & 2 &
14 & 68 & 54.37 & 21.87 & -100.46 & -15.07 & 4\\[-1ex]
14 & 68 & 54.37 & 21.87 & -100.46 & -17.13 & 2 &
14 & 68 & 54.37 & 21.87 & -100.46 & -19.32 & 4\\[-1ex]
14 & 68 & 54.37 & 48.06 & -100.46 & -52.06 & 2 &
14 & 68 & 54.37 & 48.06 & -100.46 & -54.25 & 4\\[-1ex]
14 & 79 & 60.37 & -56.32 & -87.05 & -50.79 & 4 &
14 & 79 & 60.37 & -56.32 & -87.05 & -52.97 & 2\\[-1ex]
14 & 79 & 60.37 & 34.07 & -87.05 & -35.72 & 2 &
14 & 79 & 60.37 & 34.07 & -87.05 & -37.91 & 4\\[-1ex]
14 & 91 & 60.37 & -56.32 & -159.72 & -55.04 & 4 &
14 & 91 & 60.37 & -56.32 & -159.72 & -57.22 & 2\\[-1ex]
14 & 91 & 60.37 & 34.07 & -159.72 & -39.97 & 2 &
14 & 91 & 60.37 & 34.07 & -159.72 & -42.16 & 4\\[-1ex]
18 & 114 & 57.97 & -62.65 & -99.45 & -60.95 & 4 &
18 & 114 & 57.97 & -62.65 & -99.45 & -65.20 & 4\\[-1ex]
18 & 114 & 57.97 & -62.65 & -99.45 & -67.29 & 2 &
18 & 114 & 57.97 & -62.65 & -99.45 & -67.38 & 4\\[-1ex]
18 & 114 & 57.97 & -62.65 & -99.45 & -69.36 & 4 &
18 & 114 & 57.97 & 10.16 & -99.45 & -51 & 4\\[-1ex]
18 & 114 & 57.97 & 10.16 & -99.45 & -52.97 & 2 &
18 & 114 & 57.97 & 10.16 & -99.45 & -53.06 & 4\\[-1ex]
18 & 114 & 57.97 & 10.16 & -99.45 & -55.25 & 4 &
18 & 114 & 57.97 & 10.16 & -99.45 & -59.41 & 4\\[-1ex]
18 & 86 & 57.97 & -62.65 & -146.39 & -60.95 & 4 &
18 & 86 & 57.97 & -62.65 & -146.39 & -63.13 & 4\\[-1ex]
18 & 86 & 57.97 & -62.65 & -146.39 & -65.11 & 4 &
18 & 86 & 57.97 & -62.65 & -146.39 & -65.20 & 4\\[-1ex]
18 & 86 & 57.97 & -62.65 & -146.39 & -71.54 & 2 &
18 & 86 & 57.97 & 10.16 & -146.39 & -48.81 & 4\\[-1ex]
18 & 86 & 57.97 & 10.16 & -146.39 & -51 & 4 &
18 & 86 & 57.97 & 10.16 & -146.39 & -55.16 & 4\\[-1ex]
18 & 86 & 57.97 & 10.16 & -146.39 & -55.25 & 4 &
18 & 86 & 57.97 & 10.16 & -146.39 & -57.22 & 2\\[-1ex]
24 & 124 & 55.97 & -56.59 & -140.76 & -75.33 & 4 &
24 & 124 & 55.97 & -56.59 & -140.76 & -77.52 & 2\\[-1ex]
24 & 124 & 55.97 & 16.22 & -140.76 & -63.20 & 2 &
24 & 124 & 55.97 & 16.22 & -140.76 & -65.38 & 4\\[-1ex]
24 & 88 & 55.97 & -56.59 & -103.63 & -71.08 & 4 &
24 & 88 & 55.97 & -56.59 & -103.63 & -73.27 & 2\\[-1ex]
24 & 88 & 55.97 & 16.22 & -103.63 & -58.95 & 2 &
24 & 88 & 55.97 & 16.22 & -103.63 & -61.13 & 4\\\hline
\end{array}}{betathresh}{Classes of models with different values for one-loop gauge group running coefficients and threshold corrections on $T^6/\bZ_6'$ with $v=5$.}

\clearpage

%%%%%%%%%%%%%%%%%%%%%%%%%%%%%%%%%%%%%%%%%%%%%%%%%%%%%%%%%%%%%%%%
\section{Conclusions}\label{Sec:Concl}\label{S:Conclusions}

In this article, we completed the computation of gauge threshold due to massive
string modes for intersecting fractional D6-branes. To do so, we discussed the
full dependence on open string moduli in all amplitudes and derived the
amplitude for D6-branes at one vanishing angle along a direction with $\Z_2$
action but at angles on the reming two-torus in section~\ref{sec:Thresholds}.
The divergent parts of the amplitudes provided the one-loop beta function
coefficients due to massless strings from which we were able to distinguish
symplectic and orthogonal gauge groups in appendix~\ref{App:SO-Sp}.
It turned out that for all choices of K\"ahler moduli well in the geometric
regime and a given set of Wilson lines and displacements, the gauge thresholds
of the strong interactions were strengthening the gauge
couplings at the string scale for the explicit $T^6/\Z_6'$ models in section~\ref{Ss:Ex2} while weakening them
for the $T^6/\Z_6$ model in section~\ref{Ss:Ex1}. The same phenomenon occured for the electro-weak interactions
in the $T^6/\Z_6$ model and the $T^6/\Z_6'$ model with a hidden gauge factor $Sp(6)_{h_3}$, in the latter for 
not too large volume.

In contrast to the models considered in~\cite{Cremades:2003qj} where
all three quark generations occur at the same intersection, we find
models where different generations occur at intersections of various orbifold images, e.g. the right-handed down-quarks at intersections of $a$ with
brane $c$ and $(\theta^2 c)$.
Therefore the rank of the Yukawa matrix is not restricted to one and deserves further investigation.
Something similar occurs in heterotic orbifold
constructions~\cite{Lebedev:2008un}  where one generation of quarks
arises from the bulk and two from orbifold fix points.

We found in one of the examples in section~\ref{Ss:Ex2} that hidden sector
gauge groups can become strongly coupled, which opens up a window to
supersymmetry breaking via a gaugino condensate. However, in order to discuss
the phenomenology of this model in more detail, it will be necessary to compute
Yukawa couplings on fractional branes, in particular those where chiral matter states appear at 
vanishing angle on a torus with non-trivial $\Z_2$ action, 
and instanton corrections and derive mass terms for the
non-chiral matter states and hidden sector gauginos in the model. 

From a statistical point of view, we found that the number of physically distinct
D6-branes, at least at our one-loop examination, is per given toroidal cycle
by a factor of four smaller than in our previous analysis~\cite{Gmeiner:2008xq}. This enabled us to fully classify the 768 models
with three Standard Model generations by their gauge groups and full massless
matter content as well as distinct beta function coefficients. While at tree
level only three different classes arise, including the beta function augments
the number to ten different realizations of the Standard Model sector.
Including the one-loop threshold corrections we find that there are 196
different classes of models.
This complete survey of models is also interesting from a string landscape point
of view. Eventually the results obtained here could be compared to results in
other corners of the string landscape, as a first step to other relatively
simple toroidal orientifold models, such as those
in~\cite{Blumenhagen:2004xx,Gmeiner:2005vz,Gmeiner:2006qw}. Furthermore one
could use the results to look for correlations between the coupling behavior
and other properties of the models~\cite{Gmeiner:2007qd,Gmeiner:2008qq}.

Our analysis was focussed on numerical values of the gauge couplings. In order
to obtain the standard supergravity description in terms of the holomorphic gauge kinetic functions, the K\"ahler 
and superpotential and to determine further Yukawa couplings and mass terms, it will be necessary to work out the correct
field redefinitions which mix the dilaton and  complex structure moduli from
untwisted as well as orbifold twist sectors analogously to the case for bulk branes~\cite{Lust:2004cx,Bertolini:2005qh,
Akerblom:2007uc} and rigid branes~\cite{Blumenhagen:2007ip,Angelantonj:2009yj} in the $T^6/\Z_2 \times \Z_2$ backgrounds.

In~\cite{Abel:2003ue,Abel:2008ai} and in a more stringy context
in~\cite{Cassel:2009pu}, it was realized that the existence of anomalous Abelian
gauge factors can lead to kinetic mixing. It remains to be seen if the present
models can provide for low-energy signatures which can be detected at the LHC.

%%%%%%%%%%%%%%%%%%%%%%%%%%%%%%%%%%%%%%%%%%%%%%%%%%%%%%%%%%%%%%%%
\subsection*{Acknowledgements}
We would like to thank Ralph Blumenhagen, Stephan Stieberger and especially
Maximilian Schmidt-Sommerfeld for helpful discussions.

We acknowledge the kind hospitality of CERN, the GGI and the KITP where part
of this work was performed.

The work of F.~G. is supported by the Foundation for Fundamental Research of
Matter (FOM) and the National Organisation for Scientific Research (NWO).

The work of G.H. is supported in part by the FWO - Vlaanderen, project G.0235.05,
the Federal Office for Scientific, Technical and Cultural Affairs
through the Interuniversity Attraction Poles Programme, Belgian Science Policy
P6/11-P and the National Science Foundation under Grant No. NSF PHY05-51164.

%%%%%%%%%%%%%%%%%%%%%%%%%%%%%%%%%%%%%%%%%%%%%%%%%%%%%%%%%%%%%%%%

\clearpage
\appendix
\section{Computation of massless spectra and constraints of orientifold models}\label{App:Spectra-Constraints}

%%%%%%%%%%%%%%%%%%%%%%%%%%%
\subsection{The computation of \texorpdfstring{$\Z_2$}{Z2} invariant intersection numbers}\label{AppSs:Z2-numbers} 

In this appendix, we give some details on how the $\Z_2$ invariant intersection numbers $I^{\Z_2}_{ab}$ with relative signs introduced in~\eqref{Eq:Z2-Intersections} are 
obtained.  They can be written in the factorized form
\begin{equation}\label{ApEq:Z2-inters-def}
I_{ab}^{\Z_2} = (-1)^{\tau^0_a + \tau^0_b} \, I_{ab}^{\Z_2,(1)} \, I_{ab}^{(2)} \, I_{ab}^{\Z_2,(3)},
\end{equation}
where $I_{ab}^{\Z_2,(i)}$ denote the $\Z_2$ invariant points on $T^2_i$ dressed with the corresponding Wilson lines as detailed below,
and $I_{ab}^{(2)} $ is the toroidal intersection number on the two-torus where the $\Z_2$ action is trivial (remember $\vec{v}_{\Z_2}=\frac{1}{2}(1,0,-1)$ throughout this article).  It should be noted that while in~\eqref{ApEq:Z2-inters-def} the $\Z_2$ eigenvalues $\tau^0$ appear explicitly, we define the $\Z_2$ invariant intersection numbers $I_{ab}^{\Z_2,(i \cdot j)}$
on a four-torus which appear e.g. in table~\ref{tab:Annulus-Amplitudes-thresholds} if $I_{ab}^{(2)}=0$, 
such that these are included,
\begin{equation}\label{ApEq:Z213def}
I_{ab}^{\Z_2,(1 \cdot 3)} \equiv  (-1)^{\tau^0_a + \tau^0_b} \, I_{ab}^{\Z_2,(1)} \,  I_{ab}^{\Z_2,(3)}.
\end{equation}
The orientifold action exchanges the $\Z_2$ eigenvalues, $\tau^0_{b'} = \tau^0_b+1$,such that we get a factor of $(-1)^{\tau^0_a+\tau^0_b+1}$ in equations~\eqref{ApEq:Z2-inters-def} and~\eqref{ApEq:Z213def} for $I_{ab'}^{\Z_2}$.

As discussed in~\cite{Gmeiner:2007we,Gmeiner:2007zz}, in order to avoid a double counting of orbifold image cycles in a $T^6/\Z_6$ or $T^6/\Z_6'$ background, it suffices to consider only toroidal one-cycles of the type $(n_1,m_1)=$ (odd,odd). From figure~\ref{Fig:Z3-Z6lattice}, one reads off that a cycle which passes through the origin, i.e. $(\tilde{\sigma}^1,\tilde{\sigma}^2)=(0,0)$
or $\sigma^1=0$,\footnote{By $\tilde{\sigma}^i \in \{0,1/2\}$ we denote displacements along the one-cycle $\pi_i$ as in~\cite{Gmeiner:2008xq}. Throughout
the body of this article we use the shorter notation that $\sigma^i=0$ or $1$ if a one-cycle is passing 
through the origin on the $i^{th}$ two-torus $T^2_i$ or displaced from it by $\frac{1}{2}\pi_{2i-1}$  for $m_1$ odd or by $\frac{1}{2}\pi_{2i}$ for $m_1$ even, respectively.} passes through the $\Z_2$ invariant points (1,6). On the $\Z_6$ invariant {\bf A} torus, these points transform as follows under the orbifold and orientifold action
\begin{equation}
\begin{aligned}
&\!\!\!\!\! \left(\begin{array}{c} 1 \\ 6 \end{array}\right)
\stackrel{\Z_6}{\rightarrow}
\left(\begin{array}{c} 1 \\ 4 \end{array}\right)
\stackrel{\Z_6}{\rightarrow}
\left(\begin{array}{c} 1 \\ 5 \end{array}\right)
\\
\OR & \downarrow
\\
&\!\!\!\!\!  \left(\begin{array}{c} 1 \\ 5 \end{array}\right)
\stackrel{\Z_6}{\rightarrow}
\left(\begin{array}{c} 1 \\ 6 \end{array}\right)
\stackrel{\Z_6}{\rightarrow}
\left(\begin{array}{c} 1 \\ 4 \end{array}\right)
.
\end{aligned}
\end{equation}
If the cycle is displaced from the origin by $\frac{1}{2} \pi_{1}$, i.e. $(\tilde{\sigma}^1,\tilde{\sigma}^2)=(1/2,0)$
or $\sigma^1=1$,
it passes through the $\Z_2$ invariant points (4,5), and the complete $\Z_6$ and orientifold orbit is as follows,
\begin{equation}
\begin{aligned}
&\!\!\!\!\! \left(\begin{array}{c} 4 \\ 5  \end{array}\right)
\stackrel{\Z_6}{\rightarrow}
\left(\begin{array}{c} 5 \\ 6 \end{array}\right)
\stackrel{\Z_6}{\rightarrow}
\left(\begin{array}{c} 6 \\ 4 \end{array}\right)
\\
\OR & \downarrow
\\
&\!\!\!\!\!  \left(\begin{array}{c} 4 \\ 6 \end{array}\right)
\stackrel{\Z_6}{\rightarrow}
\left(\begin{array}{c} 5 \\ 4 \end{array}\right)
\stackrel{\Z_6}{\rightarrow}
\left(\begin{array}{c} 6 \\ 5  \end{array}\right)
.
\end{aligned}
\end{equation}
In both cases, the upper entry is assigned the prefactor 1 and the lower one the factor $(-1)^{\tau^1}$
with $\tau^1=0$ if there is no Wilson line and $\tau^1=1$ in case of a discrete Wilson line.
The counting of the $\Z_2$ invariant intersection numbers with signs from relative Wilson lines can be read off
by multiplying the corresponding prefactors if branes pass through the same $\Z_2$ invariant point.
The result is summarized in table~\ref{Tab:Z2-inv-numbers-on-T1}.
%%%%%%%%%%%%%%%%%%%%%%%%%%%%%%%%%%%%%%%%%%%%%%%%%
\begin{table}[ht]
  \begin{center}
    \begin{equation*}
      \begin{array}{|c||c|c|c||c|c|c|} \hline
                                % "Uberschrift
        \multicolumn{7}{|c|}{\rule[-3mm]{0mm}{8mm}
\text{\bf $\Z_2$ invariant intersection numbers $I_{a(\theta^k b)}^{\Z_2,(1)}$ on a $\Z_6$ invariant {\bf A}-type two-torus $T^2_1$}
}\\ \hline\hline
(\sigma^1_a,\sigma^1_b) & ab & a(\theta b) & a (\theta^2 b) &  ab' & a(\theta b') & a (\theta^2 b') 
\\\hline\hline
(0,0) & 1 + (-1)^{\tau^1_a + \tau^1_b} & 1 & 1 & 1 &  1 + (-1)^{\tau^1_a + \tau^1_b} & 1 
\\
(1,0) & \emptyset & (-1)^{\tau^1_b} & (-1)^{\tau^1_a + \tau^1_b} & (-1)^{\tau^1_a + \tau^1_b} & \emptyset & (-1)^{\tau^1_b}
\\
(0,1) & \emptyset & (-1)^{\tau^1_a + \tau^1_b} & (-1)^{\tau^1_a} & (-1)^{\tau^1_a + \tau^1_b} & \emptyset & (-1)^{\tau^1_a} 
\\
(1,1) &  1 + (-1)^{\tau^1_a + \tau^1_b} & (-1)^{\tau^1_a} & (-1)^{\tau^1_b} & 1 & (-1)^{\tau^1_a} + (-1)^{\tau^1_b} & (-1)^{\tau^1_a + \tau^1_b} 
\\ \hline
     \end{array}
    \end{equation*}
  \end{center}
\caption{Counting of $\Z_2$ invariant intersection numbers with relative Wilson lines for a two-torus with $\Z_6$ symmetry. 
If the wrapping numbers of the two one-cycles are of the same type, e.g. $(n_1,m_1)=$(odd,odd), the counting gives $\pm 2$ for no 
relative Wilson line or distance and a cancellation of the two contributions in the presence of a relative Wilson line. If cycles of the same type are displaced,
they do not intersect in any $\Z_2$ invariant intersection point.
If the one-cycle wrapping numbers $(n_1,m_1)$ of brane $a$ and $(\theta^k b)$ are of different type, the $\Z_2$ invariant intersection number is $\pm 1$. }
\label{Tab:Z2-inv-numbers-on-T1}
\end{table}
%%%%%%%%%%%%%%%%%%%%%%%%%%%

On a $\Z_2$ invariant {\bf a}-type lattice as in figure~\ref{Fig:Z2-Z4lattice}, all orbifold and orientifold images $(\theta^k a)$ and $(\theta^k a')$ of a one-cycle $a$ pass through
the same $\Z_2$ fixed points, see the list below.
\begin{equation}
\begin{array}{|c||c|c|}
\hline
{\bf a}-{\rm torus} & \multicolumn{2}{|c|}{\Z_2 \; {\rm fixed \,  points}}
\\
(n_3,m_3) & \sigma^3=0 & \sigma^3=1
\\\hline\hline
\text{(odd,even)} & \left(\begin{array}{c} 1 \\ 2  \end{array}\right)  & \left(\begin{array}{c}4  \\ 3  \end{array}\right) 
\\
\text{(odd,odd)} &  \left(\begin{array}{c} 1 \\ 3 \end{array}\right)  & \left(\begin{array}{c} 2 \\ 4 \end{array}\right) 
\\
\text{(even,odd)} &  \left(\begin{array}{c} 1 \\ 4 \end{array}\right)  & \left(\begin{array}{c} 2 \\ 3 \end{array}\right) 
\\\hline
\end{array}
\end{equation}
 %%%%%%%%%%%%%%%%%%%%%%%%%%%%%%%%%%%%%%%%%%%%%%%%%
 
 The assignment of 1 and $(-1)^{\tau^3}$ for the upper and lower entry is analogous to the discussion on $T^2_1$, and the resulting 
 combinatorics of the counting of $\Z_2$ invariant intersections on $T^2_3$ is given in table~\ref{Tab:Z2-inv-numbers-on-T3}.

 %%%%%%%%%%%%%%%%%%%%%%%%%%%%%%%%%%%%%%%%%%%%%%%%%
\begin{table}[ht]
  \begin{center}
    \begin{equation*}
      \begin{array}{|c||c|c|c|} \hline
                                % "Uberschrift
        \multicolumn{4}{|c|}{\rule[-3mm]{0mm}{8mm}
\text{\bf $\Z_2$ invariant intersection numbers $I_{a(\theta^k b)}^{\Z_2,(3)}$ on the {\bf a}-type two-torus $T^2_3$}
}\\ \hline\hline
(n_3^a,m_3^a)=\text{(odd,even)} & \multicolumn{3}{|c|}{(n_3^b,m_3^b)}
\\
(\sigma^3_a,\sigma^3_b) & \text{(odd,even)} & \text{(odd,odd)} & \text{(even,odd)}
\\\hline\hline
(0,0) & 1 + (-1)^{\tau^3_a + \tau^3_b} & 1 & 1
\\
(1,0) & \emptyset & (-1)^{\tau^3_a + \tau^3_b} & (-1)^{\tau^3_b} 
\\
(0,1) & \emptyset &  (-1)^{\tau^3_a} &  (-1)^{\tau^3_a}
\\
(1,1) &  1 + (-1)^{\tau^3_a + \tau^3_b} & (-1)^{\tau^3_b}  & (-1)^{\tau^3_a + \tau^3_b} 
\\ \hline
     \end{array}
    \end{equation*}
  \end{center}
\caption{Counting of $\Z_2$ invariant intersection numbers with relative Wilson lines for a rectangular two-torus with $\Z_2$ symmetry. 
As for the $\Z_6$ invariant torus in table~\ref{Tab:Z2-inv-numbers-on-T1}, the values are even (0 or $\pm 2$) if the wrapping numbers are of the same type and $\pm 1$ otherwise. 
The remaining cases with $m_3^a=$odd are completely analogous.}
\label{Tab:Z2-inv-numbers-on-T3}
\end{table}
%%%%%%%%%%%%%%%%%%%%%%%%%%%

%%%%%%%%%%%%%%%%%%%%%%%%%%%
\subsection{The complete massless open spectrum}\label{AppSs:Spectrum}

The complete massless open-string spectrum for intersecting D6-branes in a $T^6/\Z_{2N}$ background 
has been given in~\cite{Gmeiner:2008xq}. Since the multiplicities of charged matter states serve as a guideline that both the relative and absolute prefactors of the 
 threshold corrections have been computed accurately, we list the massless open string spectrum here  again in table~\ref{NonChiralSpectrum}.
%%%%%%%%%%%%%%%%%%%%%%%%%%%%%%%%%%%%%%%%%%%%%%%%%
\begin{table}[ht]
  \begin{center}
    \begin{equation*}
      \begin{array}{|c|c|} \hline
                                % "Uberschrift
        \multicolumn{2}{|c|}{\rule[-3mm]{0mm}{8mm}
\text{\bf Chiral and non-chiral massless matter on } T^6/(\bZ_{2N} \times \OR)  
}\\ \hline\hline
\text{rep.} & \text{total number} \;\; \varphi
\\\hline\hline
({\bf Adj}_a) & 1 +\frac{1}{4} \sum_{k=1}^{N-1} \left| I_{a(\theta^k a)} +I_{a(\
theta^k a)}^{\bZ_2}\right|
\\
({\bf Anti}_a) &  \frac{1}{4} \sum_{k=0}^{N-1}\left|I_{a(\theta^k a')}
    +I_{a(\theta^k a')}^{\bZ_2} + I_a^{\OR\theta^{-k}} + I_a^{\OR\theta^{-k+N} }
 \right|
\\
({\bf Sym}_a) &  \frac{1}{4} \sum_{k=0}^{N-1}\left|I_{a(\theta^k a')}
    +I_{a(\theta^k a')}^{\bZ_2} - I_a^{\OR\theta^{-k}} - I_a^{\OR\theta^{-k+N}} 
 \right|
\\
({\bf N}_a,\overline{\bf N}_b) & \frac{1}{2} \sum_{k=0}^{N-1}\left| I_{a(\theta^k b)} +I_{a(\theta^k b)}^{\bZ_2} \right|
\\
({\bf N}_a,{\bf N}_b) & \frac{1}{2}  \sum_{k=0}^{N-1}\left| I_{a(\theta^k b')} +I_{a(\theta^k b')}^{\bZ_2} \right|
\\ \hline
     \end{array}
    \end{equation*}
  \end{center}
\caption{Chiral plus non-chiral matter states $\varphi$ in $T^6/(\bZ_{2N} \times \OR)$ models with fractional intersecting D6-branes
for generic non-vanishing angles. 
For vanishing angles, the formulae are modified as discussed in the text.}
\label{NonChiralSpectrum}
\end{table}
%%%%%%%%%%%%%%%%%%%%%%%%%%%

Modifications occur if some angle $\phi^{(i)}$ vanishes, where here again the second two-torus
is chosen to be $\Z_2$ invariant, i.e. $\vec{v}_{\Z_2}=\frac{1}{2}(1,0,-1)$:
\begin{enumerate}
\item
\underline{$(\phi^{(1)},\phi^{(2)},\phi^{(3)})=(0,0,0)$:} D6-branes wrapping the same bulk cycle carry 
no matter if there is a relative Wilson line or distance. If both vanish, there is one chiral multiplet in the 
$\Adj_a$ of $U(N_a)$ for $N_a$ identical D6-branes
and $2 \times [ ({\bf N^1}_a,\ov{\bf N^2}_a) + \cc]$ for \mbox{$\tau_{a_1}^0 = \tau_{a_2}^0 +1 \; {\rm mod} \; 2$} (i.e. fractional D6-branes with opposite $\Z_2$ eigenvalues), 
or $ 2 \times [\Anti_a + \cc]$ for $a_2 =(\theta^k a_1')$ ($\OR$ invariant bulk cycles, and the $\Z_2$ eigenvalue is exchanged by the $\OR$ projection).
%%%%%%%%%%%%%%%%%%%%%%%%%%%%%%%%%%%%%
\item
\underline{$(\phi^{(1)},\phi^{(2)},\phi^{(3)})=(\phi,0,-\phi)$:} the D6-branes are parallel along the one direction where $\Z_2$ acts trivially. 
For orientifold images, one has to distinguish if the D6-branes are parallel to some O6-planes or perpendicular to them. In case of vanishing Wilson lines and distances, 
the massless spectrum is computed from
\begin{equation}
\begin{aligned}
\varphi^{ab,\parallel T^2_2}
&\rightarrow \phantom{\frac{1}{2}}
\left|  I_{a(\theta^k b)}^{(1 \cdot 3)} +I_{a(\theta^k b)}^{\bZ_2,(1 \cdot 3)} \right|  ,
\\
\varphi^{\Adj_a, \parallel  T^2_2}  
&\rightarrow  \frac{1}{2} \left| I_{a(\theta^k a)}^{(1 \cdot 3)} 
+I_{a(\theta^k a)}^{\bZ_2,(1 \cdot 3)}\right|  ,
\\
\left.\begin{array}{c}
\varphi^{\Anti_a,\parallel T^2_2\parallel \OR\theta^{-k(+N)}}  
 \\ \varphi^{\Sym_a,\parallel T^2_2\parallel \OR\theta^{-k(+N)}}  
\end{array} \right\}  
&  \rightarrow
\frac{1 }{2}\left| I_{a(\theta^k a')}^{(1\cdot 3)} + I_{a(\theta^k a')}^{\bZ_2,(1\cdot 3)}
 \pm \left( I_{a}^{\OR\theta^{-k},(1\cdot 3)} + I_{a}^{\OR\theta^{-k+N},(1\cdot
3)} \right)\right|,
\\
\left.\begin{array}{c} \varphi^{\Anti_a,\parallel T^2_2\perp \OR\theta^{-k(+N)}} 
\\ \varphi^{\Sym_a,\parallel T^2_2\perp \OR\theta^{-k(+N)}} 
\end{array} \right\} 
& \rightarrow
\frac{1 }{2}\left| I_{a(\theta^k a')}^{(1\cdot 3)} + I_{a(\theta^k a')}^{\bZ_2,(1\cdot 3)}\right|.
\end{aligned}
\end{equation}
If the D6-branes are continuously displaced from each other or a relative Wilson line is switched on on $T^2_2$, the corresponding strings acquire a mass according to~\eqref{Eq:Open-Massformula}.\\
The open strings stretching between D6-brane $a$ and its orbifold image $(\theta a)$ in a $T^6/\Z_4$ background are of this type.
%%%%%%%%%%%%%%%%%%%%%%%%%%%%%%%%%
\item
\underline{$(\phi^{(1)},\phi^{(2)},\phi^{(3)})=(\phi,-\phi,0)$:} the D6-branes are parallel along some direction where $\Z_2$ acts non-trivially,
\begin{equation}
\begin{aligned}
\varphi^{ab,\parallel T^2_3} 
&\rightarrow  
\delta_{\sigma^3_a , \sigma^3_b} \, \delta_{\tau^3_a , \tau^3_b} \, 
\left| I_{a(\theta^k b)}^{(1 \cdot 2)}   \right| 
,
\\[1ex]
\varphi^{\Adj_a, \parallel T^2_3} 
&\rightarrow  \frac{1}{2}\left| I_{a(\theta^k a)}^{(1 \cdot 2)}   \right|
\quad {\rm for }\quad k \neq 0,
\\
\left.\begin{array}{c} \varphi^{\Anti_a,\parallel T^2_3} 
\\ \varphi^{\Sym_a,\parallel T^2_3} 
\end{array} \right\} 
&\rightarrow
\frac{1 }{2}\left| I_{a(\theta^k a')}^{(1 \cdot 2)} \pm
I_a^{\OR\theta^{-y}(1 \cdot 2)}  \right|
\\ 
&\text{\hspace{-5em}with the exponent} \quad  y=
\left\{\begin{array}{cc} 
k & a \parallel \OR\theta^{-k} \; {\rm on} \; T^2_3
\\
k+N & a \parallel \OR\theta^{-k+N} \; {\rm on} \; T^2_3
\end{array}\right.
.
\end{aligned}
\end{equation}
The parameters of discrete displacements and Wilson lines $\sigma^i$ and $\tau^i$ take values in $\{0,1\}$.
The case $(\phi^{(1)},\phi^{(2)},\phi^{(3)})=(0,\phi,-\phi)$ is completely analogous.\\
The open strings stretching from D6-brane $a$ to some orbifold image $(\theta^k a)_{k=1,2}$ in the $T^6/\Z_6'$ background
discussed in section~\ref{Ss:Ex2} are of this type.\\
\end{enumerate}

%%%%%%%%%%%%%%%%%%%%%%%%%%%%%%%%%%%%%%%%%%%%%%
\subsection{Determination of \texorpdfstring{$SO(2M)$ or $Sp(2M)$}{SO(2M) or Sp(2M)} gauge groups and their matter states}\label{App:SO-Sp}

In a $T^6/(\Z_{2N} \times \OR)$ background, there are two different kinds of D6-branes which are invariant under the orientifold projection:
either their toroidal part is parallel to two O6-planes $\OR\theta^{-k}$ and  $\OR\theta^{-k+N}$ along the two-torus where the
$\Z_2$ action is trivial, or their toroidal part is perpendicular to these two O6-planes on the same two-torus. In this appendix, we will determine the 
correct assignment of antisymmetric and symmetric representations by comparison with the general formula for beta function coefficients
in the fourth column of table~\ref{tab:Amplitudes-thresholds}. 

In ${\cal N}=1$ supersymmetric field theory, the vector multiplet in the adjoint representation $\Adj$ of some gauge group $G$ 
and chiral multiplets in various representations ${\bf R}_i$ of $G$ contribute to the one-loop beta function coefficient,
\begin{equation} 
b_G = - 3 \, C_2(\Adj) + \sum_i C_2({\bf R}_i) ,
\end{equation}
with the quadratic Casimir operators $C_2$. For an $SU(N)$ gauge group, in~\eqref{Eq:beta-SU(N)} we used the normalization for the quadratic Casimir
of the fundamental representation $C_2(\N)=\frac{1}{2}$ and the relations
\begin{equation}
\begin{aligned}
  C_2(\Adj) &= 2 N \, C_2(\N),
\\
  C_2(\Anti,/\Sym) &= (N \mp 2) \, C_2(\N).
\end{aligned}
\end{equation}

In order to compare the derivation of the beta function coefficients for $SO(2M)$ or $Sp(2M)$ gauge groups with the formulae in 
table~\ref{tab:Amplitudes-thresholds} involving intersection numbers only, the analogous relations for these groups are needed,
\begin{equation}
\begin{aligned}
C_2(\Anti_{SO/Sp(2M)}) &= (2 \, M-2)  \, C_2({\bf 2M}),
\\
C_2(\Sym_{SO/Sp(2M)}) &= (2\, M+2 )  \, C_2({\bf 2M})  .
\end{aligned}
\end{equation}

In the $T^6/\Z_6$ example in section~\ref{Ss:Ex1}, branes $x \in \{c,e\}$ are parallel to the $\OR\theta^{-4}$ plane, and the contributions to the 
beta function coefficients are 
\begin{equation}
\begin{aligned}
b_{SO/Sp(2M_x)} \supset &
\frac{1}{2}\Bigl[\left(-M_x \, I_{xx}^{\Z_2,(1 \cdot 2)} + \tilde{I}^{\OR\theta^{-1},(1 \cdot 2)}_x  \right)
\\
&
\quad+\sum_{k=1}^2 (-1)^k \left( M_x \frac{I_{x(\theta^k x)} + I_{x(\theta^k x)}^{\Z_2} }{2} -\frac{\tilde{I}_x^{\OR\theta^{-1-k}} + \tilde{I}_x^{\OR\theta^{-4-k}} }{2} \right)\Bigr]
\\
&= -  \left( 2 \, M_x + 2   \right)
+ 2 \cdot 0
,
\end{aligned}
\end{equation}
where the over-all factor $\frac{1}{2}$ arises from the fact that in table~\ref{tab:Amplitudes-thresholds}, we are implicitly summing over D6-branes and their orientifold images, but for the orientifold invariant branes $x$ no such sum appears.

The first contribution from parallel D6-branes is 
consistent with the vector and one chiral multiplet in the symmetric representation, i.e. an $Sp(2M_x)$ gauge group and one chiral multiplet in the adjoint representation,
\begin{equation}
\begin{aligned}
  -3\, C_2(\Adj_{Sp(2M_x)}) + C_2(\Adj_{Sp(2M_x)}) &= -2 \, C_2(\Adj_{Sp(2M_x)})\\
& = -2 \, (2 \, M_x +2 )  \, C_2({\bf 2M}_x),
\end{aligned}
\end{equation}
with the normalization $C_2({\bf 2M}_x)=\frac{1}{2}$.
The remaining contributions to the spectrum from the $x(\theta x) + x(\theta^2 x)$ sectors vanish since we computed in table~24 of~\cite{Gmeiner:2008xq} that the massless states are 
$\Z_2$ even, but the D6-branes $x$ are displaced on $T^2_2$ from the origin and carry a discrete Wilson line there. The resulting minus sign in the 
projection of the Chan-Paton label is responsible for the non-existence of the states at intersections of orbifold images of $x$.

In the $T^6/\Z_6'$ examples in section~\ref{Ss:Ex2}, D6-brane $c$ is parallel to the $\OR\theta^{-4}$ plane and the hidden brane $\hat{h}_1$ is parallel
to the $\OR\theta^{-1}$ plane resulting in the beta function coefficients according to table~\ref{tab:Amplitudes-thresholds}
\begin{equation}
\begin{aligned}
b_{SO/Sp(2M_c)} \supset & \frac{1}{2} \left[
\left(-M_c \, I_{cc}^{\Z_2,(1 \cdot 3)} + \tilde{I}^{\OR\theta^{-1},(1 \cdot 3)}_c  \right)
+\sum_{k=1}^2 \left(-M_c I_{c(\theta^k c)}^{(1 \cdot 2)} + \tilde{I}^{\OR\theta^{-4 +2k },(1 \cdot 2)}_c  \right)\right]
%+\left(  \right)
\\
&= -  \left( 2 \, M_c + 2   \right)
+  2 \cdot \frac{3 \,( M_c -1)}{2} ,
\\
b_{SO/Sp(2M_{\hat{h}_1})} \supset & \frac{1}{2} \left[
\left(-M_{\hat{h}_1} \, I_{\hat{h}_1\hat{h}_1}^{\Z_2,(1 \cdot 3)} + \tilde{I}^{\OR\theta^{-4},(1 \cdot 3)}_{\hat{h}_1}  \right)
+\sum_{k=1}^2 \left(-M_{\hat{h}_1} I_{\hat{h}_1(\theta^k \hat{h}_1)}^{(1\cdot 2)} + \tilde{I}_{\hat{h}_1}^{\OR\theta^{-1+2k},(1\cdot2)}  \right)
\right]
\\
&= -  \left(2 \, M_{\hat{h}_1} + 2   \right)
+  2 \cdot \frac{9 \, (M_{\hat{h}_1} -1 )}{2}
.
\end{aligned}
\end{equation}
For both D6-branes $y \in \{c,\hat{h}_1\}$, the first term indicates again an $Sp(2M_y)$ gauge group with one chiral multiplet in the adjoint representation living on parallel D6-branes, and the $y(\theta y) + y(\theta^2 y)$ sector provides three multiplets in the antisymmetric representation for $y=c$ 
and nine antisymmetrics for $y=\hat{h}_1$.

The stack $h_3$ of D6-branes belongs to the second class of $SO/Sp(2M)_{h_3}$ branes since it is orthogonal to the $\OR\theta^{-1}$ and $\OR\theta^{-4}$
planes along the $\Z_2$ invariant two-torus $T_2^2$. In this case, the beta function coefficient is computed from
\begin{equation}
\begin{aligned}
b_{SO/Sp(2M)_{h_3}} \supset & \frac{1}{2} \Bigl[
\left(- M_{h_3} I_{h_3h_3}^{\Z_2,(1 \cdot 3)} + \tilde{I}^{\OR\theta^{-1},(1 \cdot 2)}_{h_3} + \tilde{I}^{\OR\theta^{-4},(2 \cdot 3)}_{h_3}  \right)
\\
&+\sum_{k=1}^2 \left(-M_{h_3} I_{h_3(\theta^k h_3)}^{(1\cdot 2)} + \tilde{I}_{h_3}^{\OR\theta^{-1+2k},(1\cdot2)}  \right)
\Bigr]
\\
&= -  \left(2\, M_{h_3} + 4 \right)
+2 \cdot \frac{M_{h_3} - 1}{2}
.
\end{aligned}
\end{equation}
The first term is consistent with the vector in the symmetric and a chiral multiplet in the antisymmetric representation,
\begin{equation}
  - 3 \, C_2(\Adj_{Sp(2M_{h_3})}) +C_2(\Anti_{Sp(2M_{h_3})}) = -2 \,(2 M_{h_3} + 4 )C_2({\bf 2M}_{h_3}),
\end{equation}
with the normalization $C_2({\bf 2M}_{h_3})=\frac{1}{2}$.
The $h_3(\theta h_3) + h_3(\theta^2 h_3)$ sectors provide one more chiral multiplet in the 
antisymmetric representation.

%%%%%%%%%%%%%%%%%%%%%%%%%%%

\subsection{Reformulations of the RR tadpole conditions}\label{AppSs:RRtcc-rewrite}

In section~\ref{sec:notation}, we argued that several identities involving the intersection numbers $I_{ab}^{(i)}$ and 
generalizations of the (square of the) one-cycle volumes $V_{ab}^{(i)}$ hold. In this appendix, we give some evidence why these generalizations
work out correctly.

The starting point is the toroidal part. A toroidal three-cycle is fully specified by its six one-cycle wrapping numbers
\begin{equation}
\vec{X}^{\rm torus} = \left(\begin{array}{c} \vec{x}_1 \\ \vec{x}_2 \\ \vec{x}_3 \end{array}\right)
\qquad
{\rm with}
\qquad
\vec{x}_i %&
= \left(\begin{array}{c} n_i \\ m_i \end{array}\right)
,
%\end{aligned}
\end{equation}
but toroidal three-cycles in this representation are {\it not} added using the usual vector addition. 
We define the following multiplicative actions on one-cycles $\vec{x}_i$ (the first three with determinant +1, the last with -1),
\begin{equation}
\begin{aligned}
\mathcal{I} &=\left(\begin{array}{cc} 0 & 1 \\ -1 & 0  \end{array}\right)
,
\qquad
 \mathcal{V} =\left\{\begin{array}{cc}
\left(\begin{array}{cc}  \frac{R_1}{R_2} + b^2 \frac{R_2}{R_1} &  b \, \frac{R_2}{R_1} \\ b\, \frac{R_2}{R_1} & \frac{R_2}{R_1}  \end{array}\right) & {\bf a}, {\bf b}
\\
\frac{1}{\sqrt{3}} \, \left(\begin{array}{cc} 2 & 1 \\ 1 & 2  \end{array}\right) & {\bf A},  {\bf B}
\end{array}\right.
,
\\
\Theta_i  &=\left\{\begin{array}{cc}
\left(\begin{array}{cc} 0 & -1 \\ 1 & 1    \end{array}\right) & \Z_6
\\
\left(\begin{array}{cc} -1 & -1 \\ 1 & 0   \end{array}\right) & \Z_3
\\
\left(\begin{array}{cc}  -1 & 0 \\ 0 & -1  \end{array}\right) & \Z_2
\end{array}\right.
,
\qquad
\mathcal{R} =\left\{\begin{array}{cc}
\left(\begin{array}{cc} 1 & 0  \\ -2b & -1    \end{array}\right) & {\bf a}, {\bf b}
\\
\left(\begin{array}{cc}  1 & 1  \\    0 & -1 \end{array}\right) & {\bf A}
\\
\left(\begin{array}{cc}  0 & 1 \\ 1 & 0    \end{array}\right) & {\bf B}
\end{array}\right.
,
\end{aligned}
\end{equation}
where the intersection matrix $\mathcal{I}$ is independent of the lattice, $\mathcal{V}$ differs for $\Z_4$ and $\Z_6$ invariant lattices, 
$\mathcal{R}$ depends also on the orientation of each lattice and $\Theta_i$ encodes the various orbifold rotations.
These matrices give the relations
\begin{equation}
(\vec{x}^a_i)^T \mathcal{I} \vec{x}^b_i  = I_{ab}^{(i)}
,
\qquad
 (\vec{x}^a_i)^T \mathcal{V} \vec{x}^b_i  = V_{ab}^{(i)}
,
\qquad
\mathcal{R}   \vec{x}^b_i  = \vec{x}^{b'}_i 
, 
\qquad
\Theta_i^k \vec{x}^b_i = \vec{x}^{(\theta^k b)}_i
,
\end{equation}
where matrices and vectors are multiplied in the usual way.

Actions on three-cycles are described by the above components as follows,
\begin{equation}\label{Eq:Def-3-Cycle-Actions}
\begin{aligned}
\vec{\Theta} &= {\rm diag} \left(\Theta_1,\Theta_2,\Theta_3   \right),
\\
\vec{\mathcal{I}} &=  {\rm diag} \left(\mathcal{I},\mathcal{I},\mathcal{I}   \right),
\\
\vec{\mathcal{V}} &= {\rm diag} \left(\mathcal{V},\mathcal{V},\mathcal{V}   \right),
\\
\left.\begin{array}{c}
\vec{\mathcal{J}}_1 = {\rm diag} \left(\mathcal{V}, \mathcal{I},\mathcal{I}  \right),
\\
\vec{\mathcal{J}}_2 = {\rm diag} \left( \mathcal{I},\mathcal{V},\mathcal{I}   \right),
\\
\vec{\mathcal{J}}_3 = {\rm diag} \left(  \mathcal{I},\mathcal{I},\mathcal{V}  \right),
\end{array}\right\}\quad \Rightarrow \quad 
\vec{\mathcal{K}} &= \sum_{i=1}^3 \vec{\mathcal{J}}_i ,
\end{aligned}
\end{equation}
where by $\Theta_i$ we denote the fact that the orbifold action on the various tori might be different, e.g. 
in the $T^6/\Z_6'$ case, the shift vector $\vec{v}=(\frac{1}{6},\frac{1}{3},-\frac{1}{2})$ corresponds to $\vec{\Theta}={\rm diag}(\Z_6,\Z_3,\Z_2)$.

An orbifold invariant bulk cycle for $T^6/\Z_{2N}$ is (up to normalization) of the form
\begin{equation}
\vec{X}^{\rm bulk}_a = \tilde{\oplus}_{k=0}^{N-1} \vec{\Theta}^k \vec{X}^{\rm torus}_a
,
\qquad
\vec{X}^{\rm bulk}_{\OR} = \tilde{\oplus}_{k=0}^{2N-1}  \vec{X}^{\rm torus}_{\OR\theta^{-k}}
,
\end{equation}
with an unspecified prescription $\tilde{\oplus}$ for adding three-cycles.

Even though the explicit form of adding three-cycles is not known, the toroidal part of the anomaly cancellation condition~\eqref{Eq:no-Anomaly} can be written in this formalism as 
\begin{equation}\label{Eq:RR-tcc-Rewriting-Presc}
\begin{aligned}
0 &= \left( \vec{X}_a^{\rm bulk} \right)^T \, \vec{\mathcal{I}} \, \underbrace{ \left( \tilde{\oplus}_b N_b \left(  \vec{X}_b^{\rm bulk}  
\tilde{\oplus} \vec{X}_{b'}^{\rm bulk} \right) \tilde{\oplus} (-4)\vec{X}^{\rm bulk}_{\OR} 
  \right) }
\\
& \qquad\qquad\qquad\quad =0 \text{due to bulk RR tadpole condition}
\\
&=\sum_b N_b \left( \left( \vec{X}_a^{\rm bulk}\right)^T  \, \vec{\mathcal{I}}\vec{X}_b^{\rm bulk}  
+ \left( \vec{X}_a^{\rm bulk}\right)^T  \, \vec{\mathcal{I}}\vec{X}_{b'}^{\rm bulk} \right) - 4 \, \left( \vec{X}_a^{\rm bulk}\right)^T  \, \vec{\mathcal{I}}\vec{X}^{\rm bulk}_{\OR} 
,
\end{aligned}
\end{equation}
where in the last line ``$\pm$" are the usual operations of adding and subtracting numbers.
In this formulation it is natural to replace the matrix operation $\vec{\mathcal{I}}$ by any other matrix valued
transformation and still get vanishing results. The first equality in~\eqref{Eq:RRtcc-rewritten} was obtained by exchanging
$\vec{\mathcal{I}}$  in~\eqref{Eq:RR-tcc-Rewriting-Presc} with $\vec{\mathcal{K}}$ as defined in~\eqref{Eq:Def-3-Cycle-Actions}. We checked explicitly that these equations indeed hold for the examples in section~\ref{S:Examples} and the statistical ensemble in section~\ref{S:Statistics}.

The other rewritten form of the bulk RR tadpole cancellation condition in~\eqref{Eq:RRtcc-2nd-rewrite} is obtained by using the matrix $\vec{\mathcal{V}}$ defined in~\eqref{Eq:Def-3-Cycle-Actions} instead of $\vec{\mathcal{I}}$  in~\eqref{Eq:RR-tcc-Rewriting-Presc}, and the vanishing tadpole for a massless $U(1)$ factor is obtained by rewriting the bulk part of the identity~\eqref{Eq:U(1)X-no-tcc} in the present notation,
\begin{equation}
\begin{aligned}
0 &= \left( \vec{X}_X^{\rm bulk} \right)^T \vec{\mathcal{I}} \underbrace{\left( \vec{X}_{X'}^{\rm bulk} \tilde{\oplus} (-1) \vec{X}_X^{\rm bulk} \right)}
\\
&\qquad\qquad\qquad\qquad =0 \text{ for a massless } U(1)_X
\\
&= \left( \vec{X}_X^{\rm bulk} \right)^T \vec{\mathcal{I}}\vec{X}_{X'}^{\rm bulk} 
-\left( \vec{X}_X^{\rm bulk} \right)^T \vec{\mathcal{I}} \vec{X}_X^{\rm bulk},
\end{aligned}
\end{equation}
and replacing $\vec{\mathcal{I}}$ by $\vec{\mathcal{K}}$.

The above formalism can be extended to fractional D6-branes by defining similar vectors for the exceptional three-cycles,
\begin{equation}
\vec{X}^{\rm ex}_a = \left(\begin{array}{c}
y_{(ij)}^a \\ \vec{x}_2^a
\end{array}\right),
\qquad
\vec{X}^{\rm ex}_{\OR\theta^{-k}} = \left(\vec{0} \right),
\end{equation}
where the index $(ij)$ runs over the 16 $\Z_2$ fixed points on $T^2_1 \times T^2_3$, and four $y_{(ij)}^a = \pm 1$ 
for those $\Z_2$ fixed points hit by the toroidal cycle with wrapping numbers $(n_i^a,m_i^a)$, and zero for the remaining twelve. The orbifold symmetries $\Theta_i$ and orientifold involution $\mathcal{R}$ act as permutations on these 16 entries (plus an over-all sign for $\cal{R}$)
and as before on the toroidal one-cycle given by $\vec{x}_2^a$. The intersection form 
on the 16 fixed point entries is just the unit matrix $\unity_{16}$. 
The above discussion for the bulk cycles carries over directly to the exceptional cycles leading to the second equation in~\eqref{Eq:RRtcc-rewritten}.

%%%%%%%%%%%%%%%%%%%%%%%%%%%

\section{Threshold computations}\label{AppS:Thresholds}

In this appendix, we collect Jacobi theta function identities, check prefactors of tree channel amplitudes contributing to RR tadpole cancellation
on fractional D6-branes and perform the integration of Jacobi theta function derivatives needed in the gauge threshold computation.

\subsection{Jacobi theta function identities}\label{AppSs:Thetas}

The Jacobi theta functions with characteristics and the Dedekind eta function are defined by ($q \equiv e^{2 \pi i \tau}$)
\begin{equation}
\begin{aligned}
\vartheta\targ{\alpha}{\beta}(\nu,\tau)
&=\sum_{n \in \Z} q^{ \frac{(n+\alpha)^2 }{ 2}} \,  e^{2 \pi i (n+\alpha)(\nu + \beta)}
,
\\
\eta(\tau) &= q^{\frac{1}{24}} \prod_{n=1}^{\infty} \left(1-q^n \right)
.
\end{aligned}
\end{equation}
For $\alpha \in (-\frac{1}{2},\frac{1}{2}]$, the following product expansion holds,
\begin{equation}\label{Eq:App-ThetaProduct}
\frac{\vartheta\targ{\alpha}{\beta}(\nu,\tau)}{\eta (\tau)}
=e^{2\pi i \alpha ( \nu + \beta)} \, q^{\frac{\alpha^2}{2} - \frac{1}{24}}
\prod_{n=1}^{\infty}
\left(1+ e^{2 \pi i (\nu + \beta)} \, q^{n+\alpha -\frac{1}{2}} \right)
\left(1+ e^{-2 \pi i (\nu + \beta)} \, q^{n-\alpha -\frac{1}{2}} \right)
.
\end{equation}
The standard Jacobi theta functions are defined as
\begin{equation}
\begin{aligned}
&\vartheta_1 (\nu,\tau) 
= - \vartheta\targ{1/2}{1/2}(\nu,\tau), 
\qquad
\vartheta_2 (\nu,\tau) =\vartheta\targ{1/2}{0}(\nu,\tau),
\\
& \vartheta_3 (\nu,\tau) =\vartheta\targ{0}{0}(\nu,\tau),
\qquad
\vartheta_4 (\nu,\tau) =\vartheta\targ{0}{1/2}(\nu,\tau),
\end{aligned}
\end{equation}
with special relations at $\nu=0$ for the first Jacobi theta function,
\begin{equation}
\vartheta_1 ( 0,\tau) = 0 ,
\qquad
\partial_{\nu} \vartheta_1 ( 0,\tau) \equiv\vartheta_1^{\prime} ( 0,\tau)  = 2 \pi \, \eta^3 (\tau),
\qquad
\partial_{\nu}^2 \vartheta_1 ( 0,\tau) \equiv\vartheta_1^{\prime\prime} ( 0,\tau)  = 0.
\end{equation}
For the conversion from loop to tree channel, the following modular transformation properties are useful,
\begin{equation}\label{Eq:App-Theta-modular}
\begin{aligned}
\vartheta_1 (\nu,\tau) &= i \;  \frac{e^{-i \pi \nu^2 / \tau}}{\sqrt{-i\tau}}  
\; \vartheta_1 (\frac{\nu}{\tau},-\frac{1}{\tau}),
\qquad
\vartheta_3 (\nu,\tau) = \frac{e^{-i \pi \nu^2 / \tau}}{\sqrt{-i\tau}}  
\; \vartheta_3 (\frac{\nu}{\tau},-\frac{1}{\tau}),
\\
\vartheta_2 (\nu,\tau) &= \frac{e^{-i \pi \nu^2 / \tau}}{\sqrt{-i\tau}}  
\; \vartheta_4 (\frac{\nu}{\tau},-\frac{1}{\tau}),
\qquad
\vartheta_4 (\nu,\tau) = \frac{e^{-i \pi \nu^2 / \tau}}{\sqrt{-i\tau}}  
\; \vartheta_2 (\frac{\nu}{\tau},-\frac{1}{\tau}),
\\
\eta(\tau) &= \frac{1}{\sqrt{-i\tau}} \;  \eta(-\frac{1}{\tau}),
\end{aligned}
\end{equation}
and when evaluating the lattice contributions for arbitrary distances and Wilson lines, the mirror symmetry 
relations
\begin{equation}
\overline{\vartheta_i (\nu,\tau)} =\vartheta_i (\ov{\nu},-\ov{\tau}) 
\qquad
i=1\ldots 4,
\qquad\qquad
\ov{\eta(\tau) }= \eta(-\ov{\tau}) ,
\end{equation}
are needed.

The computation of gauge threshold corrections for supersymmetric D6-brane configurations, i.e.
$\sum_{i=1}^3 \phi^{(i)}=0$ with $|\phi^{(i)}| <1$, requires the following theta function identities,
\begin{enumerate}
\item 
for parallel D6-branes ($\phi^{(i)}=0$ for all $i$)
\begin{equation}
\begin{aligned}
&
\frac{\vartheta_3''(0,\tau)\left( \vartheta_3(0,\tau)\right)^3 -\vartheta_2''(0,\tau)
\left( \vartheta_2(0,\tau)\right)^3 -\vartheta_4''(0,\tau)\left( \vartheta_4(0,\tau)\right)^3  }{\eta^{12}(\tau)} =0,
\\
&\frac{\vartheta_3''(0,\tau) \vartheta_3(0,\tau) \left(\vartheta_2(0,\tau)\right)^2 
-\vartheta_2''(0,\tau) \vartheta_2(0,\tau) \left(\vartheta_3(0,\tau)\right)^2  }
{\eta^6(\tau)\left( \vartheta_4(0,\tau)\right)^2}
=4 \pi^2,
\end{aligned}
\end{equation}
%%%%%%%%%%%%%%%%
\item
for D6-branes at angles along a four-torus, e.g. $(\phi^{(1)},\phi^{(2)},\phi^{(3)})=(\underline{0,\phi,-\phi})$,
\begin{equation}
\begin{aligned}
&\vartheta_3''(0,\tau) \, \vartheta_3(0,\tau) \, 
\vartheta_3(\phi,\tau) \, \vartheta_3(-\phi,\tau)
-\vartheta_2''(0,\tau) \, \vartheta_2(0,\tau) \, 
\vartheta_2(\phi,\tau) \, \vartheta_2(-\phi,\tau)\\
&-\vartheta_4''(0,\tau) \, \vartheta_4(0,\tau) \, 
\vartheta_4(\phi,\tau) \, \vartheta_4(-\phi,\tau) 
\\
=\;&4 \pi^2 \,
\Bigl( \eta^6(\tau) \, 
\vartheta_1 (\phi,\tau) \,  \vartheta_1 (-\phi,\tau) \Bigr)
,
\\[3ex]
&\vartheta_3''(0,\tau) \, \vartheta_2(0,\tau) \,  \vartheta_3(\phi,\tau)  \vartheta_2(-\phi,\tau)
-\vartheta_2''(0,\tau) \, \vartheta_3(0,\tau) \,  \vartheta_2(\phi,\tau)  \vartheta_3(-\phi,\tau)
\\
=\;&2 \pi \left(  
\frac{\vartheta_1'}{\vartheta_1}(\phi,\tau)
+\frac{\vartheta_4'}{\vartheta_4}(-\phi,\tau)
 \right)\,
\Bigl(\eta^3(\tau) \,
\vartheta_4(0,\tau) \, \vartheta_1(\phi,\tau) \,\vartheta_4(-\phi,\tau) \Bigr)
,
\\[3ex]
&\vartheta_3''(0,\tau) \, \vartheta_3(0,\tau) \, 
\vartheta_2(\phi,\tau) \, \vartheta_2(-\phi,\tau)
-\vartheta_2''(0,\tau) \, \vartheta_2(0,\tau) \, 
\vartheta_3(\phi,\tau) \, \vartheta_3(-\phi,\tau)\\
&-\vartheta_4''(0,\tau) \, \vartheta_4(0,\tau) \, 
\vartheta_1(\phi,\tau) \, \vartheta_1(-\phi,\tau) 
\\
=\;&4 \pi^2 \,
\Bigl( \eta^6(\tau) \, 
\vartheta_4 (\phi,\tau) \,  \vartheta_4 (-\phi,\tau) \Bigr)
,
\end{aligned}
\end{equation}
%%%%%%%%%%%%%%%%
\item 
for D6-branes at angles along the whole six-torus
\begin{equation}
\begin{aligned}
& \vartheta_3''(0,\tau)\left(\prod_{i=1}^3\vartheta_3(\phi^{(i)},\tau)  \right) \\
& - \vartheta_2''(0,\tau) \left(\prod_{i=1}^3\vartheta_2(\phi^{(i)},\tau)  \right) \\
& - \vartheta_4''(0,\tau) \left(\prod_{i=1}^3\vartheta_4(\phi^{(i)},\tau)  \right) 
\\
\stackrel{ \sum_i \phi^{(i)}=0
}{=}\; & 2 \, \pi
\left(\sum_{i=1}^3 \frac{\vartheta_1^{\prime}}{\vartheta_1}(\phi^{(i)},\tau)\right)
 \left( \eta^3(\tau) \, 
\prod_{i=1}^3 \vartheta_1(\phi^{(i)},\tau) \right) 
,
\\[3ex]
&
\vartheta_3''(0,\tau) \vartheta_3(\phi^{(2)},\tau)\left(\prod_{i=1,3} \vartheta_2(\phi^{(i)},\tau) \right)\\
& - \vartheta_2''(0,\tau) \vartheta_2(\phi^{(2)},\tau)\left(\prod_{i=1,3} \vartheta_3(\phi^{(i)},\tau) \right)\\
& - \vartheta_4''(0,\tau) \vartheta_4(\phi^{(2)},\tau) \left(\prod_{i=1,3} \vartheta_1(\phi^{(i)},\tau) \right)
\\
\stackrel{\sum_i \phi^{(i)}=0
}{=}\; &2 \, \pi
\left(
\frac{\vartheta_1^{\prime}}{\vartheta_1}(\phi^{(2)},\tau) + 
\sum_{i=1,3} \frac{\vartheta_4^{\prime}}{\vartheta_4}(\phi^{(i)},\tau)
\right)
 \left( \eta^3(\tau)\, 
\vartheta_1(\phi^{(2)},\tau) \,\prod_{i=1,3}\vartheta_4(\phi^{(i)},\tau) \right)
,
\end{aligned}
\end{equation}
%%%%%%%%%%%%%%%%
\end{enumerate}
%%%%%%%%%%%%%%
where again primes denote derivatives with respect to the first argument.
The first identity for each case belongs to the purely toroidal contribution, the second (and third) one is required 
for the $\Z_2$ twisted part. 

The resulting ratios of Jacobi theta functions and their derivatives  can be expanded as follows,
\begin{equation}\label{Eq:App-ThetaPrime}
\begin{aligned}
\frac{\vartheta_1^{\prime}}{\vartheta_1}(\nu,\tau) &=\pi \cot (\pi \nu) + 4 \pi \, \sum_{n,k=1}^{\infty} \sin (2\pi \nu k) \, q^{nk},
\\
\frac{\vartheta_4^{\prime}}{\vartheta_4}(\nu,\tau) &=4 \pi \, \sum_{n,k=1}^{\infty} \sin (2\pi \nu k) \, q^{(n-\frac{1}{2})k}.
\end{aligned}
\end{equation}

The vacuum RR tadpole cancellation conditions are obtained using the asymptotics
\begin{equation}
\frac{\vartheta_2(0,\tau)}{\eta^3(\tau)} 
\stackrel{\tau \rightarrow i \infty}{\longrightarrow} 2, 
\qquad
\frac{\vartheta_3}{\vartheta_4}(\nu,\tau)
\stackrel{\tau \rightarrow i \infty}{\longrightarrow} 1,
\qquad
\frac{\vartheta_2}{\vartheta_1}(\nu,\tau)
\stackrel{\tau \rightarrow i \infty}{\longrightarrow} \cot (\pi \nu).
\end{equation}

In detail, the tree channel oscillator contributions in the RR sector and their asymptotics are listed in Tables~\ref{tab:tadosc1} and~\ref{tab:tadosc2}
in dependence of the relative angles and the loop channel insertion which becomes a twist sector in the tree channel. 
For parallel D6-branes, some lattice contribution has to be added. These contributions are identical
to~\eqref{Eq:Tree-Lattice-Sum} for the annulus and~\eqref{Eq:lattice-M-general} for the M\"obius strip. Their asymptotics is 1, as can be seen from the expansion~\eqref{Eq:Tree-Lattice-Sum-2}.
The last columns of Tables~\ref{tab:tadosc1} and~\ref{tab:tadosc2} demonstrate that upon taking into account also the RR tadpole contributions
of the closed string tree level Klein bottle amplitude, equation~\eqref{Eq:RRtcc-2nd-rewrite} is obtained.
\mathtabfix{
\begin{array}{|c|c|c||c|c||c|}\hline
\multicolumn{6}{|c|}{\text{\bf Annulus}}
\\\hline
\frac{\rm Angle}{\pi} &  \rotatebox{90}{\!\!\!{\rm insertion}} & \begin{array}{c} \pi^2 \times \\ {\rm prefactor} \end{array}
& \begin{array}{c}{\rm RR-sector} \\ {\rm oscillators} \end{array}  & {\rm asymptotics}
 &   \rotatebox{90}{\!\!\!$\begin{array}{c} \pi^2 \times \\ {\rm prefactor} \; \times\\ {\rm asymptotics}\end{array}$}
\\\hline
(0,0,0) & \unity 
& -\frac{N_a N_b}{32} \prod_{i=1}^3 V_{ab}^{(i)}
& \left(\frac{\vartheta_2(0,\tau)}{\eta^3(\tau)}   \right)^4 & 16
& -\frac{ N_a N_b \prod_{i=1}^3 V_{ab}^{(i)}}{2}
\\
& \Z_2 
&  -\frac{I_{ab}^{\Z_2,(1 \cdot 3)} N_a N_b V_{ab}^{(2)}}{8}
& \left(\frac{\vartheta_2(0,\tau)}{\eta^3(\tau)} \frac{\vartheta_3}{\vartheta_4}(0,\tau) \right)^2 & 4
&  -\frac{I_{ab}^{\Z_2,(1 \cdot 3)} N_a N_b V_{ab}^{(2)}}{2}
\\\hline
(\phi,-\phi,0) & \unity 
&-\frac{I_{ab}^{\Z_2,(1 \cdot 3)} N_a N_b V_{ab}^{(2)}}{8}
&  \left(\frac{\vartheta_2(0,\tau)}{\eta^3(\tau)} \right)^2 \frac{\vartheta_2}{-\vartheta_1}(\phi,\tau)
\frac{\vartheta_2}{-\vartheta_1}(-\phi,\tau) & 4 \cot(\pi \phi)  \cot(-\pi \phi) 
&-\frac{N_a N_b \prod_{i=1}^3 V_{ab}^{(i)}}{2}
\\
(\phi,0,-\phi) & \Z_2 
& -\frac{I_{ab}^{\Z_2,(1 \cdot 3)} N_a N_b V_{ab}^{(2)}}{8}
&  \left(\frac{\vartheta_2(0,\tau)}{\eta^3(\tau)} \right)^2 
\frac{\vartheta_3}{\vartheta_4}(\phi,\tau) \frac{\vartheta_3}{\vartheta_4}(-\phi,\tau) & 4 
&-\frac{I_{ab}^{\Z_2,(1 \cdot 3)} N_a N_b V_{ab}^{(2)}}{2}
\\
(0,\phi,-\phi) & \Z_2 
& \frac{N_aN_bI_{ab}^{\Z_2}}{4} 
& \frac{\vartheta_2(0,\tau)}{\eta^3(\tau)} \frac{\vartheta_3}{\vartheta_4}(0,\tau) 
\frac{\vartheta_2}{-\vartheta_1}(\phi,\tau) \frac{\vartheta_3}{\vartheta_4}(-\phi,\tau) & -2 \cot (\pi\phi)
& -\frac{N_aN_bI_{ab}^{\Z_2,(1 \cdot 3)}  V_{ab}^{(2)}}{2}
\\\hline
(\phi^1,\phi^2,\phi^3) & \unity 
& \frac{N_aN_bI_{ab}}{4}
& \frac{\vartheta_2(0,\tau)}{\eta^3(\tau)} 
\prod_{i=1}^3 \frac{\vartheta_2}{-\vartheta_1}(\phi^i,\tau)
& - 2 \prod_{i=1}^3 \cot (\pi\phi^i)
& -\frac{N_aN_b\prod_{i=1}^3 V_{ab}^{(i)}}{2}
\\
& \Z_2 
& \frac{N_aN_bI_{ab}^{\Z_2}}{4}
&\frac{\vartheta_2(0,\tau)}{\eta^3(\tau)} \frac{\vartheta_2}{-\vartheta_1}(\phi^2,\tau)
\prod_{i=1,3}\frac{\vartheta_3}{\vartheta_4}(\phi^i,\tau)
& -2 \cot (\pi \phi^2)
& -\frac{N_aN_bI_{ab}^{\Z_2,(1 \cdot 3)} V_{ab}^{(2)}}{2}
\\\hline
\end{array}
}{tadosc1}{Annulus contributions to the RR tadpole cancellation conditions. `insertion' in the second column refers to the loop channel interpretation and 
the prefactors in the third column are the same as in table~\protect\ref{tab:Annulus-Amplitudes-thresholds} except for the factor $N_a$ which is absent for the threshold correction to a $SU(N_a)$ gauge coupling. The $\Z_2$ action leaves the second torus invariant, $\vec{v}_{\Z_2}=\frac{1}{2}
(1,0,-1)$.
}
%%%%%%%%%%%%%%%%%%%%%%%%%%%%%%%%%%%%%%%%%%%%%%%%%%%

For the annulus we have $\tau=2il$. The M\"obius strip contributions are obtained from the amplitudes with $\unity$ insertion by setting
$\tau=2il-\frac{1}{2}$. In addition there is a factor of -4 coming from the charge of the orientifold planes.

%%%%%%%%%%%%%%%%%%%%%%%%%%%%%%%%%%%%%%%%%%%%%%%%%%%
\mathtabfix{
\begin{array}{|c|c|c||c|c||c|}\hline
\multicolumn{6}{|c|}{\text{\bf M\"obius  strip}}
\\\hline
\frac{\rm Angle}{\pi} &  \rotatebox{90}{\!\!\!{\rm insertion}} & \begin{array}{c} \pi^2 \times \\ {\rm prefactor} \end{array}
& \begin{array}{c}{\rm RR-sector} \\{\rm oscillators}\end{array} & {\rm asymptotics}
 &   \rotatebox{90}{\!\!\!$\begin{array}{c}\pi^2 \times \\ {\rm prefactor} \; \times\\ {\rm asymptotics}\end{array}$}
\\\hline
(0,0,0) & \!\!\!\OR\theta^{-k}\!\!\! 
& \frac{N_a \prod_{i=1}^3 \tilde{V}_{a,\OR\theta^{-k}}^{(i)}  }{8}
& \left(\frac{\vartheta_2(0,\tau)}{\eta^3(\tau)}   \right)^4 & 16
& 2 \, N_a \prod_{i=1}^3 \tilde{V}_{a,\OR\theta^{-k}}^{(i)}  
\\\hline
(\phi,-\phi,0) &  \!\!\!\OR\theta^{-k}\!\!\! 
& \frac{N_a \tilde{I}_a^{\OR\theta^{-k},(1 \cdot 2)} \tilde{V}_{a,\OR\theta^{-k}}^{(3)}}{2}  
& \left(\frac{\vartheta_2(0,\tau)}{\eta^3(\tau)} \right)^2 \frac{\vartheta_2}{-\vartheta_1}(\phi,\tau)
\frac{\vartheta_2}{-\vartheta_1}(-\phi,\tau) & 4 \cot(\pi \phi)  \cot(-\pi \phi) 
& 2 \, N_a \, \prod_{i=1}^3 \tilde{V}_{a,\OR\theta^{-k}}^{(i)}  
\\\hline
(\phi^1,\phi^2,\phi^3) & \!\!\!\OR\theta^{-k}\!\!\!
& - N_a \tilde{I}_a^{\OR\theta^{-k}}
& \frac{\vartheta_2(0,\tau)}{\eta^3(\tau)} 
\prod_{i=1}^3 \frac{\vartheta_2}{-\vartheta_1}(\phi^i,\tau)
& - 2 \prod_{i=1}^3 \cot (\pi\phi^i)
& 2\, N_a \, \prod_{i=1}^3 \tilde{V}_{a,\OR\theta^{-k}}^{(i)}  
\\\hline
\end{array}
}{tadosc2}{M\"obius strip contributions to the RR tadpole cancellation conditions. The prefactors in the third column are the same as in table~\protect\ref{tab:Mobius-Amplitudes-thresholds}.}
%%%%%%%%%%%%%%%%%%%%%%%%%%%%%%%%%%%%%%%%%%%%%%%%%%%

%%%%%%%%%%%%%%%%%%%%%%%%%%%%%%%%%%%%%%%%%%%%%%%%%%%%%%%%%%%%%%%

\subsection{Technical details of threshold computation: integrals of Jacobi theta functions}\label{App:Technical}

%%%%%%%%%%%%%%%%%%%%%%%%%%%%%%%%%%%%%%%%%%%%%%%%%%%%%%%%%%%
%\subsubsection{Integration of Jacobi theta functions}\label{AppSs:Integration-Jacobi}

The integral over lattice contributions has been performed in detail in section~\ref{Ss:Lattice}. 
Since our choice of regularization differs slightly from the one used in~\cite{Akerblom:2007np}, and the intermediate steps in the computation of the relevant integrals
over Jacobi theta functions and their derivative are different, we briefly present them here.

The integrals over oscillator contributions are performed as follows. For the {\it annulus} and $i=1,4$, 
\begin{equation}\label{Eq:App-Theta-Integrals}
\begin{aligned}
\int\limits_0^{\infty} dl l^{\varepsilon}  \frac{\vartheta_i^{\prime}}{\vartheta_i}(\nu,2il ) &=\pi \cot (\pi \nu) \, \delta_{i1} 
\int\limits_0^{\infty} dl l^{\varepsilon}\\
&\quad  + 4 \pi \, \sum_{n,k=1}^{\infty} \sin (2\pi \nu k) \, 
\int\limits_0^{\infty} dl l^{\varepsilon}
e^{-4\pi l (n-\frac{\delta_{i4}}{2} )k}
\\
&=\pi \cot (\pi \nu) \, \delta_{i1} 
\int\limits_0^{\infty} dl l^{\varepsilon}
\\
& \quad
+\frac{1}{(4 \pi)^{\varepsilon}} \, \sum_{k=1}^{\infty} \frac{{\rm sin} (2 \pi \nu \, k)}{k^{1+\varepsilon}}
\sum_{n=1}^{\infty}
\frac{1}{(n-\frac{\delta_{i4}}{2})^{1+\varepsilon}}
\int\limits_0^{\infty} dx x^{\varepsilon}e^{-x}
%%%%%
\\
&=
\pi \cot (\pi \nu) \, \delta_{i1} 
\int\limits_0^{\infty} dl l^{\varepsilon}
\\
& \quad
+\frac{1}{(4 \pi)^{\varepsilon}} \, \sum_{k=1}^{\infty} \frac{{\rm sin} (2 \pi \nu \, k)}{k^{1+\varepsilon}}
\left[\delta_{i1} + (2^{1+\varepsilon}-1)\, \delta_{i4} \right] \zeta(1+\varepsilon)
\Gamma(1+\varepsilon)
\\
&= \pi \cot (\pi \nu) \, \delta_{i1} 
\int\limits_0^{\infty} dl 
\\
& \quad +
\frac{1}{\varepsilon}  \sum_{k=1}^{\infty} \frac{{\rm sin} (2 \pi \nu \, k)}{k} 
- \sum_{k=1}^{\infty} \frac{{\rm sin} (2 \pi \nu \, k) \, {\rm ln}(4 \pi k)}{k} 
\\
& \quad
+
2\, {\rm ln} (2) \, \delta_{i4}  \sum_{k=1}^{\infty} \frac{{\rm sin} (2 \pi \nu \, k)}{k} 
+ {\cal O}(\varepsilon)
,
\end{aligned}
\end{equation}
with the Gamma and Riemann zeta function
\begin{equation}\label{Eq:App-Gamma-Zeta}
\begin{aligned}
\Gamma(1+\varepsilon) & \equiv \int\limits_0^{\infty} x^\varepsilon e^{-x} dx
=  1-\gamma \varepsilon + \left( \frac{\pi^2}{12}+\frac{\gamma^2}{2} \right)  \varepsilon^2 + {\cal O}(\varepsilon^3) 
,
\\
\zeta(1+\varepsilon) & \equiv \sum_{k=1}^{\infty} \frac{1}{k^{1+\varepsilon}} 
= \frac{1}{\varepsilon} +\gamma -\gamma_1 \varepsilon + \frac{\gamma_2}{2}\varepsilon^2 + {\cal O}(\varepsilon^3) 
,
\end{aligned}
\end{equation}
and $\gamma \simeq 0.5772$ the Euler Mascheroni constant. 

Up to this point, all expressions are periodic in $\nu$ mod 1. In the range $ 0 < |\nu| < 1$, 
the sums in~\eqref{Eq:App-Theta-Integrals} are Fourier sine expansions,
\begin{equation}
\begin{aligned}
 \sum_{k=1}^{\infty} \frac{{\rm sin} (2 \pi \nu \, k)}{k} 
&= \pi \left( \frac{{\rm sgn}(\nu)}{2} - \nu \right)
,
%%%%%%%%%%%%
\\
\sum_{k=1}^{\infty} \frac{{\rm sin} (2 \pi \nu \, k) \, {\rm ln}(4 \pi k)}{k} 
&=\pi \left(\frac{1}{2} {\rm ln} \left(\frac{\Gamma(|\nu|)}{\Gamma(1-|\nu|)}\right)^{{\rm sgn}(\nu)} -[ {\rm ln} 2 -\gamma][\nu-\frac{{\rm sgn}(\nu)}{2}] \right)
,
\end{aligned}
\end{equation}
leading to ($0< |\nu| <1$)
\begin{equation}\label{Eq:App-Theta-Final-Integrals}
\begin{aligned}
\frac{1}{\pi} \int\limits_0^{\infty} dl l^{\varepsilon}  \frac{\vartheta_i^{\prime}}{\vartheta_i}(\nu,2il ) 
&=\cot (\pi \nu) \, \delta_{i1}  \int\limits_0^{\infty} dl 
+ \left( \frac{1}{\varepsilon} +\gamma - \ln 2  \right) \left( \frac{{\rm sgn}(\nu)}{2} -\nu \right)
\\
&\quad-\frac{1}{2} \ln \left(\frac{\Gamma(|\nu|)}{\Gamma(1-|\nu|)}\right)^{{\rm sgn}(\nu)} 
+ \delta_{i4} \ln 2 \left({\rm sgn} (\nu) - 2 \, \nu  \right)
.
\end{aligned}
\end{equation}
In order to extract the one-loop running due to massless strings, the dimensional regularization has to agree with the
one for the lattice contributions in section~\ref{Ss:Lattice}, i.e.
\begin{equation}
\ln \left( \frac{M_{\rm string}}{\mu}  \right)^2  \equiv \frac{1}{\varepsilon} +\gamma - \ln 2.
\end{equation}
The first line in~\eqref{Eq:App-Theta-Final-Integrals} consists now of a tadpole term and the contribution to the beta function coefficient
from massless string states, whereas the second line provides the contribution of the massive strings to the gauge thresholds.

The {\it M\"obius strip} has as the second argument in the Jacobi theta functions $2il-\frac{1}{2}$ leading to
\begin{equation}\label{Eq:App-Theta-Integrals-MS}
\begin{aligned}
\int\limits_0^{\infty} dl l^{\varepsilon}  \frac{\vartheta_1^{\prime}}{\vartheta_1}(\nu,2il-\frac{1}{2} ) &=\pi \cot (\pi \nu) \, \
\int\limits_0^{\infty} dl l^{\varepsilon}
\\
&\quad  + 4 \pi \, \sum_{n,k=1}^{\infty} (-1)^{kn} \, \sin (2\pi \nu k) \, 
\int\limits_0^{\infty} dl l^{\varepsilon}
e^{-4\pi l nk}
\\
&=\pi \cot (\pi \nu) \int\limits_0^{\infty} dl l^{\varepsilon}
\\
&\quad  + 
\frac{1}{(4 \pi)^{\varepsilon}} \, \sum_{k=1}^{\infty}  \frac{{\rm sin} (2 \pi \nu \, k)}{k^{1+\varepsilon}}
\sum_{n=1}^{\infty}(-1)^{kn} 
\frac{1}{n^{1+\varepsilon}}
\Gamma(1+\varepsilon)
\\
&= 
\pi \cot (\pi \nu) \int\limits_0^{\infty} dl
\\
&\quad + \frac{1}{2 \, \varepsilon}  \,\left[
 \sum_{k=1}^{\infty} \frac{{\rm sin} (2 \pi \nu \, k)}{k}
%%%%%
+ \sum_{k=1}^{\infty} \, (-1)^k \,  \frac{{\rm sin} (2 \pi \nu \, k)}{k} \right]
\\
&\quad  - \frac{\ln(2)}{2} \,\left[ \sum_{k=1}^{\infty} \frac{{\rm sin} (2 \pi \nu \, k)}{k}
%%%%
- \sum_{k=1}^{\infty}\, (-1)^k \,  \frac{{\rm sin} (2 \pi \nu \, k)}{k}\right] 
\\
&\quad - \frac{1}{2} \, \,\left[ 
 \sum_{k=1}^{\infty} \frac{{\rm sin} (2 \pi \nu \, k)  \ln(4\pi k)}{k}
%%%%%
+ \sum_{k=1}^{\infty} \, (-1)^k \,  \frac{{\rm sin} (2 \pi \nu \, k) \ln(4\pi k)}{k} 
\right]
\\
&\quad+{\mathcal O}(\varepsilon)
.
\end{aligned}
\end{equation}
In order to rewrite this expression, two more Fourier sine expansions are needed,
again valid in the range $0 < |\nu| <1$,
\begin{equation}
\begin{aligned}
 \sum_{k=1}^{\infty}  (-1)^k \, \frac{{\rm sin} (2 \pi \nu \, k)}{k} 
&= -\pi \,  \nu  + \pi \, {\rm sgn}(\nu) \, H \left(|\nu|-\frac{1}{2} \right)
,
%%%%%%%%%%%%
\\
\sum_{k=1}^{\infty} (-1)^k \, \frac{{\rm sin} (2 \pi \nu \, k) \, {\rm ln}(4 \pi k)}{k} 
&=\frac{\pi}{2} \, {\rm ln} \, \frac{\Gamma(\nu +\frac{1}{2} -\sgn(\nu) H(|\nu|-\frac{1}{2}))}{\Gamma(\frac{1}{2}-
\nu +\sgn (\nu) H(|\nu|-\frac{1}{2}))}
\\
&
-\pi [ {\rm ln} 2 -\gamma]%
\left[ \nu - {\rm sgn}(\nu) \, H \left(|\nu|-\frac{1}{2} \right) \right]
,
\end{aligned}
\end{equation}
where the Heavyside step function 
\begin{equation}\label{EqApp:Heavyside}
H (x) =\left\{\begin{array}{cc}
1 & 0 < x
\\
\frac{1}{2} & x=0
\\
0 & x < 0
\end{array}\right.
\end{equation}
has been introduced.

For the range $0<|\nu|<1$, the M\"obius strip oscillator contribution is thus of the form 
\begin{equation}
\begin{aligned}
\frac{1}{\pi} \,\int_0^{\infty} dl l^{\varepsilon}  \frac{\vartheta_1^{\prime}}{\vartheta_1}(\nu,2il-\frac{1}{2} ) 
&= \cot (\pi \nu) \, \int_0^{\infty} dl\\
&\quad +  \left( \frac{1}{\varepsilon} +\gamma - \ln 2  \right) \left( \frac{{\rm sgn}(\nu)}{4} \left[ 1 + 2 \, H(|\nu|-\frac{1}{2})
\right]-\nu \right)
\\
&\quad - \frac{\ln 2}{4} \, \sgn(\nu) \left[ 1 - 2 \, H(|\nu|-\frac{1}{2})\right]
\\
&\quad -\frac{1}{4} \, \ln \left(\frac{\Gamma(|\nu|)}{\Gamma(1-|\nu|)}\right)^{{\rm sgn}(\nu)} 
\\
&\quad
-\frac{1}{4} \, \ln \, \frac{\Gamma(\nu+\frac{1}{2} -\sgn(\nu)  H(|\nu|-\frac{1}{2}))}{\Gamma(-\nu+\frac{1}{2} +\sgn(\nu)  H(|\nu|-\frac{1}{2}))}
+{\mathcal O}(\varepsilon)
.
\end{aligned}
\end{equation}
The first line contains the tadpole, the second one the beta function coefficient and the remaining three lines the contribution to the gauge thresholds.

%%%%%%%%%%%%%%%%%%%%%%%%%%%%%%%%%%%%%%%%%%%%%%
\section{Tables of gauge threshold computations}\label{AppS:Tables}

In this appendix, we collect the computation of the annulus and M\"obius strip diagrams which contribute to the gauge threshold. 
The case for one vanishing angle is explained in detail in sections~\ref{Ss:ZeroAngle} and~\ref{Ss:MS-ZeroAngle}.

%%%%%%%%%%%%%%%%%%%%%%%%%%%%%%%%%%%%%%%%%%%%%%%%%%%%%%%%%%%
\renewcommand{\arraystretch}{1.3}
\mathsidetabfix{
\begin{array}{|c|c|c|c|c|} \hline
                                % "Uberschrift
        \multicolumn{5}{|c|}{\rule[-3mm]{0mm}{8mm}
\text{\bf Annulus amplitudes for gauge thresholds of } SU(N_a)}
\\\hline\hline
\pi(\phi_{ab}^{(1)},\phi_{ab}^{(2)},\phi_{ab}^{(3)}) & \rotatebox{90}{\!\!\!{\rm insertion}} 
& \begin{array}{c} \pi^2 \times \\ {\rm prefactor} \end{array}
& A^{\rm insertion}_{\rm compact}(\alpha,\beta;\{\phi^{(i)}\};2il) 
& \sum\limits_{(\alpha,\beta)} (-1)^{2(\alpha+\beta)}
\frac{\vartheta^{\prime\prime} \targ{\alpha}{\beta}
(0,2il)}{\eta^3(2il)} \frac{A^{\rm insertion}_{\rm compact} (\{\phi^{(i)}\};2il)}{\pi^2}\Big|_{\sum\limits_i \phi_i=0}
\\\hline\hline
(0,0,0) & \unity & -\frac{N_b \left(\prod\limits_{i=1}^3 V_{ab}^{(i)} \right)}{32} & 
\left(\prod\limits_{i=1}^3 \tilde{\mathcal L}_{ab}^{(i)} (l)\right) 
\left(\frac{\vartheta\targ{\alpha}{\beta} (0,2il)}{\eta^3 (2il)}
\right)^3
& 0 
\\\hline
& \Z_2 &  -\frac{I_{ab}^{\Z_2,(1 \cdot 3)} N_b V_{ab}^{(2)}}{8}  & \tilde{\mathcal L}_{ab}^{(2)} (l)
\frac{\vartheta\targ{\alpha}{\beta} (0,2il)}{\eta^3 (2il)}
\left(\frac{\vartheta\targ{\alpha + 1/2}{\beta}}
{\vartheta\targ{0}{1/2}}(0,2il)
\right)^2 
& 4 \, \tilde{\mathcal L}_{ab}^{(2)} (l)
\\\hline\hline
\pi (\phi,-\phi,0) & \unity & -\frac{ N_b I_{ab}^{(1 \cdot 2)}V_{ab}^{(3)}}{8}
&\tilde{\mathcal L}_{ab}^{(3)} (l) \frac{\vartheta\targ{\alpha}{\beta}(0,2il)}
{\eta^3(2il)}
\frac{\vartheta\targ{\alpha}{\beta}}
{\vartheta\targ{1/2}{1/2}} (\phi,2il)
\frac{\vartheta\targ{\alpha}{\beta}}
{\vartheta\targ{1/2}{1/2}}(-\phi,2il)
& 4 \, \tilde{\mathcal L}_{ab}^{(3)} (l)
\\\hline
\pi (\phi,0,-\phi) & \Z_2 &  -\frac{N_b  I_{ab}^{\Z_2,(1\cdot 3)}V_{ab}^{(2)}}{8} 
& \tilde{\mathcal L}_{ab}^{(2)} (l)
\frac{\vartheta\targ{\alpha}{\beta} (0,2il)}{\eta^3 (2il)}
\frac{\vartheta\targ{\alpha + 1/2}{\beta}}
{\vartheta\targ{0}{1/2}}(\phi,2il)
\frac{\vartheta\targ{\alpha + 1/2}{\beta}}
{\vartheta\targ{0}{1/2}}(-\phi,2il)
& 4 \, \tilde{\mathcal L}_{ab}^{(2)} (l) 
\\\hline
\pi (0,\phi,-\phi) & \Z_2 &  \frac{N_b  I_{ab}^{\Z_2} }{4}
&\frac{\vartheta\targ{\alpha + 1/2}{\beta}}
{\vartheta\targ{0}{1/2}}(0,2il)
\frac{\vartheta\targ{\alpha}{\beta}}
{\vartheta\targ{1/2}{1/2}} (\phi,2il)
\frac{\vartheta\targ{\alpha + 1/2}{\beta}}
{\vartheta\targ{0}{1/2}}(-\phi,2il)
& -\frac{2}{\pi}\left(\frac{\vartheta_1'}{\vartheta_1}(\phi,2il)
+\frac{\vartheta_4'}{\vartheta_4}(-\phi,2il)\right)
\\\hline\hline
\pi (\phi^{(1)},\phi^{(2)},-\sum_{k=1}^2\phi^{(k)}) & \unity &  \frac{N_b I_{ab}}{4}
& \prod\limits_{i=1}^3\frac{\vartheta\targ{\alpha}{\beta}}
{\vartheta\targ{1/2}{1/2}} (\phi^{(i)},2il)
&-\frac{2}{\pi} \sum\limits_{i=1}^3 \frac{\vartheta_1'}{\vartheta_1}(\phi^{(i)},2il)
\\\hline
& \Z_2 & \frac{N_b I_{ab}^{\Z_2}}{4}
& \frac{\vartheta\targ{\alpha}{\beta}}
{\vartheta\targ{1/2}{1/2}} (\phi^{(2)},2il)
\prod\limits_{i=1,3}\frac{\vartheta\targ{\alpha+1/2}{\beta}}
{\vartheta\targ{0}{1/2}} (\phi^{(i)},2il)
& -\frac{2}{\pi}\left( 
\frac{\vartheta_1'}{\vartheta_1}(\phi^{(2)},2il)
+\sum\limits_{i=1,3} \frac{\vartheta_4'}{\vartheta_4}(\phi^{(i)},2il)
\right)
%%%%%%%%%%%%%%%%%%%%%%%%%%%%%%
\\\hline
\end{array}
}{Annulus-Amplitudes-thresholds}{The annulus contributions to the gauge threshold calculations. In the first column, the relative 
angles are listed, in the second the loop channel insertion, in the third the combinatorial prefactor consisting of intersection numbers
and additional factors arising in the transformation from loop to tree channel. These prefactors and the compact contributions listed in the fourth
column are the same for the vacuum RR tadpole calculation in table~\protect\ref{tab:tadosc1} except for the missing factor $N_a$. Finally, in the last column the sum over spin structures for all, compact and non-compact, oscillator contributions is performed. }
%%%%%%%%%%%%%%%%%%%%%%%%%%%%%%%%%%%%%%%%%%%%%%%%%

%%%%%%%%%%%%%%%%%%%%%%%%%%%%%%%%%%%%%%%%%%%%%%%%%%%%%%%%%%%
\mathsidetabfix{
\begin{array}{|c|c|c|c|c|} \hline
                                % "Uberschrift
        \multicolumn{5}{|c|}{\rule[-3mm]{0mm}{8mm}
\text{\bf M\"obius amplitudes for gauge thresholds of } SU(N_a)}
\\\hline\hline
\!\!\pi(\phi^{(1)}_{a,\OR\theta^{-k}},\phi^{(2)}_{a,\OR\theta^{-k}},\phi^{(3)}_{a,\OR\theta^{-k}})\!\
! 
&\rotatebox{90}{\!\!\!{\rm insertion}}  & \begin{array}{c} \pi^2 \times \\   {\rm prefactor} \end{array}
& M^{\rm insertion}_{\rm compact}(\alpha,\beta;\{\phi^{(i)}\};2il-\frac{1}{2})  
& \sum\limits_{(\alpha,\beta)}   (-1)^{2(\alpha+\beta)}
 \frac{\vartheta^{\prime\prime}\targ{\alpha}{\beta}
(0,2il-\frac{1}{2})}{\eta^3(2il-\frac{1}{2})}  \frac{M^{\rm insertion}_{\rm compact} (\{\phi^{(i)}\}
;2il-\frac{1}{2})}{\pi^2}\Big|_{\sum\limits_i \phi_i=0}
\\\hline
(0,0,0) & \OR\theta^{-k} & \frac{ \prod\limits_{i=1}^3 \tilde{V}_{a,\OR\theta^{-k}}^{(i)} }{8}
& \left(\prod\limits_{i=1}^3 \tilde{\mathcal L}_{a,\OR\theta^{-k}}^{(i)} (l) \right) 
\left(\frac{\vartheta\targ{\alpha}{\beta} (0,2il-\frac{1}{2})}{\eta^3 (2il-\frac{1}{2})}\right)^3
& 0
\\\hline
\pi(\phi,-\phi,0)  & \OR\theta^{-k} 
& \frac{ \tilde{I}_a^{\OR\theta^{-k},(1\cdot 2)} \tilde{V}_{a,\OR\theta^{-k}}^{(3)} }{2} &  
\begin{array}{l}
\tilde{\mathcal L}_{a,\OR\theta^{-k}}^{(3)} (l) \frac{\vartheta\targ{\alpha}{\beta}(0,2il-\frac{1}{2})}{\eta^3(2il-\frac{1}{2})}\\
\frac{\vartheta\targ{\alpha}{\beta}}
{\vartheta\targ{1/2}{1/2}} (\phi,2il-\frac{1}{2})
\frac{\vartheta\targ{\alpha}{\beta}}
{\vartheta\targ{1/2}{1/2}}(-\phi,2il-\frac{1}{2})
\end{array}
& 4 \, \tilde{\mathcal L}_{a,\OR\theta^{-k}}^{(3)} (l)
\\\hline
\pi(\phi^{(1)},\phi^{(2)},-\sum\limits_{i=1}^2 \phi^{(i)})  &  \OR\theta^{-k} & -   \tilde{I}_a^{\OR\theta^{-k}} & 
\prod\limits_{i=1}^3\frac{\vartheta\targ{\alpha}{\beta}}
{\vartheta\targ{1/2}{1/2}} (\phi^{(i)},2il-\frac{1}{2})
& -\frac{2}{\pi} \sum\limits_{i=1}^3 \frac{\vartheta_1'}{\vartheta_1}(\phi^{(i)},2il-\frac{1}{2})
\\\hline
\end{array}
}{Mobius-Amplitudes-thresholds}{M\"obius strip amplitudes contributing to the gauge  thresholds. The notation is the same as 
in table~\protect\ref{tab:Annulus-Amplitudes-thresholds}, and up to the factor $N_a$ which is absent for an $SU(N_a)$ gauge threshold, the prefactors are identical to those in table~\protect\ref{tab:tadosc2}.}

%%%%%%%%%%%%%%%%%%%%%%%%%%%%%%%%%%%%%%%%%%%%%%%%%

%%%%%%%%%%%%%%%%%%%%%%%%%%%%%%%%%%%%%%%%%%%%%%%%%%%%%%%%%%%%%%%%%%%%%%%
\mathsidetabfix{
\begin{array}{|c||c||c|c|c|} \hline
                                % "Uberschrift
\multicolumn{5}{|c|}{\rule[-3mm]{0mm}{8mm}
\text{\bf Gauge threshold contributions to $SU(N_a)$ per sector}}
\\\hline\hline
{\rm Annulus:} \frac{\rm Angle}{\pi} &  \rotatebox{90}{\!\!\!{\rm insertion}} 
& {\rm tadpole} & \ln \left(\frac{M_{\rm string}}{\mu}  \right)^2
& \Delta
\\\hline\hline
(0,0,0) & \unity &  - & - & -
\\\hline
& \Z_2 & -\frac{ I^{\Z_2,(1 \cdot 3)}_{ab} N_b \, V_{ab}^{(2)}}{2}  & 
\begin{array}{c}
\!\!\!\!
(-1)^{\tau_0^{ab}+1} \, 2 N_b \,\prod_{i=1}^3  \delta_{\sigma^i_{ab},0} \, \delta_{\tau^i_{ab},0} 
\!\!\!\!
\\
= -\frac{ I^{\Z_2,(1 \cdot 3)}_{ab} N_b \delta_{\sigma^2_{ab},0} \delta_{\tau^2_{ab},0}}{2}
\end{array}
&
%%%%%%%%%%%%%%%%%%%%%%%%%
\begin{array}{c}
\frac{ I^{\Z_2,(1 \cdot 3)}_{ab} N_b}{2} \,\delta_{\sigma^2_{ab},0} \delta_{\tau^2_{ab},0} \, \ln \left( 2 \pi v_2 V^{(
2)} \,  \eta^4 (i v_2) \right)
\\
+ \frac{ I^{\Z_2,(1 \cdot 3)}_{ab}  N_b}{2} \,\left(1-\delta_{\sigma^2_{ab},0} \delta_{\tau^2_{ab},0} \right) \,\ln\left|
e^{-\pi (\sigma^2_{ab})^2 v_2/4}\frac{\vartheta_1 (\frac{\tau^2_{ab}}{2}-i\frac{\sigma^2_{ab}}{2} \, v_2,
i v_2)}{\eta (i v_2)}\right|^2
\end{array}
%%%%%%%%%%%%%%%%%%%%%%%%%
\\\hline\hline
(\phi,-\phi,0) & \unity & -\frac{N_b}{2} \, V_{ab}^{(3)} I_{ab}^{(1 \cdot 2)}
& 
 -\frac{N_b}{2} \,\delta_{\sigma^3_{ab},0} \, \delta_{\tau^3_{ab},0} \, I_{ab}^{(1 \cdot 2)}
&
%%%%%%%%%%%%%%%%%%%%%%%%%
\begin{array}{c}
\frac{N_b}{2} \, I_{ab}^{(1 \cdot 2)} \, \delta_{\sigma^3_{ab},0} \delta_{\tau^3_{ab},0} \, \ln \left( 2 \pi v_3 V^{(3)
} \,  \eta^4 (i v_3) \right)
\\
+ \frac{N_b}{2} \,  I_{ab}^{(1 \cdot 2)} \, \left(1-\delta_{\sigma^3_{ab},0} \delta_{\tau^3_{ab},0} \right) \,\ln\left|
e^{-\pi (\sigma^3_{ab})^2 v_3/4}\frac{\vartheta_1 (\frac{\tau^3_{ab}}{2}-i\frac{\sigma^3_{ab}}{2} v_3,
i v_3)}{\eta (i v_3)}\right|^2
\end{array}
%%%%%%%%%%%%%%%%%%%%%%%%%
\\\hline
(\phi,0,-\phi) & \Z_2 & -\frac{N_b}{2}V_{ab}^{(2)}  I_{ab}^{\Z_2,(1\cdot 3)} 
& -\frac{N_b}{2}\delta_{\sigma^2_{ab},0} \, \delta_{\tau^2_{ab},0} \,  I_{ab}^{\Z_2,(1\cdot 3)} 
&
%%%%%%%%%%%%%%%%%%%%%%%%%
\begin{array}{c}
\frac{N_b}{2}I_{ab}^{\Z_2,(1\cdot 3)} \, \delta_{\sigma^2_{ab},0} \delta_{\tau^2_{ab},0} \, \ln \left( 2 \pi v_2 V^{(2)
} \,  \eta^4 (i v_2) \right)
\\
+\frac{N_b}{2}I_{ab}^{\Z_2,(1\cdot 3)} \, \left(1-\delta_{\sigma^2_{ab},0} \delta_{\tau^2_{ab},0} \right) \,\ln\left|
e^{-\pi (\sigma^2_{ab})^2 v_2/4}\frac{\vartheta_1 (\frac{\tau^2_{ab}}{2}-i\frac{\sigma^2_{ab}}{2} v_2,
i v_2)}{\eta (i v_2)}\right|^2
\end{array}
%%%%%%%%%%%%%%%%%%%%%%%%%
\\\hline
(0,\phi,-\phi) & \Z_2 & -\frac{N_b V_{ab}^{(2)} I_{ab}^{\Z_2,(1 \cdot 3)}}{2} 
& - & \frac{N_b I_{ab}^{\Z_2}}{2} \ln (2) \left( \sgn(\phi) - 2 \, \phi\right) 
\\\hline\hline
%%%%%%%%%%%%%%%%%%%%%%%%%
\!\!(\phi^{(1)},\phi^{(2)},-\sum_{k=1}^2\phi^{(k)})\!\! & \unity &  
-\frac{N_b}{2} \, \sum_{i=1}^3 V_{ab}^{(i)} I_{ab}^{(j\cdot k)}
& -\frac{N_b}{2} \, I_{ab} \frac{\sum_{i=1}^3 \sgn(\phi^{(i)}_{ab})}{2} 
& \frac{N_b}{2} \, \frac{I_{ab}}{2} \sum_{i=1}^3 \ln \left(\frac{\Gamma(|\phi^{(i)}|)}{\Gamma(1-|\phi
^{(i)}|)}\right)^{\sgn(\phi^{(i)}_{ab})}
\\\hline
& \Z_2 & -\frac{N_b}{2} \, V_{ab}^{(2)} I_{ab}^{\Z_2,(1 \cdot 3)} 
& -\frac{N_b}{2} \,I_{ab}^{\Z_2} \frac{\sum_{i=1}^3\sgn(\phi^{(i)}_{ab})}{2}
& 
\begin{array}{c}
\frac{N_b}{2} \, \frac{I_{ab}^{\Z_2}}{2} \sum_{i=1}^3 \ln \left(\frac{\Gamma(|\phi^{(i)}|)}{\Gamma(1-
|\phi^{(i)}|)}\right)^{\sgn(\phi^{(i)}_{ab})}
\\
- \frac{N_b}{2} \,I_{ab}^{\Z_2} \ln (2) \sum_{j=1,3} \left(\sgn(\phi^{(j)}_{ab}) -2 \phi^{(j)}  \right)
\end{array}
\\\hline\hline\hline
%%%%%%%%%%%%%%%%%%%%%%%%%%%%%%%%%%%%%%%%%%%%%%%%%%%%%%%%%%%%%%%
%%%%%%%%%%%%%%%%%%%%%%%%%%%%%%%%%%%%%%%%%%%%%%%%%%%%%%%%%%%%%%%
\text{M\"obius:} \frac{\rm Angle}{\pi} & \multicolumn{4}{c|}{} 
\\\hline
(0,0,0) & \OR\theta^{-m} &  - & - & -
\\\hline
(\phi,-\phi,0)  & \OR\theta^{-m} & 2 \, \tilde{V}_{a\OR\theta^{-m}}^{(3)}  \tilde{I}_a^{\OR\theta^
{-m},(1\cdot 2)}
& \delta_{\sigma^3_{aa'},0} \, \delta_{\tau^3_{aa'},0} \,  \tilde{I}_a^{\OR\theta^{-m},(1\cdot 2)}
&
%%%%%%%%%%%%%%%%%%%%%%%%%
\!\!\!\!\!
\begin{array}{c}
- \tilde{I}_a^{\OR\theta^{-m},(1\cdot 2)}\, \delta_{\sigma^3_{aa'},0} \delta_{\tau^3_{aa'},0} \, \ln \left( 4 \pi \tilde{v}_3 \tilde{V}^{(3)} \,  \eta^4 (i \tilde{v}_3) \right)
\\
-\tilde{I}_a^{\OR\theta^{-m},(1\cdot 2)}\, \left(1-\delta_{\sigma^3_{aa'},0} \delta_{\tau^3_{aa'},0} \right) \,\ln\left| e^{-\pi(\sigma^3_{aa'})^2 \tilde{v}_3/4}\frac{\vartheta_1 (\frac{\tau^3_{aa'}}{2}-i\frac{\sigma^3_{aa'}}{2} \tilde{v}_3,i \tilde{v}_3)}{\eta (i \tilde{v}_3)}\right|^2
\end{array}
%%%%%%%%%%%%%%%%%%%%%%%%%
\\\hline
\!\!\!\!\!
\begin{array}{c}
(\phi^{(1)},\phi^{(2)},-\sum_{i=1}^2 \phi^{(i)})  
\\
0<|\phi^{(i)}|,|\phi^{(j)}| \leq |\phi^{(k)}|<1
\\
\sgn(\phi^{(i)}) = \sgn(\phi^{(j)})\\
 \neq \sgn(\phi^{(k)})
\end{array}
\!\!\!\!\!\!%
&  \OR\theta^{-m} 
&  2 \, \sum_{i=1}^{3} \tilde{V}_{a,\OR\theta^{-m}}^{(i)} \tilde{I}_a^{\OR\theta^{-m},(j \cdot k)}
&  \tilde{I}_a^{\OR\theta^{-m}} \left[ H( |\phi^{(k)}|-\frac{1}{2}) -\frac{1}{2} \right] \cdot \sgn(\phi^{(k)})
&
\!\!\!\!\!\!\!
\begin{array}{c}
\tilde{I}_a^{\OR\theta^{-m}} \, \ln(2) \, \sgn(\phi_k) \cdot \left[H( |\phi^{(k)}|-\frac{1}{2}) +\frac{1}{2}  \right]
\\
-\frac{\tilde{I}_a^{\OR\theta^{-m}}}{2} \sum_{n=1}^3 \ln  \left(
\frac{\Gamma(|\phi^{(n)}|)}{\Gamma(1-|\phi^{(n)}|)}\right)^{\sgn(\phi^{(n)})}
\\
-\frac{\tilde{I}_a^{\OR\theta^{-m}}}{2} \sum_{n=1}^3 
\ln\left(\frac{\Gamma(\phi^{(n)}+\frac{1}{2} - \sgn(\phi^{(n)}) \cdot H( |\phi^{(n)}|-\frac{1}{2}))}{\Gamma(\frac{1}{2}-\phi^{(n)} + \sgn(\phi^{(n)}) \cdot H ( |\phi^{(n)}|-\frac{1}{2}) ) }
\right)
\end{array}
%\\%%%%%%%%%%%%%%%%%%%%%%%%%%%
\\\hline
    \end{array}
}{Amplitudes-thresholds}{
Contributions per sector to the threshold calculation. The first column contains the relative angles between two branes or branes and orientifold planes,
the second one specifies if a untwisted or twisted contribution is computed.
The third column contains the prefactors of the tadpole term $\int_{0}^{\infty} dl$, the fourth
the contribution to the beta function coefficient due to massless string states and the last one the gauge threshold correction that arises from massive 
strings.
}

%%%%%%%%%%%%%%%%%%%%
\clearpage
%%%%%%%%%%%%%%%%%%%%
%%%%%%%%%%%%%%%%%%%%%%%%%%%%
\addcontentsline{toc}{section}{References}	 
\bibliographystyle{utphys}	 
\bibliography{refs_threshold}	 

\end{document}